\documentclass[aps,amsmath]{revtex4}
\usepackage{graphicx}  
\usepackage{dcolumn}   
\usepackage{amssymb}   
\usepackage{rotating}  
\usepackage{appendix}
\usepackage{color}
\def\MET{{\mbox{$E\kern-0.57em\raise0.19ex\hbox{/}_{T}$}}}
\def\met{{\mbox{$E\kern-0.57em\raise0.19ex\hbox{/}_{T}$}}}
\def\DZ{D0 }
\def\DZero{D0 }
\def\Dzero{D0 }

\def\ifb{~fb$^{-1}$}
\def\pp{$p\bar{p}$}
\def\bb{$b\bar{b}$}
\def\cc{$c\bar{c}$}
\def\ttbar{$t\bar{t}$}
\def\WH{$WH\rightarrow \ell\nu b\bar{b}$}

\def\lmet{$WH\rightarrow \ell\kern-0.45em\raise0.19ex\hbox{/} \nu b\bar{b}$}
\def\ZH{$ZH\rightarrow \nu\bar{\nu} b\bar{b}$}
\def\ZHll{$ZH\rightarrow \ell^+ \ell^- b\bar{b}$}

\def\vww{$VH \rightarrow \ell^{\pm}\ell'^{\pm} + X$}
\def\vhe{$VH \rightarrow e^{\pm} \nu_e \mu^{\pm} \nu_{\mu} + X$}

\def\hww{$H\rightarrow W^+ W^-$}
\def\hbb{$H\rightarrow b\bar{b}$}

\def\tevE{$\sqrt{s}=1.96$~TeV}

\begin{document}

\rightline{FERMILAB-CONF-12-065-E}
\rightline{CDF Note 10806}
\rightline{\DZ Note 6303}
\vskip0.3in

\title{Combined CDF and \DZ Search for Standard Model Higgs Boson Production with up to 10.0 fb$^{-1}$ of Data\\[1.5cm]}

\author{The TEVNPH Working Group\footnote{The Tevatron
New-Phenomena and Higgs Working Group can be contacted at
TEVNPHWG@fnal.gov. More information can be found at http://tevnphwg.fnal.gov/.}
 }
\affiliation{\vskip0.3cm for the CDF and \DZ Collaborations\\ \vskip0.2cm
\today}
\begin{abstract}
\vskip0.3in
We combine results from CDF and D0 on direct searches for the standard model (SM)
Higgs boson ($H$) in \pp~collisions at the Fermilab Tevatron at $\sqrt{s}=1.96$~TeV.
Compared to the previous Tevatron Higgs boson search combination more data have been 
added, additional channels have been incorporated, and some previously used channels
have been reanalyzed to gain sensitivity.  With up to 10\ifb\ of luminosity analyzed, 
the 95\% C.L. median expected upper limits on Higgs boson production are factors of 
0.94, 1.10, and 0.49 times the values of the SM cross section for Higgs bosons of 
mass $m_{H}=$115~GeV/$c^2$, 125~GeV/$c^2$,and 165~GeV/$c^2$, respectively.  We 
exclude, at the 95\% C.L., two regions: $100<m_H<106$~GeV/$c^2$, and  $147<m_{H}<179$~GeV/$c^{2}$.   
We expect to exclude the regions $100<m_{H}<119$~GeV/$c^{2}$ 
and $141<m_{H}<184$~GeV/$c^{2}$.   There is an excess of data events with respect 
to the background estimation in the mass range $115<m_{H}<135$~GeV/$c^{2}$ which 
causes our limits to not be as stringent as expected.  At $m_H=120$~GeV/$c^{2}$, 
the $p$-value for a background fluctuation to produce this excess is $\sim$3.5$\times$10$^{-3}$, 
corresponding to a local significance of 2.7 standard deviations.  The global significance for 
such an excess anywhere in the full mass range is approximately 2.2 standard deviations.  We 
also combine separately searches for $H \to b\bar{b}$ and $H \to W^+W^-$, and find 
that the excess is concentrated in the $H\to b\bar{b}$ channel, although the results 
in the $H\to W^+W^-$ channel are also consistent with the possible presence of a low-mass 
Higgs boson.  

\vskip 1cm
\center {\em Preliminary Results}
\end{abstract}

\maketitle

\newpage
\section{Introduction} 

Understanding the mechanism for electroweak symmetry breaking, specifically by 
testing for the presence or absence of the standard model (SM) Higgs boson,
has been a major goal of particle physics for many years, and is a central part of the
Fermilab Tevatron physics program. Both the CDF and \Dzero collaborations
have performed new combinations~\cite{CDFHiggs,DZHiggs} of multiple
direct searches for the SM Higgs boson.  The new searches include more
data, additional channels, and improved analysis techniques compared to 
previous analyses. Precision electroweak data, including the recently updated measurements of the $W$-boson mass from
the CDF and \Dzero Collaborations~\cite{CDFMW,DZMW}, yield an
indirect constraint on the allowed mass of the Higgs boson, $m_H < 152$~GeV/$c^2$~\cite{lepewwg}, at
95\% confidence level (C.L.).
The Large
Electron Positron Collider (LEP) has excluded Higgs
boson masses below 114.4~GeV/$c^2$~\cite{lep}, and the LHC experiments, 
ATLAS and CMS, now limit the SM Higgs boson to have a mass between 115.5 and 127
GeV/$c^2$~\cite{atlas,cms} at the 95\% C.L.  Both LHC experiments report local
$\sim 3$ standard deviation (s.d.) excesses at approximately 125 GeV/$c^2$. 
The sensitivities of the new combinations presented here significantly exceeds 
those of previous Tevatron combinations~\cite{prevhiggs,WWPRLhiggs}, 
providing sensitivity within the allowed Higgs boson mass range. 

In this note, we combine the most recent results of all such
searches in \pp~collisions at~\tevE\ in the Higgs boson mass range from
100--200 GeV/$c^2$.  The analyses combined
here seek signals of Higgs bosons produced in association with
a vector boson ($q\bar{q}\rightarrow W/ZH$), through gluon-gluon
fusion ($gg\rightarrow H$), and through vector boson fusion (VBF)
($q\bar{q}\rightarrow q^{\prime}\bar{q}^{\prime}H$) corresponding
to integrated luminosities up to 10.0\ifb~at CDF and up to
9.7\ifb~at D0.  The Higgs boson decay modes studied are 
$H\rightarrow b{\bar{b}}$, $H\rightarrow W^+W^-$, $H\rightarrow ZZ$, 
$H\rightarrow\tau^+\tau^-$ and $H\rightarrow \gamma\gamma$. 
For Higgs boson masses greater than 125 GeV$/c^2$, $H\rightarrow W^+W^-$ modes with 
leptonic decay provide the greatest sensitivity~\cite{keung,glover,hhg,dittmar}, while below 125 GeV$/c^2$ sensitivity comes 
mainly from ($q\bar{q}\rightarrow W/ZH$) where $H$ decays to $b\bar{b}$ and 
the $W$ or $Z$ decays leptonically~\cite{glashow,marciano,hhg}. The dominant decay mode for 
a low mass Higgs boson is $H \to b\bar{b}$, and thus measurements of this process 
provide constraints on possible Higgs boson phenomenology that are complementary to those provided by the LHC.

To simplify the combination, the searches are separated into 
mutually exclusive final states referred to
as ``analysis sub-channels'' in this note.  Listings of these analysis sub-channels
are provided in Tables~\ref{tab:cdfacc} and~\ref{tab:dzacc}.  The selection
procedures for each analysis are detailed in Refs.~\cite{cdfWH}
through~\cite{dzHgg}, and are briefly described below.

\section{Acceptance, Backgrounds, and Luminosity}  

Event selections are similar for the corresponding CDF and D0 analyses, 
consisting typically of a preselection followed by the use of a multivariate 
analysis technique with a final discriminating variable to separate signal 
and background.  For the case of \WH, an isolated lepton ($\ell=$ electron 
or muon) and two or three jets are required, with one or more of the jets 
being $b$-tagged, i.e., identified as containing a weakly-decaying $b$ 
hadron.  Selected events must also display a significant imbalance in 
transverse momentum (referred to as missing transverse energy or \met).  
Events with more than one isolated lepton are rejected.


For the D0 \WH\ analyses, the data are split by lepton type and jet multiplicity (two or three jet
sub-channels), and on the number of $b$-tagged jets.
Orthogonal selections corresponding to events with exactly one tight $b$-tagged jet (TST),
 exactly two loose but not tight $b$-tagged jets (LDT) and exactly two tight $b$-tagged jets (TDT) are made. 
 Every event is placed into one of these
   mutually exclusive categories. 
As with other D0 analyses targeting the \hbb\ decay,  
a boosted decision tree based $b$-tagging algorithm, which builds and improves upon the previous neural network $b$-tagger~\cite{Abazov:2010ab}, is used. For example, the loose $b$-tagging criterion corresponds to an identification efficiency of $\approx 80\%$ for true $b$-jets for a mis-identification rate of $\approx 10\%$. 
The outputs of boosted decision trees, trained
separately for each sample (i.e. jet multiplicity, lepton flavor and $b$-tag category) and for
each Higgs boson mass, are used as the final discriminating variables.


For the CDF \WH\ analyses, events are analyzed in two and three jet
sub-channels separately, and in each of these samples the events
are grouped into various lepton and $b$-tag categories.  Events are
broken into separate analysis categories based on the quality of the
identified lepton.  Separate categories are used for events with a
high quality muon or central electron candidate, an isolated track
or identified loose muon in the extended muon coverage, a forward
electron candidate, and a loose central electron or isolated track
candidate.  The final two lepton categories, which provide some 
acceptance for lower quality electrons and single prong tau decays, 
are used only in the case of two-jet events. Within the lepton 
categories there are five $b$-tagging categories considered for 
two-jet events: two tight $b$-tags (TT), one tight $b$-tag and one 
loose $b$-tag (TL), a single tight $b$-tag (Tx), two loose $b$-tags
(LL), and a single loose $b$-tag.  For three jet categories only 
the TT and TL $b$-tagging categories are considered.  The tight and 
loose $b$-tag definitions are taken for the first time from a neural 
network tagging algorithm~\cite{cdfHobit} based on sets of kinematic 
variables sensitive to displaced decay vertices and tracks within 
jets with large transverse impact parameters relative to the 
hard-scatter vertices.  Using an operating point which gives an 
equivalent rate of false tags, the new algorithm improves upon 
previous $b$-tagging efficiencies by $\sim$20\%.  A Bayesian neural 
network discriminant is trained at each Higgs boson mass within the 
test range for each of the specific categories (defined by lepton 
type, $b$-tagging type, and number of jets) to separate signal from
backgrounds.   

For the \ZH\ analyses, the selection is similar to the $WH$ selection,
except all events with isolated leptons are rejected and stronger multijet
background suppression techniques are applied.  Both the CDF and D0 analyses
use a track-based missing transverse momentum calculation as a discriminant
against false \met. In addition both CDF and D0 utilize multivariate
techniques, a boosted decision tree at D0 and a neural network at CDF, to
further discriminate against the multijet background before $b$-tagging.
There is a sizable fraction of the \WH\ signal in which the lepton is
undetected that is selected in the \ZH\ samples,  so these analyses are
also referred to as $VH \rightarrow \met b \bar{b}$.  The CDF analysis
uses three non-overlapping categories of $b$-tagged events (SS, SJ and 1S).
These categories are based on two older CDF $b$-tagging algorithms, an 
algorithm for reconstructing displaced, secondary vertices of $b$-quark 
decays (S) and an algorithm for assigning a likelihood for tracks within 
a jet to have originated from a displaced vertex (J).  The D0 analysis 
requires exactly two jets. The $b$-tagging criteria have been re-optimized 
to reduce the loss in sensitivity due to systematic uncertainties. The 
$b-$tagger output values for each of the two jets are added to form an event 
$b$-tag, the value of which is used to define two high purity samples: the 
medium $b$-tag sample (MS) and the tight $b$ tag-sample (TS). After applying 
a multijet veto, these samples have a signal-to-background ratio of 0.3\% 
and 1.5\% respectively. Boosted decision trees, trained separately for the 
different $b$-tagging categories and at each test mass, are used as the 
final discriminant. Overall, the sensitivity has been improved by $\approx$ 
25\% with respect to the previous result.  The CDF analysis uses a second
layer of neural network discriminants for separating signal from backgrounds.
 

The \ZHll\ analyses require two isolated leptons and at least two jets.
D0's \ZHll\ analyses separate events into non-overlapping samples of
events with either one tight $b$-tag (TST) 
or one tight and one loose $b$-tag (TLDT).
 CDF has incorporated its new neural network 
$b$-tagging algorithm in this analysis and uses four out of the five
$WH$ tagging categories (TT, TL, Tx, and LL).  CDF now also separates
events with two or three jets into independent analysis channels.  To increase 
signal acceptance D0 loosens the selection criteria for one of the 
leptons to include an isolated track not reconstructed in the muon 
detector ($\mu\mu_{trk}$) or an electron from the inter-cryostat region 
of the D0 detector ($ee_{ICR}$).  Combined with the dielectron ($ee$) and 
dimuon ($\mu\mu$) analyses, these provide four orthogonal analyses, 
and each uses 9.7 fb$^{-1}$ of data in this combination. CDF uses 
neural networks to select loose dielectron and dimuon candidates.
D0 applies a kinematic fit to optimize reconstruction, while CDF 
corrects jet energies for \met\ using a neural network approach.  D0 
uses random forests of decision trees to provide the final variables 
for setting limits.  CDF utilizes a multi-layer discriminant based 
on neural networks where separate discriminant functions are used to 
define four separate regions of the final discriminant function.

For the \hww~analyses, signal events are characterized by large \met~and
two opposite-signed, isolated leptons.  The presence of neutrinos in the
final state prevents the accurate reconstruction of the candidate Higgs
boson mass.
D0 selects events containing electrons and/or muons, dividing the data sample
into three final states: $e^+e^-$, $e^\pm \mu^\mp$, and $\mu^+\mu^-$. Each final state is further
subdivided according to the number of jets in the event: 0, 1, or 2 or more (``2+'') jets.
The dimuon and dielectron channels use boosted decision trees to reduce the dominant Drell-Yan background. 
Decays involving tau
leptons are included in two orthogonal ways. A dedicated
analysis ($\mu\tau_{\rm{had}}$) using 7.3 fb$^{-1}$
of data studying the final state involving a muon and a hadronic tau decay plus up to one jet is included in the
Tevatron combination. Final states involving
other tau decays and mis-identified hadronic tau decays are included in the $e^+e^-$, $e^\pm \mu^\mp$,
and $\mu^+\mu^-$ final state analyses.
CDF separates the \hww\ events in five non-overlapping samples, split
into ``high $s/b$'' and ``low $s/b$'' categories defined by lepton
types and the number of reconstructed jets: 0, 1, or 2+ jets.  The sample
with two or more jets is not split into low $s/b$ and high $s/b$ lepton
categories due to the smaller statistics in this channel.  A sixth CDF
channel is the low dilepton mass ($m_{\ell^+\ell^-}$) channel, which
accepts events with $m_{\ell^+\ell^-}<16$~GeV$/c^2$.  CDF has further improved 
its analysis of the low dilepton mass channel by reducing the $\Delta R$
cut applied to dilepton pairs down to 0.1, which increases Higgs signal
acceptance in this channel $\sim$10\%.   

The division of events into categories based on the number of reconstructed
jets allows the analysis discriminants to separate differing contributions 
of signal and background processes more effectively.  The signal production 
mechanisms considered are $gg\rightarrow H\rightarrow W^+W^-$, 
$WH/ZH\rightarrow jjW^+W^-$, and vector-boson fusion.  The relative fractions 
of the contributions from each of the three signal processes and background 
processes, notably $W^+W^-$ production and $t{\bar{t}}$ production, are very 
different in the different jet categories.  Dividing our data into these
categories provides more statistical discrimination, but introduces the need 
to evaluate the systematic uncertainties carefully in each jet category.  A 
discussion of these uncertainties is found in Section~\ref{sec:signal}.

The D0 $e^+e^-$, $e^\pm \mu^\mp$, and $\mu^+\mu^-$ final state channels use
boosted decision trees as the final discriminants; for categories with non-zero 
jet multiplicity $b$-tagging information is included. The $\mu\tau_{\rm{had}}$ 
channel uses neural networks as the final discriminant.  CDF uses neural-network 
outputs, including likelihoods constructed from calculated matrix-element 
probabilities as additional inputs for the 0-jet bin.

D0 includes a \vww\ analysis in which the associated vector boson and the $W$ boson 
from the Higgs boson decay are required to decay leptonically, giving like-sign 
dilepton final states.  Previously the three final $e^\pm e^\pm$, $e^\pm \mu^\pm$, 
and $\mu^{\pm}\mu^{\pm}$ were considered. In this combination, only the most sensitive 
$e^\pm \mu^\pm$ final state is included.  The combined output of two 
decision trees, trained against the instrumental and diboson backgrounds respectively,
is used as the final discriminant. For the first time however, D0 includes tri-lepton 
analyses to increase the sensitivity to associated production and other decay modes, 
such as $H \rightarrow ZZ$. The $ee\mu$, $\mu \mu e$ and $\tau\tau\mu$ final states 
are considered. The $ee\mu$ and $\mu \mu e$ final states use boosted decision trees 
as the final discriminants.  The $\tau\tau\mu$ final states are sub-divided according 
to the jet multiplicity to improve the sensitivity and a kinematic variable based on 
the event $P_T$ used as the discriminating variable.

CDF also includes a separate analysis of events with same-sign leptons to incorporate 
additional potential signal from associated production events in which the two leptons 
(one from the associated vector boson and one from a $W$ boson produced in the Higgs 
boson decay) have the same charge.  CDF additionally incorporates three tri-lepton 
channels to include additional associated production contributions in which leptons 
result from the associated $W$ boson and the two $W$ bosons produced in the Higgs 
boson decay or where an associated $Z$ boson decays into a dilepton pair and a third 
lepton is produced in the decay of either of the $W$ bosons resulting from the Higgs 
boson decay.  In the latter case, CDF separates the sample into one jet and two or
more jet sub-channels to take advantage of the fact that the Higgs boson candidate
mass can be reconstructed from the invariant mass of the two jets, the lepton, and
the missing transverse energy.  CDF also includes for the first time a new tri-lepton 
channel focusing on $WH$ production in which one of the three leptons is reconstructed 
as a hadronic tau.

CDF includes a search for $H \rightarrow ZZ$ using four lepton events.  In addition 
to the simple four-lepton invariant mass discriminant used previously for separating 
potential Higgs boson signal events from the non-resonant $ZZ$ background, the \met\
in these events is now used as a second discriminating variable to better identify 
four lepton signal contributions from $ZH \rightarrow ZWW$ and $ZH \rightarrow Z\tau
\tau$ production.  CDF has also updated its opposite-sign channels in which one of 
the two lepton candidates is a hadronic tau.  Events are separated into $e$-$\tau$
and $\mu$-$\tau$ channels.  The final discriminants are obtained from boosted decision 
trees which incorporate both hadronic tau identification and kinematic event variables 
as inputs.  

D0 also includes channels in which one of the $W$
bosons in the $H \rightarrow W^+W^-$ process decays leptonically and the other
decays hadronically.  Electron and muon final states are studied separately.
Random forests are used for the final discriminants. 

CDF includes an updated, generic analysis searching for Higgs bosons decaying
to tau lepton pairs incorporating contributions from direct $gg \rightarrow H$
production, associated $WH$ or $ZH$ production, and vector boson fusion production.
CDF also includes an analysis of events that contain one or more reconstructed 
leptons ($\ell$ = $e$ or $\mu$) in addition to a tau lepton pair focusing on 
associated production where $H \rightarrow \tau \tau$ and additional leptons 
are produced in the decay of the $W$ or $Z$ boson.  For these searches multiple 
Support Vector Machine (SVM)~\cite{svm} classifiers are obtained using separate 
trainings for the signal against each of the primary backgrounds.  In the generic 
search, events with either one or two jets are separated into two independent 
analysis channels.  The final discriminant for setting limits is obtained using 
the minimum score of four SVM classifiers obtained from trainings against the 
primary backgrounds ($Z \rightarrow \tau \tau$, $t \bar{t}$, multi-jet, and 
$W$+jet production).  In the extended analysis events are separated into five 
separate analysis channels ($\ell \ell \ell$, $e \mu \tau_{\rm{had}}$, $\ell 
\ell \tau_{\rm{had}}$, $\ell \tau_{\rm{had}} \tau_{\rm{had}}$, and $\ell \ell 
\ell \ell$). The four lepton category includes $\tau_{\rm{had}}$ candidates.  The 
final discriminants are likelihoods based on outputs obtained from independent 
SVM trainings against each of the primary backgrounds ($Z$+jets, $t\bar{t}$, and 
dibosons).  These channels are included in the combination only for lower Higgs 
masses to avoid overlap with other search channels.

The D0 $\ell^\pm \tau^{\mp}_{\rm{had}}jj$ analyses likewise include direct $gg 
\rightarrow H$ production, associated $WH$ or $ZH$ production, and vector boson fusion 
production.  Decays of the Higgs boson to tau, $W$ and $Z$ boson pairs are 
considered. A final state consisting of one leptonic tau decay, one hadronic 
tau decay and two jets is required. Both muonic and electronic sub-channels 
are considered. Recent improvements include increased trigger efficiencies. The 
output of boosted decision trees is used as the final discriminant.

CDF incorporates an updated all-hadronic analysis based on the older CDF 
$b$-tagging algorithms, which results in two sub-channels (SS and SJ). Both 
$WH/ZH$ and VBF production contribute to the $jjb{\bar{b}}$ final state.  
Events with either four or five reconstructed jets are selected, and at least 
two must be $b$-tagged.  The large QCD multijet backgrounds are modeled from 
the data by applying a measured mistag probability to the non $b$-tagged jets 
in events containing a single $b$-tag.  Neural network discriminants based on 
kinematic event variables including those designed to separate quark and gluon 
jets are used to obtain the final limits.

D0 and CDF both contribute analyses searching for Higgs bosons decaying into
diphoton pairs.  The CDF analysis looks for a signal peak in the diphoton
invariant mass spectrum above the smooth background originating from QCD production.  
Events are separated into four independent analysis channels 
based on the photon candidates contained within the event: two central candidates 
(CC), one central and one plug candidate (CP), one central and one central 
conversion candidate (C$^\prime$C), or one plug and one central conversion candidate 
(C$^\prime$P).  In the D0 analysis the contribution of jets misidentified as photons 
is reduced by combining information sensitive to differences in the energy 
deposition from these particles in the tracker, calorimeter and central preshower 
in a neural network (ONN). The output of boosted decision trees, rather than the 
diphoton invariant mass, is used as the final discriminating variable. Previously, 
the transverse energies of the leading two photons along with the azimuthal 
opening angle between them and the diphoton invariant mass and transverse momentum 
were used as input variables. Additional variables, including the ONN output value 
for the two photons have been included, resulting in a sizeable improvement in
sensitivity of $\approx 20\%$.

CDF incorporates three non-overlapping sets of analysis channels searching for 
the process $t \bar{t} H \rightarrow t \bar{t} b \bar{b}$.  One set of channels 
selects events with a reconstructed lepton, large missing transverse energy, and 
four or more reconstructed jets.  Events containing four, five, and six or more 
jets are are analyzed separately and further sub-divided into five $b$-tagging 
categories based on the older CDF tagging algorithms (three tight $b$-tags (SSS), 
two tight and one loose $b$-tags (SSJ), one tight and two loose $b$-tags (SJJ), 
two tight $b$-tags (SS), and one tight and one loose $b$-tags (SJ)).  Neural network 
discriminants trained at each mass point are used to set limits.  A second set of 
channels selects events with no reconstructed lepton.  These events are separated 
into two categories, one containing events with large missing transverse energy 
and five to nine reconstructed jets and another containing events with low missing 
transverse energy and seven to ten reconstructed jets.  Events in these two channels 
are required to have a minimum of two $b$-tagged jets based on an independent neural 
network tagging algorithm.  Events with three or more $b$-tags are analyzed in 
separate channels from those with exactly two tags.  Two stages of neural network 
discriminants are used (the first to help reject large multijet backgrounds and the 
second to separate potential $t\bar{t}H$ signal events from $t\bar{t}$ background 
events).

For both CDF and D0, events from QCD multijet (instrumental) backgrounds are
typically measured in independent data samples using several different methods.
For CDF, backgrounds from SM processes with electroweak gauge bosons or top
quarks were generated using \textsc{PYTHIA}~\cite{pythia}, \textsc{ALPGEN}~\cite{alpgen},
\textsc{MC@NLO}~\cite{MC@NLO}, and \textsc{HERWIG}~\cite{herwig} programs.
For D0, these backgrounds were generated using \textsc{PYTHIA}, \textsc{ALPGEN},
and \textsc{COMPHEP}~\cite{comphep}, with \textsc{PYTHIA} providing parton-showering
and hadronization for all the generators.  These background processes were normalized
using either experimental data or next-to-leading order calculations (including
\textsc{MCFM}~\cite{mcfm} for the $W+$ heavy flavor process).  All Monte Carlo samples
are passed through detailed GEANT-based simulations~\cite{geant} of the CDF and D0 detectors.

Tables~\ref{tab:cdfacc} and~\ref{tab:dzacc} summarize, for CDF and D0 respectively,
the integrated luminosities, the Higgs boson mass ranges over which the searches are performed,
and references to further details for each analysis.


\begin{table}[h]
\caption{\label{tab:cdfacc}Luminosity, explored mass range and references
for the different processes and final states ($\ell$ = $e$ or $\mu$) for 
the CDF analyses.  The generic labels ``$2\times$'', ``$3\times$'', and 
``$4\times$'' refer to separations based on lepton categories.}
\begin{ruledtabular}
\begin{tabular}{lccc} \\
Channel & Luminosity  & $m_H$ range & Reference \\
        & (fb$^{-1}$) & (GeV/$c^2$) &           \\ \hline
$WH\rightarrow \ell\nu b\bar{b}$ 2-jet channels \ \ \ 4$\times$(TT,TL,Tx,LL,Lx)                                                                            & 9.45 & 100-150 & \cite{cdfWH} \\
$WH\rightarrow \ell\nu b\bar{b}$ 3-jet channels \ \ \ 3$\times$(TT,TL)                                                                                     & 9.45 & 100-150 & \cite{cdfWH} \\
$ZH\rightarrow \nu\bar{\nu} b\bar{b}$ \ \ \ (SS,SJ,1S)                                                                                                     & 9.45 & 100-150 & \cite{cdfmetbb} \\
$ZH\rightarrow \ell^+\ell^- b\bar{b}$ 2-jet channels \ \ \ 2$\times$(TT,TL,Tx,LL)                                                                          & 9.45 & 100-150 & \cite{cdfZH} \\
$ZH\rightarrow \ell^+\ell^- b\bar{b}$ 3-jet channels \ \ \ 2$\times$(TT,TL,Tx,LL)                                                                          & 9.45 & 100-150 & \cite{cdfZH} \\
$H\rightarrow W^+ W^-$ \ \ \ 2$\times$(0 jets,1 jet)+(2 or more jets)+(low-$m_{\ell\ell}$)                                                                 & 9.7  & 110-200 & \cite{cdfHWW} \\
$H\rightarrow W^+ W^-$ \ \ \ ($e$-$\tau_{\rm{had}}$)+($\mu$-$\tau_{\rm{had}}$)                                                                             & 9.7  & 130-200 & \cite{cdfHWW2} \\
$WH\rightarrow WW^+ W^-$ \ \ \ (same-sign leptons)+(tri-leptons)                                                                                           & 9.7  & 110-200 & \cite{cdfHWW} \\
$WH\rightarrow WW^+ W^-$ \ \ \ tri-leptons with 1 $\tau_{\rm{had}}$                                                                                        & 9.7  & 130-200 & \cite{cdfHWW2} \\
$ZH\rightarrow ZW^+ W^-$ \ \ \ (tri-leptons with 1 jet)+(tri-leptons with 2 or more jets)                                                                  & 9.7  & 110-200 & \cite{cdfHWW} \\
$H\rightarrow ZZ$ \ \ \ four leptons                                                                                                                       & 9.7  & 120-200 & \cite{cdfHZZ} \\
$H$ + $X\rightarrow \tau^+ \tau^-$ \ \ \ (1 jet)+(2 jets)                                                                                                  & 8.3  & 100-150 & \cite{cdfHtt} \\
$WH \rightarrow \ell \nu \tau^+ \tau^-$/$ZH \rightarrow \ell^+ \ell^- \tau^+ \tau^-$ \ \ \ $\ell$-$\tau_{\rm{had}}$-$\tau_{\rm{had}}$                      & 6.2  & 100-150 & \cite{cdfVHtt} \\
$WH \rightarrow \ell \nu \tau^+ \tau^-$/$ZH \rightarrow \ell^+ \ell^- \tau^+ \tau^-$ \ \ \ ($\ell$-$\ell$-$\tau_{\rm{had}}$)+($e$-$\mu$-$\tau_{\rm{had}}$) & 6.2  & 100-125 & \cite{cdfVHtt} \\
$WH \rightarrow \ell \nu \tau^+ \tau^-$/$ZH \rightarrow \ell^+ \ell^- \tau^+ \tau^-$ \ \ \ $\ell$-$\ell$-$\ell$                                            & 6.2  & 100-105 & \cite{cdfVHtt} \\
$ZH \rightarrow \ell^+ \ell^- \tau^+ \tau^-$ \ \ \ four leptons including $\tau_{\rm{had}}$ candidates                                                     & 6.2  & 100-115 & \cite{cdfVHtt} \\
$WH+ZH\rightarrow jjb{\bar{b}}$ \ \ \  (SS,SJ)                                                                                                             & 9.45 & 100-150 & \cite{cdfjjbb} \\
$H \rightarrow \gamma \gamma$ \ \ \  (CC,CP,CC-Conv,PC-Conv)                                                                                               & 10.0 & 100-150 & \cite{cdfHgg} \\
$t\bar{t}H \rightarrow W W b\bar{b} b\bar{b}$ (lepton) \ \ \  (4jet,5jet,$\ge$6jet)$\times$(SSS,SSJ,SJJ,SS,SJ)                                             & 9.45 & 100-150 & \cite{cdfttHLep} \\
$t\bar{t}H \rightarrow W W b\bar{b} b\bar{b}$ (no lepton) \ \ \  (low met,high met)$\times$(2 tags,3 or more tags)                                         & 5.7  & 100-150 & \cite{cdfttHnoLep} \\
\end{tabular}
\end{ruledtabular}
\end{table}

\vglue 0.5cm

\begin{table}[h]
\caption{\label{tab:dzacc}Luminosity, explored mass range and references
for the different processes
and final states ($\ell=e, \mu$) for the D0 analyses.
}
\begin{ruledtabular}
\begin{tabular}{lccc} \\
Channel & Luminosity  & $m_H$ range & Reference \\
        & (fb$^{-1}$) & (GeV/$c^2$) &           \\ \hline
$WH\rightarrow \ell\nu b\bar{b}$ \ \ \ (TST,LDT,TDT)$\times$(2,3 jet)             & 9.7  & 100-150 & \cite{dzWHl} \\
$ZH\rightarrow \nu\bar{\nu} b\bar{b}$ \ \ \ (MS,TS)   & 9.5  & 100-150 & \cite{dzZHv2} \\
%
$ZH\rightarrow \ell^+\ell^- b\bar{b}$ \ \ \ (TST,TLDT)$\times$($ee$,$\mu\mu$,$ee_{ICR}$,$\mu\mu_{trk}$) & 9.7  & 100-150 & \cite{dzZHll1} \\
%
$H$+$X$$\rightarrow$$ \ell^\pm \tau^{\mp}_{\rm{had}}jj$  \ \ \      & 4.3-6.2  & 105-200 & \cite{dzVHt2} \\
%
$VH \rightarrow e^\pm \mu^\pm + X $ \ \ \  & 9.7  & 115-200 & \cite{dzWWW2} \\
$H\rightarrow W^+ W^- \rightarrow \ell^\pm\nu \ell^\mp\nu$ \ \ \ (0,1,2+ jet)     & 8.6-9.7  & 115-200 & \cite{dzHWW}\\
%
$H\rightarrow W^+ W^- \rightarrow \mu\nu \tau_{\rm{had}}\nu$ \ \ \      & 7.3  & 115-200 & \cite{dzVHt2}\\
$H\rightarrow W^+ W^- \rightarrow \ell\bar{\nu} jj$      & 5.4  & 130-200 & \cite{dzHWWjj}\\
%
$VH \rightarrow \ell\ell\ell + X $ 									       & 9.7  & 100-200 & \cite{dzlll} \\
$VH \rightarrow \tau \tau \mu + X $ 									       & 7.0  & 115-200 & \cite{dzttl} \\
$H \rightarrow \gamma \gamma$                                 & 9.7  & 100-150 & \cite{dzHgg} \\
\end{tabular}
\end{ruledtabular}
\end{table}

\section{Signal Predictions}
\label{sec:signal}

In order to predict the kinematic distributions of Higgs boson signal events, CDF and D0
use the \textsc{PYTHIA}~\cite{pythia} Monte Carlo program, with
\textsc{CTEQ5L} and \textsc{CTEQ6L1}~\cite{cteq} leading-order (LO)
parton distribution functions.    We scale these Monte Carlo predictions to 
the most recent higher-order calculations of inclusive cross sections, and differential
cross sections, such as in the Higgs boson $p_T$ spectrum and the number of associated jets, as described below.
The $gg\rightarrow H$ production cross section we use is
calculated at next-to-next-to leading order (NNLO) in QCD with a next-to-next-to leading log (NNLL)
resummation of soft gluons; the calculation also includes two-loop electroweak effects and
handling of the running $b$ quark mass~\cite{anastasiou,grazzinideflorian}.
The numerical values in Table~\ref{tab:higgsxsec} are updates~\cite{grazziniprivate}
of these predictions with $m_t$ set to 173.1~GeV/$c^2$~\cite{tevtop09},
and with a treatment of the massive top and bottom loop corrections up to
next-to-leading-order (NLO) + next-to-leading-log (NLL) accuracy. The
factorization and renormalization scale choice for this calculation is $\mu_F=\mu_R=m_H$.
These calculations are refinements of the earlier NNLO calculations of the $gg\rightarrow H$
production cross section~\cite{harlanderkilgore2002,anastasioumelnikov2002,ravindran2003}.
Electroweak corrections were computed in Refs.~\cite{actis2008,aglietti}. Soft gluon
resummation was introduced in the prediction of the $gg\rightarrow H$ production cross
section in Ref.~\cite{catani2003}.  The $gg\rightarrow H$ production cross section depends strongly
on the gluon parton density function, and the accompanying value
of $\alpha_s(q^2)$.  The cross sections used here are calculated
with the MSTW~2008 NNLO PDF set~\cite{mstw2008}, as recommended by the PDF4LHC working group~\cite{pdf4lhc}.  The inclusive
Higgs boson production cross sections are
listed in Table~\ref{tab:higgsxsec}.

For analyses that consider inclusive $gg\rightarrow H$ production but do not split it into separate channels
based on the number of reconstructed jets, we use the inclusive uncertainties from the simultaneous variation
of the factorization and renormalization scale up and down by a factor of two.  We use the
prescription of the PDF4LHC working group for evaluating PDF uncertainties on the inclusive production
cross section.  QCD scale uncertainties that affect the cross section via their impacts on the PDFs are included
as a correlated part of the total scale uncertainty.  The remainder of the PDF uncertainty is treated as
uncorrelated with the QCD scale uncertainty.

For analyses seeking $gg\rightarrow H$ production that
divide events into categories based on the number of reconstructed jets, we employ a new
approach for evaluating the impacts of the scale uncertainties.  Following the recommendations of 
Ref.~\cite{bnlaccord,lhcdifferential},
we treat the QCD scale uncertainties obtained from the NNLL inclusive~\cite{grazzinideflorian,anastasiou}, NLO one or
more jets~\cite{anastasiouwebber}, and NLO two or more jets~\cite{campbellh2j}
cross section calculations as uncorrelated with one another.  We then obtain
QCD scale uncertainties for the exclusive $gg\rightarrow H+0$~jet, 1~jet, and 2~or more jet categories
by propagating the uncertainties on the inclusive cross section predictions through the subtractions
needed to predict the exclusive rates.  For example, the $H$+0~jet cross section is obtained by
subtracting the NLO $H+1$~or more jet cross section from the inclusive NNLL+NNLO cross section.
We now assign three separate, uncorrelated scale uncertainties
which lead to correlated and anticorrelated uncertainty contributions between exclusive jet categories.
The procedure in Ref.~\cite{anastasiouwebber} is used
to determine PDF model uncertainties.  These are obtained
separately for each jet bin and treated as 100\% correlated
between jet bins and between D0 and CDF.

The scale choice affects the $p_T$ spectrum of the Higgs boson when
produced in gluon-gluon fusion, and this effect changes the acceptance
of the selection requirements and also the shapes of the distributions
of the final discriminants.  The effect of the acceptance change is
included in the calculations of Ref.~\cite{anastasiouwebber} and
Ref.~\cite{campbellh2j}, as the experimental requirements are simulated
in these calculations. The effects on the final discriminant shapes
are obtained by reweighting the $p_T$ spectrum of the Higgs boson
production in the Monte Carlo simulations to higher-order calculations.
The Monte Carlo signal simulation used by CDF and D0 is provided by
the LO generator {\sc pythia}~\cite{pythia} which includes a parton
shower and fragmentation and hadronization models.   We reweight the
Higgs boson $p_T$ spectra in our {\sc pythia} Monte Carlo samples to
that predicted by {\sc hqt}~\cite{hqt} when making predictions of
differential distributions of $gg\rightarrow H$ signal events. To
evaluate the impact of the scale uncertainty on our differential
spectra, we use the {\sc resbos}~\cite{resbos} generator, and apply
the scale-dependent differences in the Higgs boson $p_T$ spectrum to
the {\sc hqt} prediction, and propagate these to our final
discriminants as a systematic uncertainty on the shape, which is
included in the calculation of the limits.

We include all significant Higgs boson production modes in the high-mass
search.   Besides gluon-gluon fusion through virtual quark loops
(ggH), we include Higgs boson production in association with a $W$
or $Z$ vector boson (VH), and vector boson  fusion (VBF). For the low-mass searches,
we target the $WH$, $ZH$, VBF, and  $t{\bar{t}}H$ production
modes with specific searches, including also those signal components
not specifically targeted but which fall in the acceptance nonetheless.
Our $WH$ and $ZH$ cross sections are from Ref.~\cite{djouadibaglio}.  
This calculation starts with the NLO calculation of
{\sc v2hv}~\cite{v2hv} and includes NNLO QCD contributions~\cite{vhnnloqcd}, as well
as one-loop electroweak corrections~\cite{vhewcorr}.
A similar calculation
of the $WH$ cross section is available in Ref.~\cite{grazziniferrera}.
We use the VBF cross section computed at NNLO in QCD in Ref.~\cite{vbfnnlo}.
Electroweak corrections to the VBF production cross section are computed
with the {\sc hawk} program~\cite{hawk}, and are small and negative (2-3\%)
in the Higgs boson mass range considered here.  We include these corrections in the VBF
cross sections used for this result.  The $t{\bar{t}}H$ production cross
sections we use are from Ref.~\cite{tth}.

The Higgs boson decay branching ratio
predictions used for this result are those of Ref.~\cite{lhcxs,lhcdifferential}.  In this calculation,
the partial decay widths for all Higgs boson decays except to pairs of $W$ and $Z$ bosons
are computed with \textsc{HDECAY}~\cite{hdecay}, and the $W$ and $Z$ pair decay widths are
computed with {\sc Prophecy4f}~\cite{prophecy4f}.
The relevant decay branching ratios are listed in Table~\ref{tab:higgsxsec}.
The uncertainties on the predicted branching ratios from uncertainties in $m_b$, $m_c$, 
$\alpha_s$, and missing higher-order effects are presented in Ref.~\cite{dblittlelhc,denner}.  


\begin{sidewaystable}
\begin{center}
\caption{
The production cross sections and decay branching fractions for the SM
Higgs boson assumed for the combination.}
\vspace{0.2cm}
\label{tab:higgsxsec}
\begin{tabular}{|c|c|c|c|c|c|c|c|c|c|c|c|}\hline
$m_H$ & $\sigma_{gg\rightarrow H}$ & $\sigma_{WH}$ & $\sigma_{ZH}$ & $\sigma_{VBF}$ & $\sigma_{t{\bar{t}}H}$  &
$B(H\rightarrow b{\bar{b}})$ & $B(H\rightarrow c{\bar{c}})$ & $B(H\rightarrow \tau^+{\tau^-})$ & $B(H\rightarrow W^+W^-)$ & $B(H\rightarrow ZZ)$ & $B(H\rightarrow\gamma\gamma)$ \\
(GeV/$c^2$) & (fb)  & (fb)    & (fb)    & (fb)   & (fb)     & (\%)   & (\%)    & (\%)  & (\%)   & (\%)     & (\%) \\ \hline
\hline
   100 &  1821.8    &  281.1  & 162.7   &  97.3  &  8.0   & 79.1   & 3.68     & 8.36    & 1.11   & 0.113  & 0.159   \\
   105 &  1584.7    &  238.7  & 139.5   &  89.8  &  7.1   & 77.3   & 3.59     & 8.25    & 2.43   & 0.215  & 0.178   \\
   110 &  1385.0    &  203.7  & 120.2   &  82.8  &  6.2   & 74.5   & 3.46     & 8.03    & 4.82   & 0.439  & 0.197   \\
   115 &  1215.9    &  174.5  & 103.9   &  76.5  &  5.5   & 70.5   & 3.27     & 7.65    & 8.67   & 0.873  & 0.213   \\
   120 &  1072.3    &  150.1  &  90.2   &  70.7  &  4.9   & 64.9   & 3.01     & 7.11    & 14.3   & 1.60   & 0.225   \\
   125 &   949.3    &  129.5  &  78.5   &  65.3  &  4.3   & 57.8   & 2.68     & 6.37    & 21.6   & 2.67   & 0.230   \\
   130 &   842.9    &  112.0  &  68.5   &  60.5  &  3.8   & 49.4   & 2.29     & 5.49    & 30.5   & 4.02   & 0.226   \\
   135 &   750.8    &   97.2  &  60.0   &  56.0  &  3.3   & 40.4   & 1.87     & 4.52    & 40.3   & 5.51   & 0.214   \\
   140 &   670.6    &   84.6  &  52.7   &  51.9  &  2.9   & 31.4   & 1.46     & 3.54    & 50.4   & 6.92   & 0.194   \\
   145 &   600.6    &   73.7  &  46.3   &  48.0  &  2.6   & 23.1   & 1.07     & 2.62    & 60.3   & 7.96   & 0.168   \\
   150 &   539.1    &   64.4  &  40.8   &  44.5  &  2.3   & 15.7   & 0.725    & 1.79    & 69.9   & 8.28   & 0.137   \\
   155 &   484.0    &   56.2  &  35.9   &  41.3  &  2.0   & 9.18   & 0.425    & 1.06    & 79.6   & 7.36   & 0.100   \\
   160 &   432.3    &   48.5  &  31.4   &  38.2  &  1.8   & 3.44   & 0.159    & 0.397   & 90.9   & 4.16   & 0.0533  \\
   165 &   383.7    &   43.6  &  28.4   &  36.0  &  1.6   & 1.19   & 0.0549   & 0.138   & 96.0   & 2.22   & 0.0230  \\
   170 &   344.0    &   38.5  &  25.3   &  33.4  &  1.4   & 0.787  & 0.0364   & 0.0920  & 96.5   & 2.36   & 0.0158  \\
   175 &   309.7    &   34.0  &  22.5   &  31.0  &  1.3   & 0.612  & 0.0283   & 0.0719  & 95.8   & 3.23   & 0.0123  \\
   180 &   279.2    &   30.1  &  20.0   &  28.7  &  1.1   & 0.497  & 0.0230   & 0.0587  & 93.2   & 6.02   & 0.0102  \\
   185 &   252.1    &   26.9  &  17.9   &  26.9  &  1.0   & 0.385  & 0.0178   & 0.0457  & 84.4   & 15.0   & 0.00809 \\
   190 &   228.0    &   24.0  &  16.1   &  25.1  &  0.9   & 0.315  & 0.0146   & 0.0376  & 78.6   & 20.9   & 0.00674 \\
   195 &   207.2    &   21.4  &  14.4   &  23.3  &  0.8   & 0.270  & 0.0125   & 0.0324  & 75.7   & 23.9   & 0.00589 \\
   200 &   189.1    &   19.1  &  13.0   &  21.7  &  0.7   & 0.238  & 0.0110   & 0.0287  & 74.1   & 25.6   & 0.00526 \\ \hline
\end{tabular}   
\end{center}    
\end{sidewaystable}

\section{Distributions of Candidates} 

All analyses provide binned histograms of the final discriminant variables
for the signal and background predictions, itemized separately for each
source, and the observed data.
The number of channels combined is large, and the number of bins
in each channel is large.  Therefore, the task of assembling
histograms and visually checking whether the expected and observed limits are
consistent with the input predictions and observed data is difficult.
We therefore provide histograms that aggregate all channels' signal,
background, and data together.  In order to preserve most of the
sensitivity gain that is achieved by the analyses by binning the data
instead of collecting them all together and counting, we aggregate the
data and predictions in narrow bins of signal-to-background ratio,
$s/b$.  Data with similar $s/b$ may be added together with no loss in
sensitivity, assuming similar systematic uncertainties on the predictions.
The aggregate histograms do not show the effects of systematic
uncertainties, but instead compare the data with the central
predictions supplied by each analysis.

The range of $s/b$ is quite large in each analysis, and so
$\log_{10}(s/b)$ is chosen as the plotting variable.  Plots of the
distributions of $\log_{10}(s/b)$ are shown for Higgs boson masses
of 115, 125, and 165~GeV/$c^2$ in Figure~\ref{fig:lnsb}, demonstrating 
agreement with background over five orders of magnitude.   These
distributions can be integrated from the high-$s/b$ side downwards,
showing the sums of signal, background, and data for the most pure
portions of the selection of all channels added together.  
The integrals of the $\approx 100$ highest $s/b$ events are shown in 
Figure~\ref{fig:integ}, plotted as functions of the number of 
signal events expected.  The most significant
candidates are found in the bins with the highest $s/b$; an excess
in these bins relative to the background prediction drives the Higgs
boson cross section limit upwards, while a deficit drives it downwards.
The lower-$s/b$ bins show that the modeling of the rates and kinematic
distributions of the backgrounds is very good.  The integrated plots
show an excess consistent with signal for the analyses seeking 
a Higgs boson mass of 125~GeV/$c^2$, and a deficit of events in the 
highest-$s/b$ bins for the analyses seeking
a Higgs boson of mass 165~GeV/$c^2$.

We also show the distributions of the data after subtracting the
expected background, and compare that with the expected signal yield
for a Standard Model Higgs boson, after collecting all bins in all
channels sorted by $s/b$.  These background-subtracted distributions
are shown in Figure~\ref{fig:bgsub} for Higgs boson masses of 115, 125, and
165~GeV/$c^2$.  These graphs also show the
remaining uncertainty on the background prediction after fitting the
background model to the data within the systematic uncertainties on
the rates and shapes in each contributing channel.

\begin{figure}[t]
\begin{centering}
\includegraphics[width=0.4\textwidth]{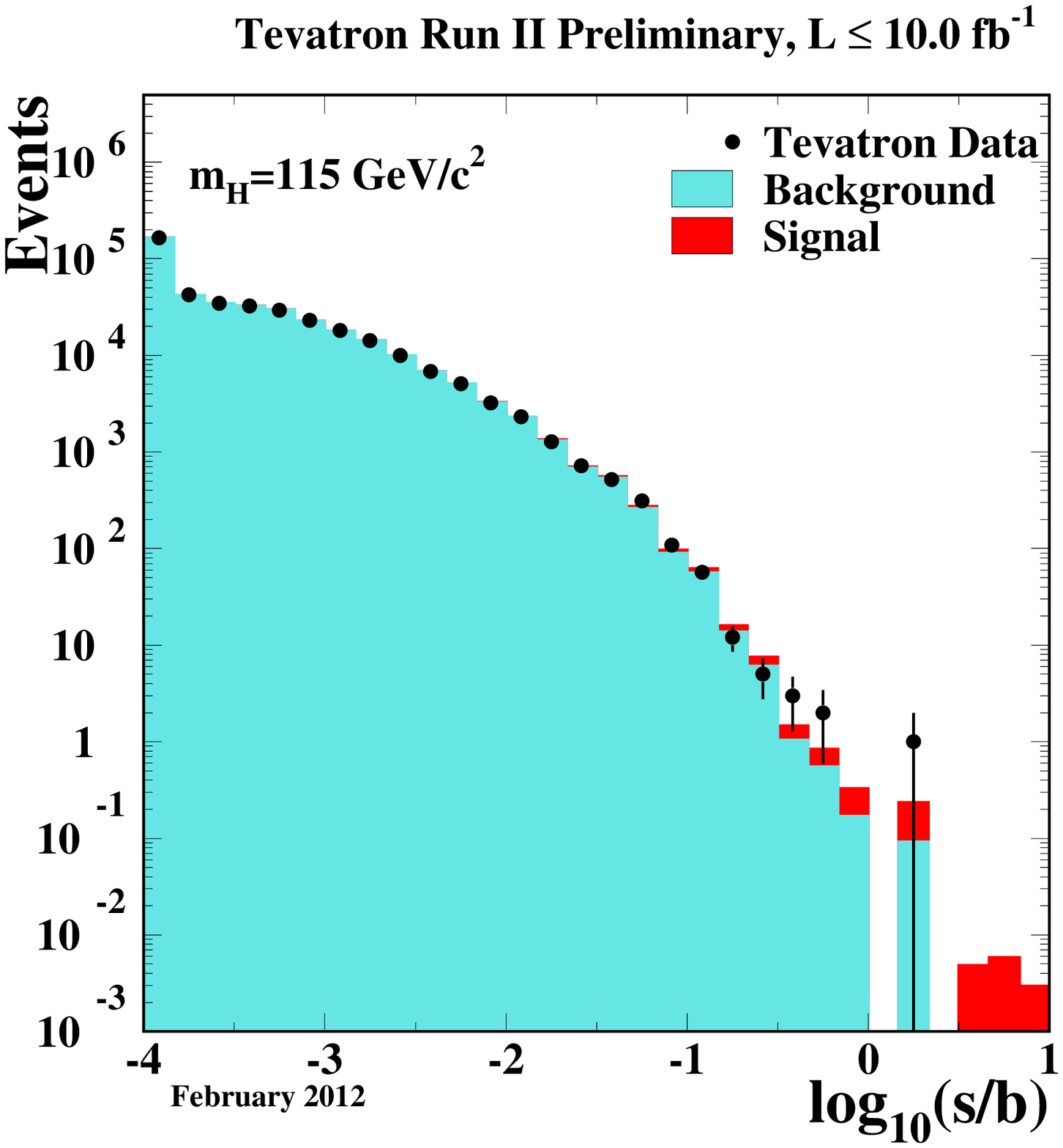}\includegraphics[width=0.4\textwidth]{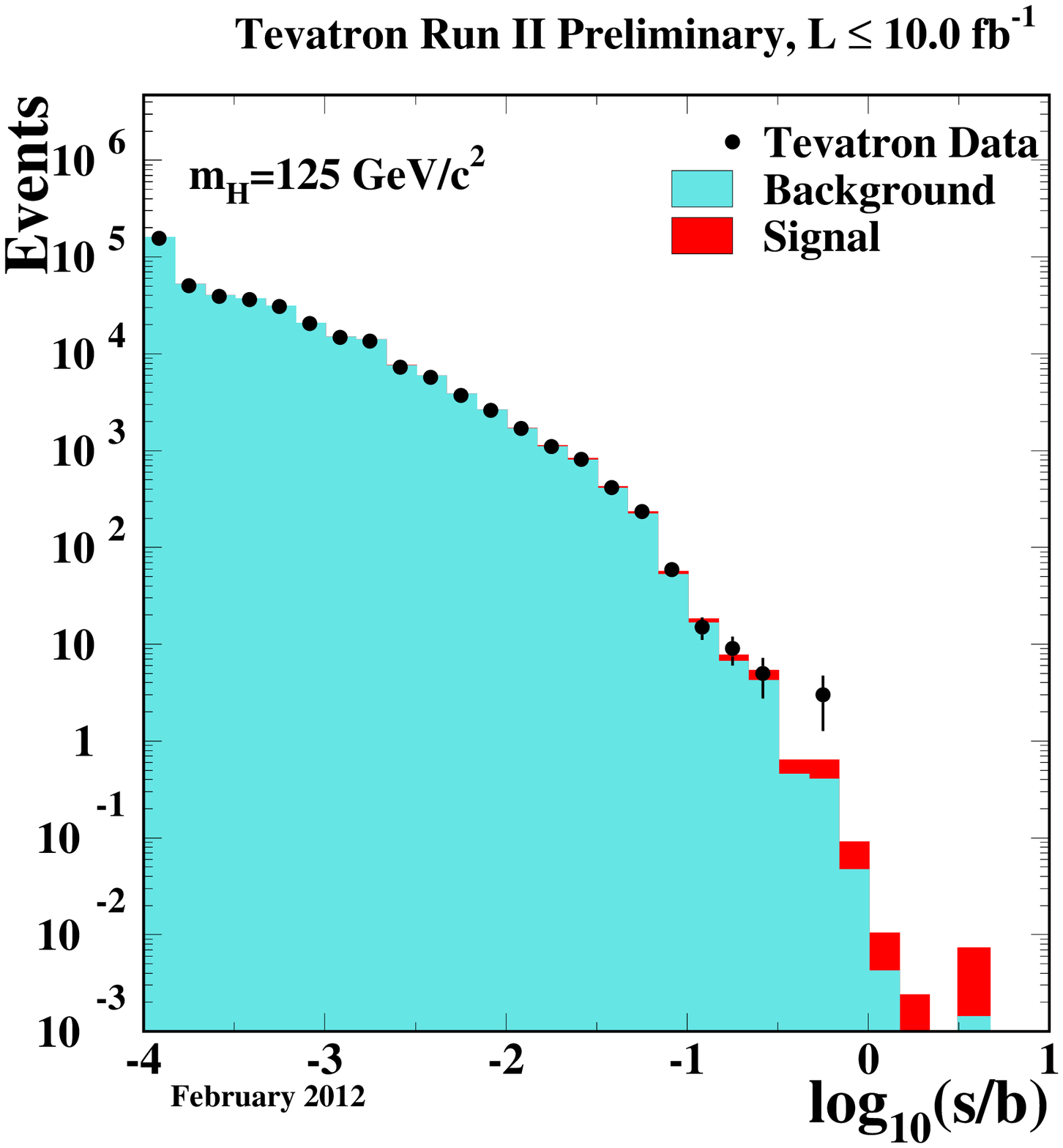}
\includegraphics[width=0.4\textwidth]{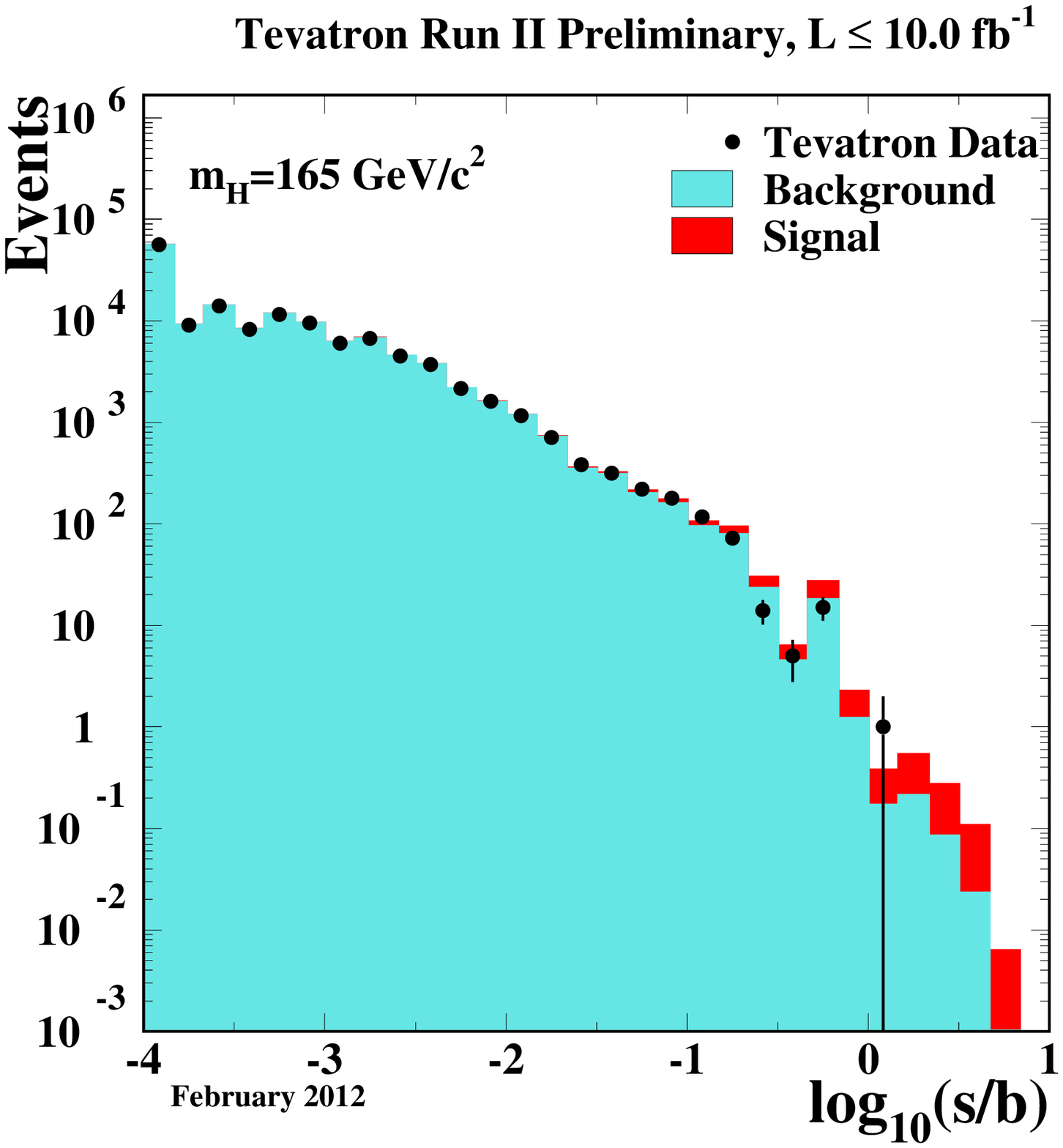}
\caption{
\label{fig:lnsb} Distributions of $\log_{10}(s/b)$, for the data from all
contributing channels from CDF and D0, for Higgs boson masses of 115, 125, and
165~GeV/$c^2$.  The data are shown with points, and the expected signal
is shown stacked on top of the backgrounds.  Underflows and overflows are
collected into the leftmost and rightmost bins. }
\end{centering}
\end{figure}

\begin{figure}[t]
\begin{centering}
\includegraphics[width=0.4\textwidth]{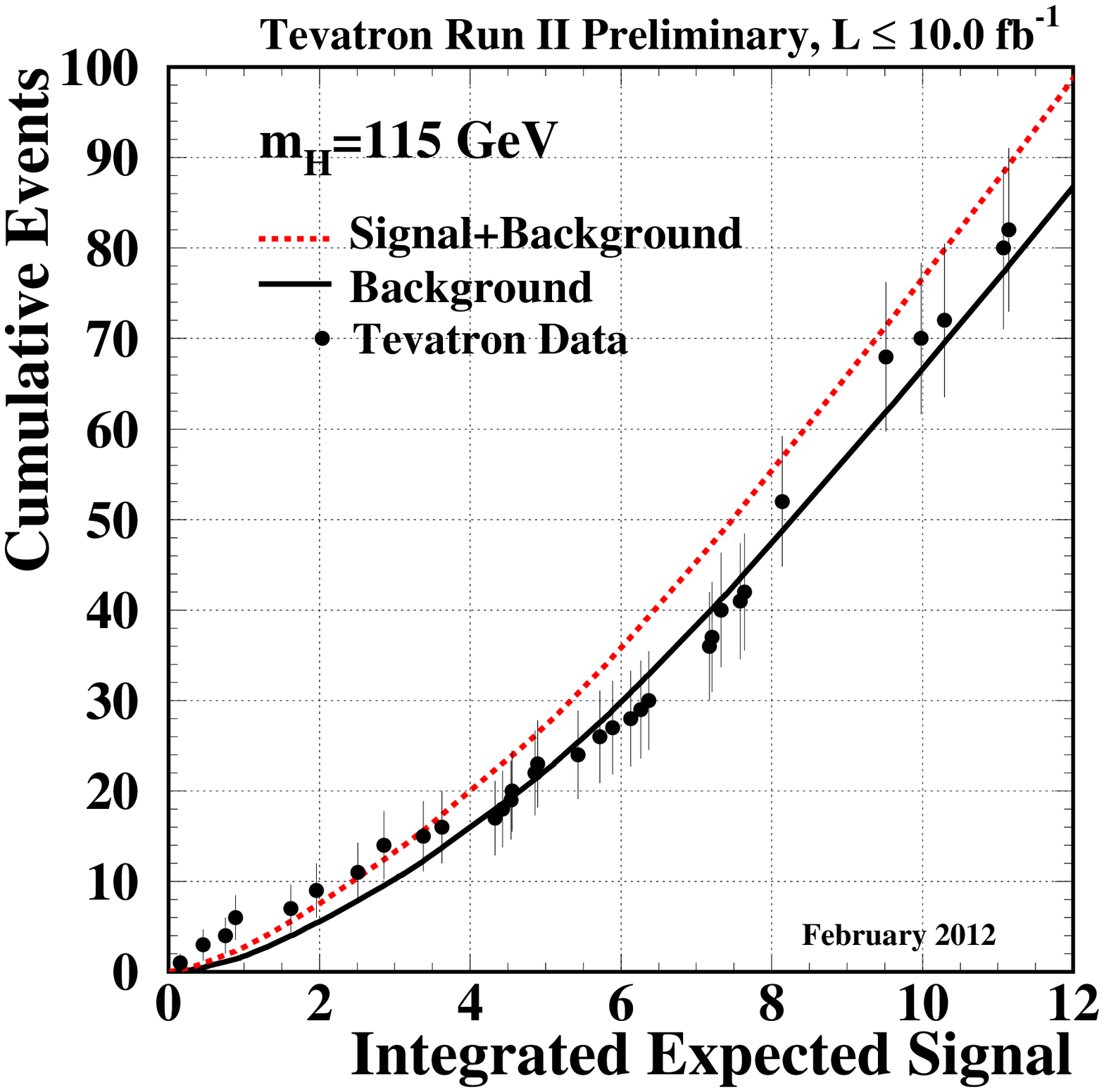}\includegraphics[width=0.4\textwidth]{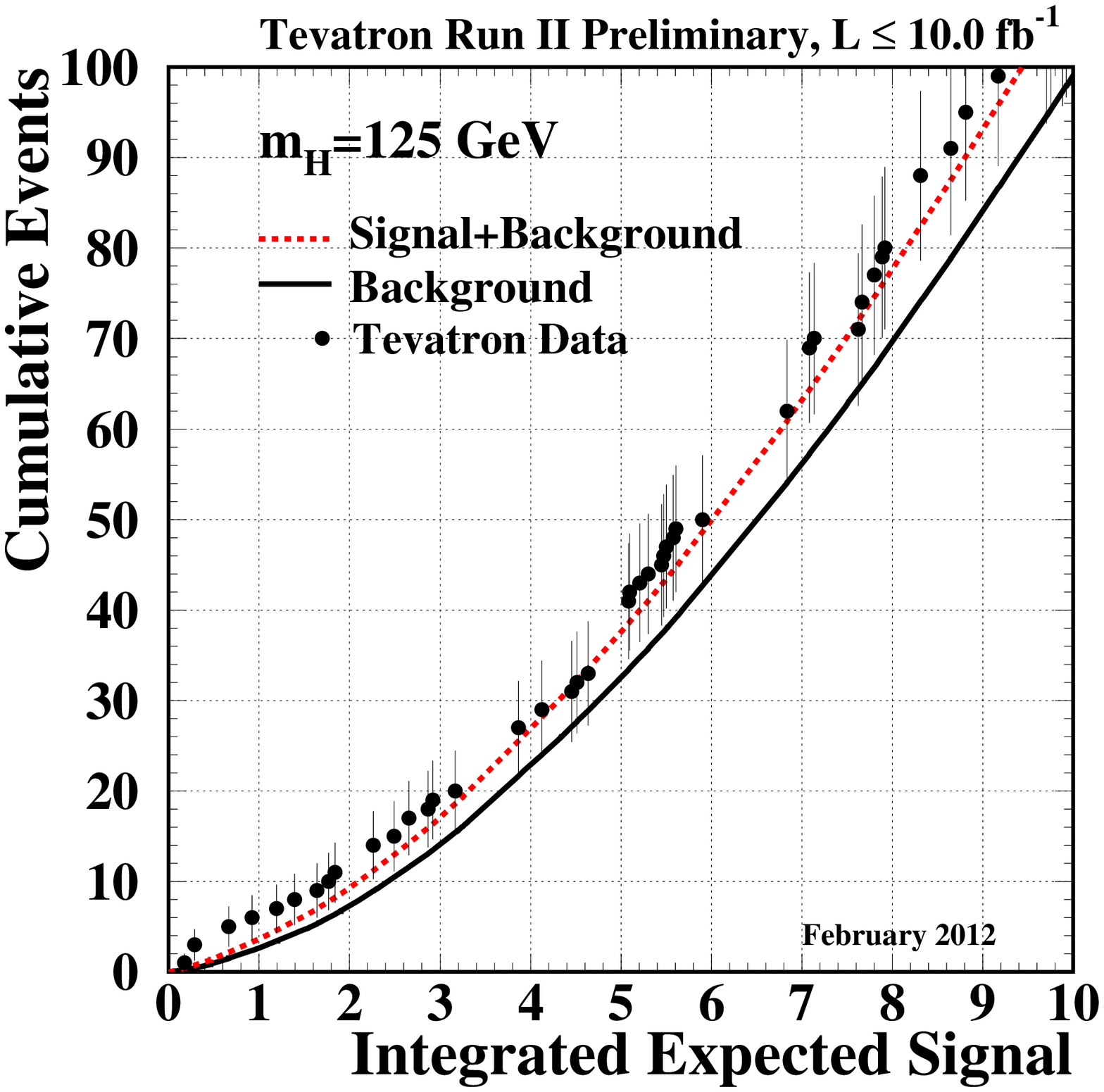}
\includegraphics[width=0.4\textwidth]{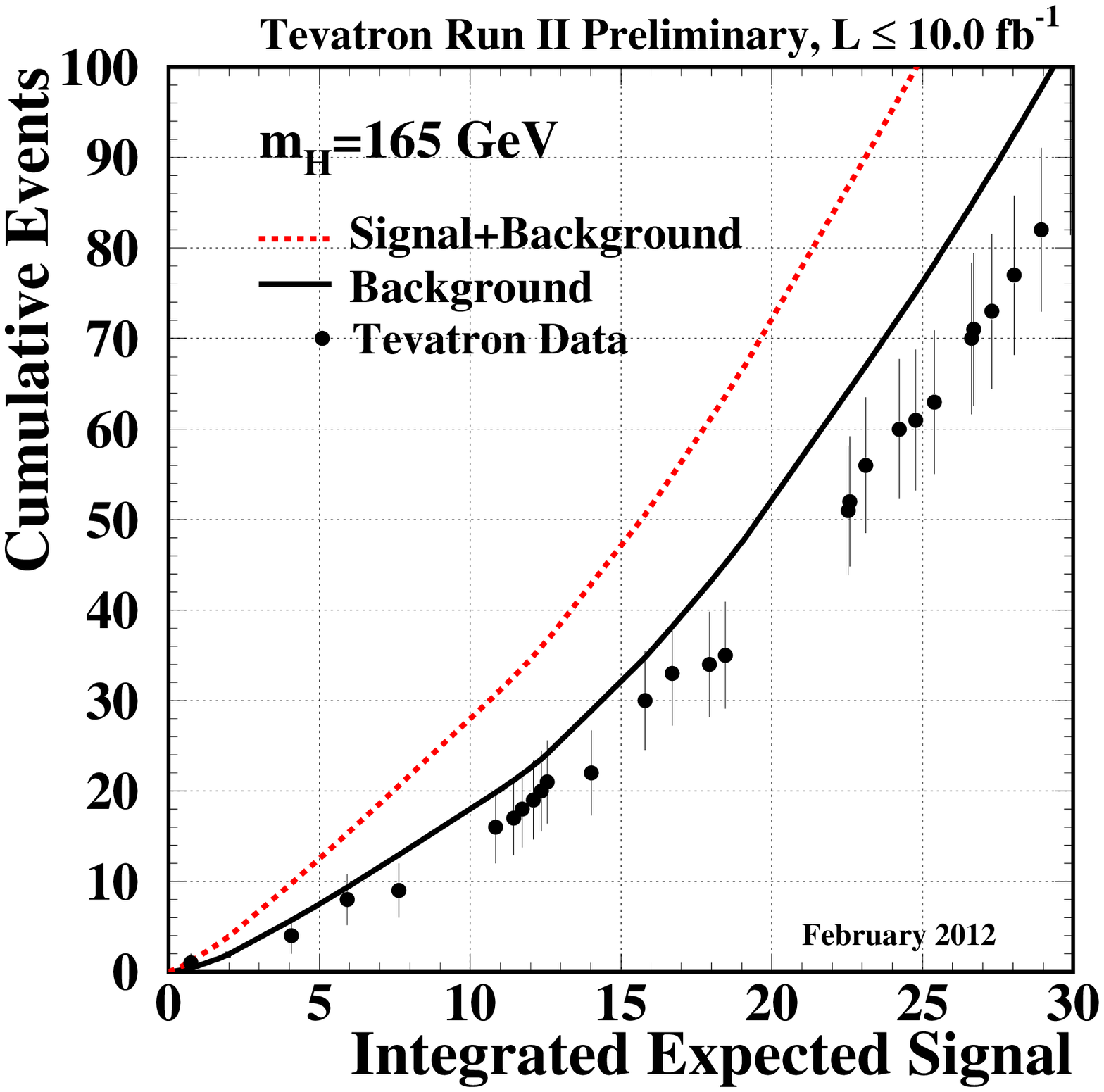}
\caption{
\label{fig:integ} Integrated distributions of $s/b$, starting at the high
$s/b$ side, for Higgs boson masses of 115, 125, and 165~GeV/$c^2$.  The total
signal+background and background-only integrals are shown separately, along
with the data sums.  Data are only shown for bins that have data events in
them.}
\end{centering}
\end{figure}

\begin{figure}[t]
\begin{centering}
\includegraphics[width=0.45\textwidth]{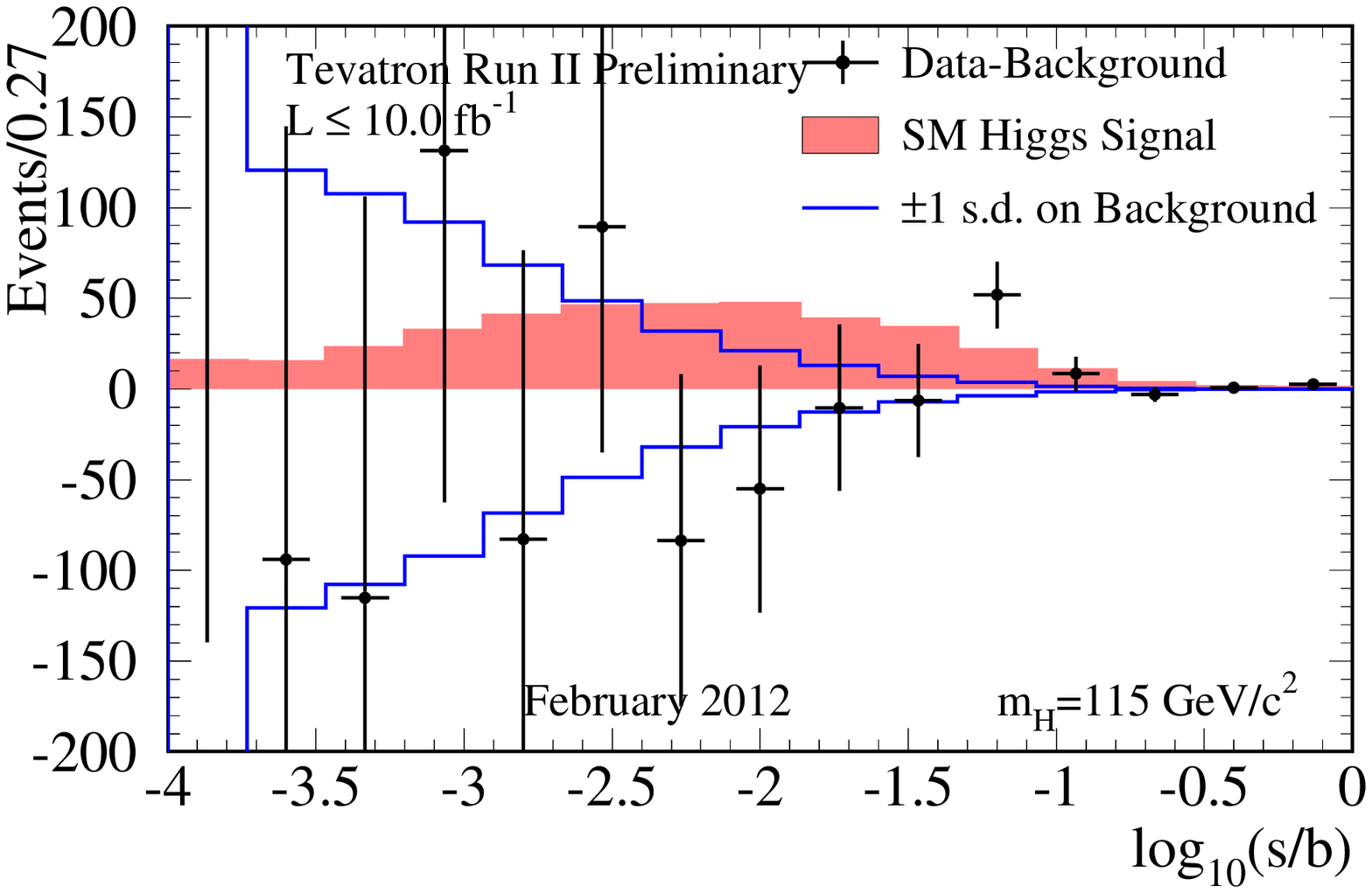}\includegraphics[width=0.45\textwidth]{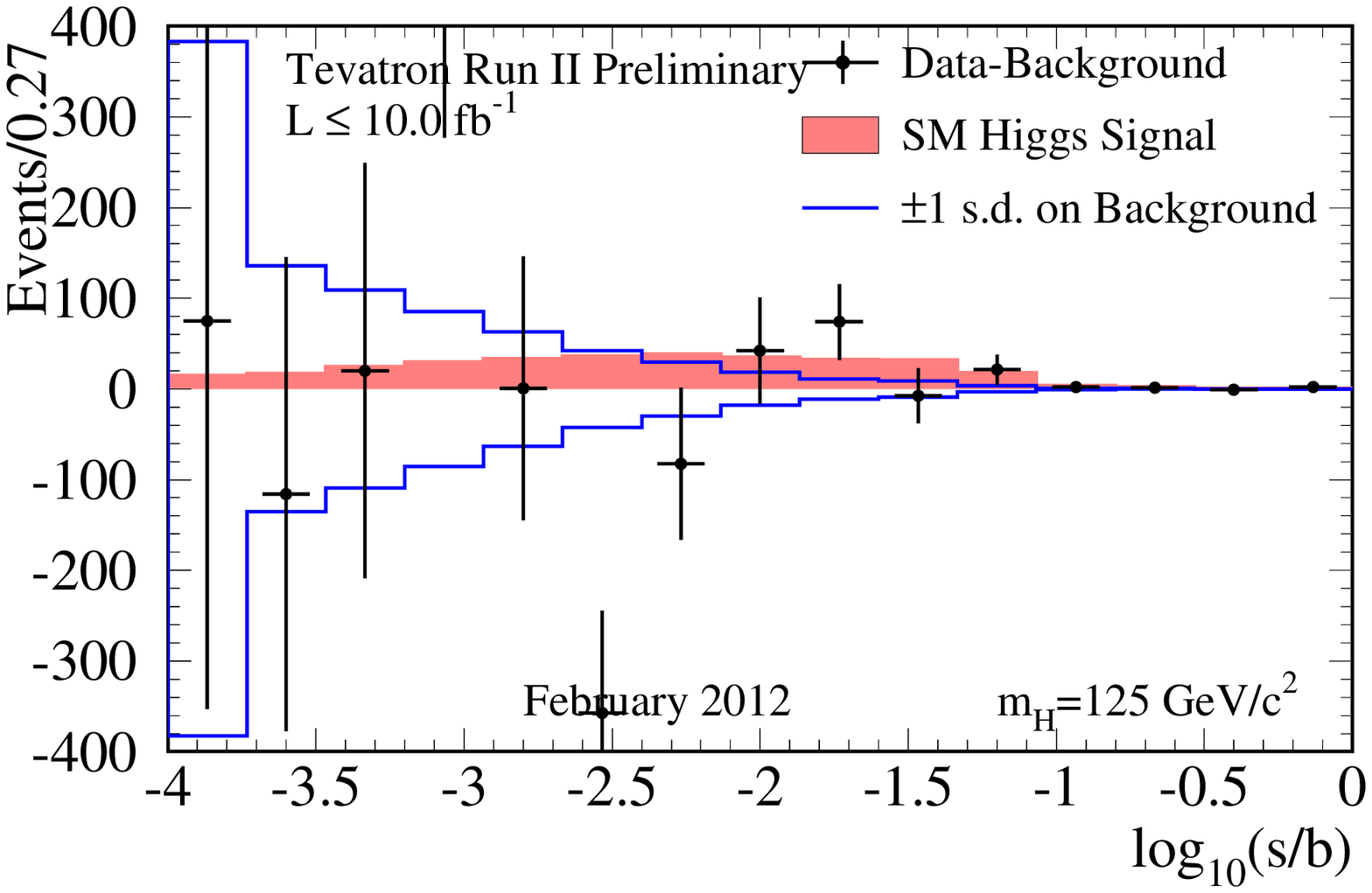}
\includegraphics[width=0.45\textwidth]{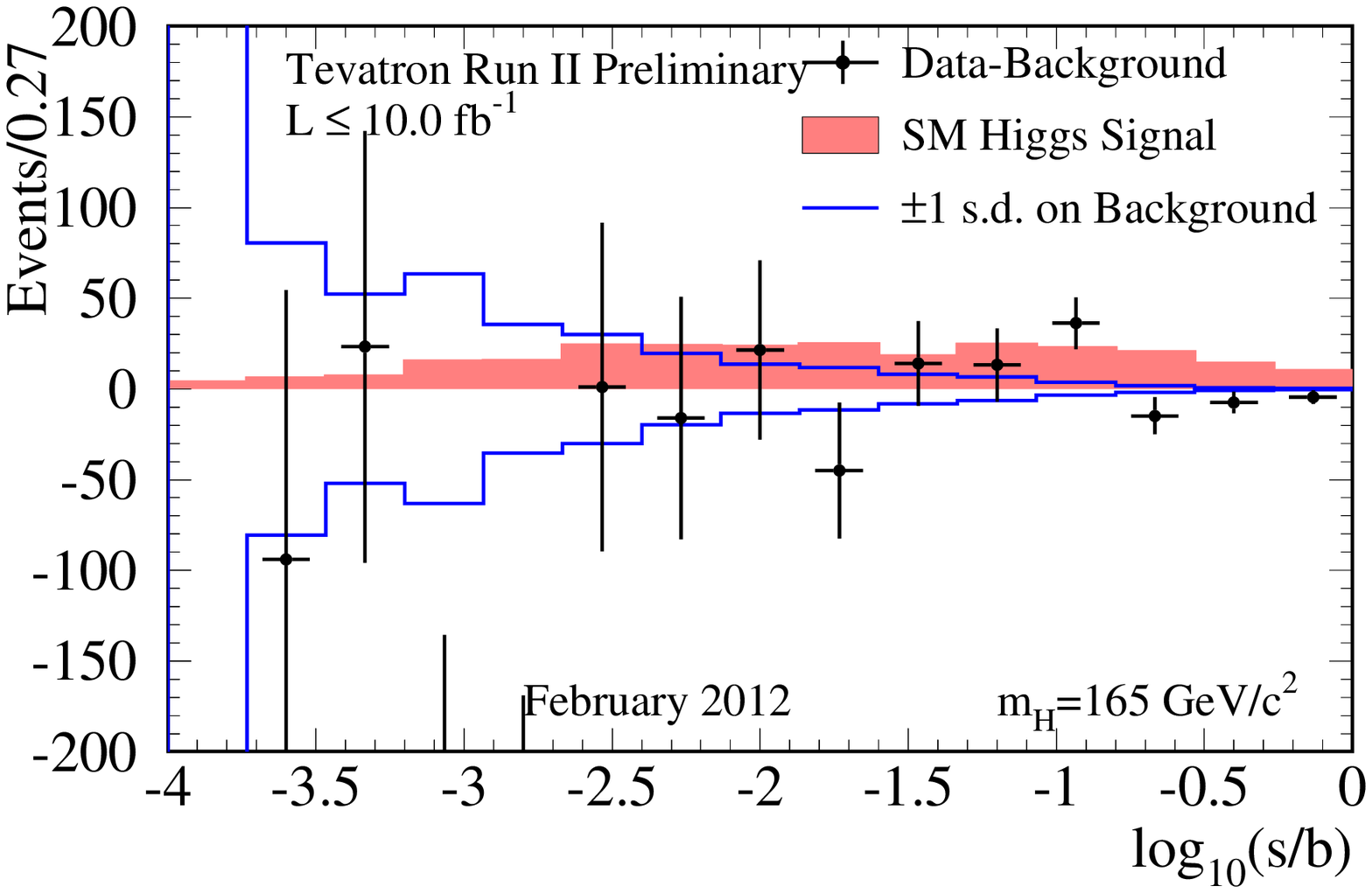}
\caption{
\label{fig:bgsub} Background-subtracted data distributions for all channels, summed in bins of $s/b$,
for Higgs boson masses of 115, 125, and 165~GeV/$c^2$.  The background has been fit, within its systematic
uncertainties and assuming no Higgs boson signal is present, to the data.  
The points with error bars indicate the background-subtracted data; the
sizes of the error bars are the square roots of the predicted background in each bin.  The unshaded
(blue-outline) histogram shows the systematic uncertainty on the best-fit background model, and the
shaded histogram shows the expected signal for a Standard Model Higgs boson.}
\end{centering}
\end{figure}

\section{Combining Channels} 

To gain confidence that the final result does not depend on the
details of the statistical formulation,
we perform two types of combinations, using
Bayesian and  Modified Frequentist approaches, which yield limits on the Higgs boson
production rate that agree within 10\% at each
value of $m_H$, and within 1\% on average.
Both methods rely on distributions in the final discriminants, and not just on
their single integrated values.  Systematic uncertainties enter on the
predicted number of signal and background events as well
as on the distribution of the discriminants in
each analysis (``shape uncertainties'').
Both methods use likelihood calculations based on Poisson
probabilities.

\subsection{Bayesian Method}

Because there is no experimental information on the production cross section for
the Higgs boson, in the Bayesian technique~\cite{CDFHiggs} we assign a flat prior
for the total number of selected Higgs boson events.  For a given Higgs boson mass, the
combined likelihood is a product of likelihoods for the individual
channels, each of which is a product over histogram bins:

\begin{equation}
{\cal{L}}(R,{\vec{s}},{\vec{b}}|{\vec{n}},{\vec{\theta}})\times\pi({\vec{\theta}})
= \prod_{i=1}^{N_C}\prod_{j=1}^{N_b} \mu_{ij}^{n_{ij}} e^{-\mu_{ij}}/n_{ij}!
\times\prod_{k=1}^{n_{np}}e^{-\theta_k^2/2}
\end{equation}

\noindent where the first product is over the number of channels
($N_C$), and the second product is over $N_b$ histogram bins containing
$n_{ij}$ events, binned in  ranges of the final discriminants used for
individual analyses, such as the dijet mass, neural-network outputs,
or matrix-element likelihoods.
 The parameters that contribute to the
expected bin contents are $\mu_{ij} =R \times s_{ij}({\vec{\theta}}) + b_{ij}({\vec{\theta}})$
for the
channel $i$ and the histogram bin $j$, where $s_{ij}$ and $b_{ij}$
represent the expected background and signal in the bin, and $R$ is a scaling factor
applied to the signal to test the sensitivity level of the experiment.
Truncated Gaussian priors are used for each of the nuisance parameters
$\theta_k$, which define the
sensitivity of the predicted signal and background estimates to systematic uncertainties. These
can take the form of uncertainties on overall rates, as well as the shapes of the distributions
used for combination.   These systematic uncertainties can be far larger
than the expected SM Higgs boson signal, and are therefore important in the calculation of limits.
The truncation is applied so that no prediction of any signal or background in any bin is negative.
The posterior density function is then integrated over all parameters (including correlations) except for $R$,
and a 95\% credibility level upper limit on $R$ is estimated
by calculating the value of $R$ that corresponds to 95\% of the area
of the resulting distribution.  This posterior density function may also be used to estimate the best-fit
value of $R$ by finding that value which maximizes the posterior density.  The fitted uncertainties are given
by the shortest interval containing 68\% of the integrated posterior density.  These values are compared with
those obtained from a profile likelihood fit to $R$, maximizing over the values of the nuisance parameters,
and give good agreement.

\subsection{Modified Frequentist Method}

The Modified Frequentist technique relies on the ${\rm CL}_{\rm s}$ method, using
a log-likelihood ratio (LLR) as test statistic~\cite{DZHiggs}:
\begin{equation}
LLR = -2\ln\frac{p({\mathrm{data}}|H_1)}{p({\mathrm{data}}|H_0)},
\end{equation}
where $H_1$ denotes the test hypothesis, which admits the presence of
SM backgrounds and a Higgs boson signal, while $H_0$ is the null
hypothesis, for only SM backgrounds and 'data' is either an ensemble of pseudo-experiment
data constructed from the expected signal and backgrounds, or the
actual observed data.  The probabilities $p$ are
computed using the best-fit values of the nuisance parameters for each
pseudo-experiment, separately for each of the two hypotheses, and include the
Poisson probabilities of observing the data multiplied by Gaussian
priors for the values of the nuisance parameters.  This technique
extends the LEP procedure~\cite{pdgstats} which does not involve a
fit, in order to yield better sensitivity when expected signals are
small and systematic uncertainties on backgrounds are
large~\cite{pflh}.

The ${\rm CL}_{\rm s}$ technique involves computing two $p$-values, ${\rm CL}_{\rm s+b}$ and ${\rm CL}_{\rm b}$.
The latter is defined by
\begin{equation}
1-{\rm CL}_{\rm b} = p(LLR\le LLR_{\mathrm{obs}} | H_0),
\end{equation}
where $LLR_{\mathrm{obs}}$ is the value of the test statistic computed for the
data. $1-{\rm CL}_{\rm b}$ is the probability of observing a signal-plus-background-like outcome
without the presence of signal, i.e. the probability
that an upward fluctuation of the background provides  a signal-plus-background-like
response as observed in data.
The other $p$-value is defined by
\begin{equation}
{\rm CL}_{\rm s+b} = p(LLR\ge LLR_{\mathrm{obs}} | H_1),
\end{equation}
and this corresponds to the probability of a downward fluctuation of the sum
of signal and background in
the data.  A small value of ${\rm CL}_{\rm s+b}$ reflects inconsistency with  $H_1$.
It is also possible to have a downward fluctuation in data even in the absence of
any signal, and a small value of ${\rm CL}_{\rm s+b}$ is possible even if the expected signal is
so small that it cannot be tested with the experiment.  To minimize the possibility
of  excluding  a signal to which there is insufficient sensitivity
(an outcome  expected 5\% of the time at the 95\% C.L., for full coverage),
we use the quantity ${\rm CL}_{\rm s}={\rm CL}_{\rm s+b}/{\rm CL}_{\rm b}$.  If ${\rm CL}_{\rm s}<0.05$ for a particular choice
of $H_1$, that hypothesis is deemed to be excluded at the 95\% C.L. In an analogous
way, the expected ${\rm CL}_{\rm b}$, ${\rm CL}_{\rm s+b}$ and ${\rm CL}_{\rm s}$ values are computed from the median of the
LLR distribution for the background-only hypothesis.

Systematic uncertainties are included  by fluctuating the predictions for
signal and background rates in each bin of each histogram in a correlated way when
generating the pseudo-experiments used to compute ${\rm CL}_{\rm s+b}$ and ${\rm CL}_{\rm b}$.

An alternate computation of the $p$-value $1-{\rm CL}_{\rm b}$ is to use the fitted value of $R$ as a test statistic
instead of LLR.  This method is nearly as optimal as using LLR in our searches, and has been applied in the single top
quark observation~\cite{d0singletop}.  The background-only $p$-value is the probability of obtaining the fitted
cross section observed in the data or more, assuming that a signal is absent.  We use this method to quote our
$p$-values and significances.

\subsection{Systematic Uncertainties} 

Systematic uncertainties differ
between experiments and analyses, and they affect the rates and shapes of the predicted
signal and background in correlated ways.  The combined results incorporate
the sensitivity of predictions to  values of nuisance parameters,
and include correlations between rates and shapes, between signals and backgrounds,
and between channels within experiments and between experiments.
More on these issues can be found in the
individual analysis notes~\cite{cdfWH} through~\cite{dzHgg}.  Here we
discuss only the largest contributions and correlations between and
within the two experiments.

\subsubsection{Correlated Systematics between CDF and D0}

The uncertainties on the measurements of the integrated luminosities are 6\%
(CDF) and 6.1\% (D0).
Of these values, 4\% arises from the uncertainty
on the inelastic \pp~scattering cross section, which is correlated
between CDF and D0.
CDF and D0 also share the assumed values and uncertainties on the production cross sections
for top-quark processes (\ttbar~and single top) and for electroweak processes
($WW$, $WZ$, and $ZZ$).  In order to provide a consistent combination, the values of these
cross sections assumed in each analysis are brought into agreement.  We use
$\sigma_{t\bar{t}}=7.04^{+0.24}_{-0.36}~{\rm (scale)}\pm 0.14{\rm (PDF)}\pm 0.30{\rm (mass)}$,
following the calculation of Moch and Uwer~\cite{mochuwer}, assuming
a top quark mass $m_t=173.1\pm 1.2$~GeV/$c^2$~\cite{tevtop09},
and using the MSTW2008nnlo PDF set~\cite{mstw2008}.  Other
calculations of $\sigma_{t\bar{t}}$ are similar~\cite{otherttbar}.

For single top, we use the 
next-to-next-to-next-to-leading-order (NNNLO) at next-to-leading logarithmic (NLL) 
$t$-channel calculation of Kidonakis~\cite{kid1},
which has been updated using the MSTW2008nnlo PDF set~\cite{mstw2008}~\cite{kidprivcomm}.
For the $s$-channel process we use~\cite{kid2}, again based on the MSTW2008nnlo PDF set.
Both of the cross section values below are the sum of the single $t$ and single ${\bar{t}}$
cross sections, and both assume $m_t=173.1\pm 1.2$ GeV$/c^2$.
\begin{equation}
\sigma_{t-{\rm{chan}}} = 2.10\pm 0.027~{\rm{(scale)}} \pm 0.18~{\rm{(PDF)}}  \pm 0.045~{\rm{(mass)}}~{\rm {pb}}.
\end{equation}
\begin{equation}
\sigma_{s-{\rm{chan}}} = 1.05\pm 0.01~{\rm{(scale)}} \pm 0.06~{\rm{(PDF)}}  \pm 0.03~{\rm{(mass)}}~{\rm {pb}}.
\end{equation}
Other calculations of $\sigma_{\rm{SingleTop}}$ are
similar for our purposes~\cite{harris}.

MCFM~\cite{mcfm} has been used to compute the NLO cross sections for $WW$, $WZ$,
and $ZZ$ production~\cite{dibo}.  Using a scale choice $\mu_0=M_V^2+p_T^2(V)$ and
the MSTW2008 PDF set~\cite{mstw2008}, the cross section for inclusive $W^+W^-$
production is
\begin{equation}
\sigma_{W^+W^-} = 11.34~^{+0.56}_{-0.49}~{\rm{(scale)}}~^{+0.35}_{-0.28}~{\rm(PDF)}~{\rm{pb}}
\end{equation}
and the cross section for inclusive $W^\pm Z$ production is
\begin{equation}
\sigma_{W^\pm Z} = 3.22~^{+0.20}_{-0.17}~{\rm{(scale)}}~^{+0.11}_{-0.08}~{\rm(PDF)}~{\rm{pb}}
\end{equation}
The calculation is done using $Z\to\ell^+\ell^-$ and therefore necessarily includes contributions from 
$\gamma^*\to\ell^+\ell^-$.  The cross sections quoted above have the requirement
$75\leq m_{\ell^+\ell^-}\leq 105$~GeV/$c^2$ for the leptons from the neutral current
exchange.  The same dilepton invariant mass requirement is applied to both
sets of leptons in determining the $ZZ$ cross section which is
\begin{equation}
\sigma_{ZZ} = 1.20~^{+0.05}_{-0.04}~{\rm{(scale)}}~^{+0.04}_{-0.03}~{\rm(PDF)}~{\rm{pb}}
\end{equation}
For the diboson cross section calculations, $|\eta_{\ell}|<5$ for all calculations.
Loosening this requirement to include all leptons leads to $\sim$+0.4\% change in
the predictions.  Lowering the factorization and renormalization scales by a factor
of two increases the cross section, and raising the scales by a factor of two
decreases the cross section.  The PDF uncertainty has the same fractional impact on
the predicted cross section independent of the scale choice.  All PDF uncertainties
are computed as the quadrature sum of the twenty 68\% C.L. eigenvectors provided with
MSTW2008 (MSTW2008nlo68cl).

In many analyses, the dominant background yields are calibrated with data control
samples.  Since the methods of measuring the multijet (``QCD'') backgrounds differ
between CDF and D0, and even between analyses within the collaborations, there is
no correlation assumed between these rates.  Similarly, the large uncertainties on
the background rates for $W$+heavy flavor (HF) and $Z$+heavy flavor are considered
at this time to be uncorrelated.
The calibrations of fake leptons,
unvetoed $\gamma\rightarrow e^+e^-$ conversions, $b$-tag efficiencies and mistag
rates are performed by each collaboration using independent data samples and
methods, and are therefore also treated as uncorrelated.

\subsubsection{Correlated Systematic Uncertainties for CDF}
The dominant systematic uncertainties for the CDF analyses are shown in the
Appendix in Tables~\ref{tab:cdfsystwh2jet} and~\ref{tab:cdfsystwh3jet} for
the \WH\ channels, in Table~\ref{tab:cdfvvbb1} for the $WH,ZH\rightarrow\MET
b{\bar{b}}$ channels, in Tables~\ref{tab:cdfllbb1} and~\ref{tab:cdfllbb2}
for the $ZH\rightarrow\ell^+\ell^-b{\bar{b}}$ channels, in
Tables~\ref{tab:cdfsystww0}, \ref{tab:cdfsystww4}, and~\ref{tab:cdfsystww5}
for the $H \rightarrow W^+W^-\rightarrow \ell^{\prime \pm}\nu \ell^{\prime
\mp}\nu$ channels, in Table~\ref{tab:cdfsystwww} for the $WH \rightarrow
WWW \rightarrow\ell^{\prime \pm}\ell^{\prime \pm}$ and $WH\rightarrow
WWW \rightarrow \ell^{\pm}\ell^{\prime \pm} \ell^{\prime \prime \mp}$
channels, in Table~\ref{tab:cdfsystzww} for the $ZH \rightarrow ZWW
\rightarrow \ell^{\pm}\ell^{\mp} \ell^{\prime \pm}$ channels, In
Table~\ref{tab:cdfsystH4l} for the $H \rightarrow 4 \ell$ channel, in
Tables~\ref{tab:cdfsystttHLJ}, \ref{tab:cdfsysttthmetjets}, and
\ref{tab:cdfsysttthalljets} for the $t\bar{t}H \rightarrow W^+ b W^- \bar{b}
b\bar{b}$ channels, in Table~\ref{tab:cdfsysttautau} for the $H \rightarrow
\tau^+\tau^-$ channels, in Table~\ref{tab:cdfsystVtautau} for the $WH
\rightarrow \ell \nu \tau^+ \tau^-$ and $ZH \rightarrow \ell^+ \ell^- \tau^+
\tau^-$ channels, in Table~\ref{tab:cdfallhadsyst} for the $WH/ZH$ and VBF
$\rightarrow jjb{\bar{b}}$ channels, and in Table~\ref{tab:cdfsystgg}
for the $H \rightarrow \gamma \gamma$ channel.  Each source induces a
correlated uncertainty across all CDF channels' signal and background
contributions which are sensitive to that source.  For \hbb, the largest
uncertainties on signal arise from measured $b$-tagging efficiencies,
jet energy scale, and other Monte Carlo modeling.  Shape dependencies of
templates on jet energy scale, $b$-tagging, and gluon radiation (``ISR''
and ``FSR'') are taken into account for some analyses (see tables).
For \hww, the largest uncertainties on signal acceptance originate from
Monte Carlo modeling.  Uncertainties on background event rates vary
significantly for the different processes.  The backgrounds with the
largest systematic uncertainties are in general quite small. Such
uncertainties are constrained by fits to the nuisance parameters, and
they do not affect the result significantly.  Because the largest
background contributions are measured using data, these uncertainties
are treated as uncorrelated for the \hbb~channels.  The differences in
the resulting limits when treating the remaining uncertainties as either
correlated or uncorrelated is less than $5\%$.

\subsubsection{Correlated Systematic Uncertainties for D0 }
The dominant systematic uncertainties for the D0 analyses are shown in the Appendix,
in Tables~\ref{tab:d0systwh1}, \ref{tab:d0sysVHtau}, \ref{tab:d0vvbb}, \ref{tab:d0llbb1},
\ref{tab:d0systww}, \ref{tab:d0systwwtau}, \ref{tab:d0systwww-em}, \ref{tab:d0systttm}, \ref{tab:d0systlll},
 \ref{tab:d0lvjj},
and \ref{tab:d0systgg}.
  Each source induces a correlated
uncertainty across all D0 channels sensitive to that source. Wherever appropriate the
impact of systematic effects on both the rate and shape of the predicted signal and
background is included.  For the low mass, \hbb~analyses, significant sources of
uncertainty include the measured $b$-tagging rate and the normalization of the $W$
and $Z$ plus heavy flavor backgrounds. For the \hww and $VH \rightarrow leptons +X$ analyses, significant
sources of uncertainty are the measured efficiencies for selecting leptons. For analyses
involving jets the determination of the jet energy scale, jet resolution and the
multijet background contribution are significant sources of uncertainty. Significant
sources for all analyses are the uncertainties on the luminosity and the cross sections
for the simulated backgrounds.  All systematic uncertainties arising from the same
source are taken to be correlated among the different backgrounds and between signal
and background.

\begin{figure}[t]
\begin{centering}
\includegraphics[width=14.0cm]{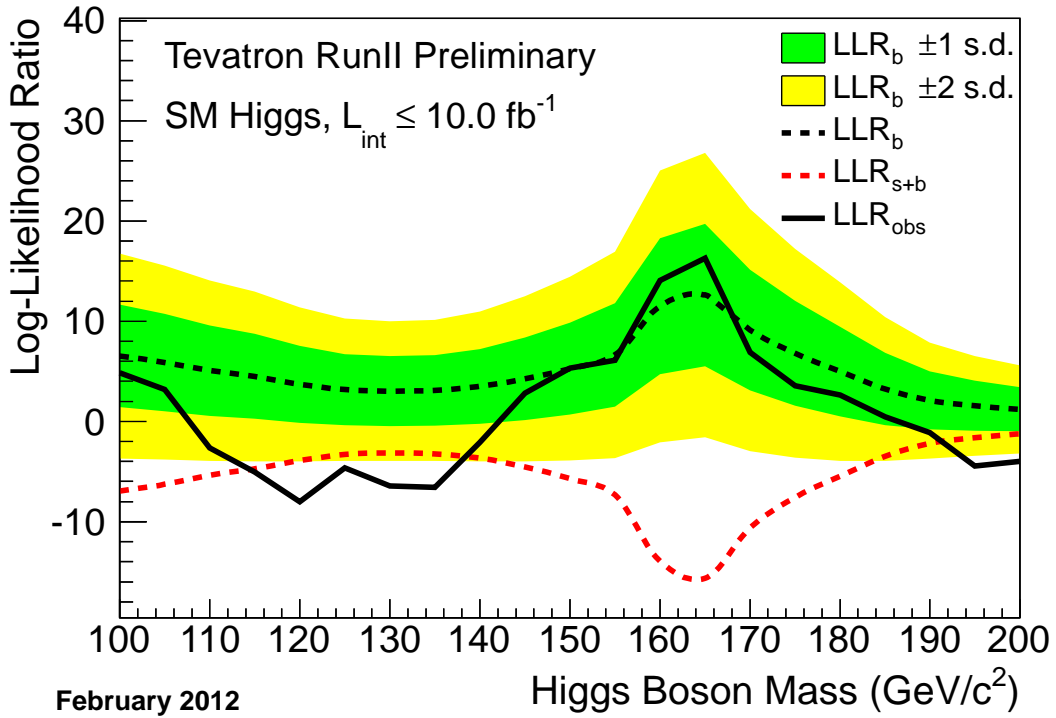}
\caption{
\label{fig:comboLLR} {
Distributions of the log-likelihood ratio (LLR) as a function of Higgs boson mass obtained with
the ${\rm CL}_{\rm s}$ method for the combination of all CDF and D0 analyses. The green and yellow
bands correspond to the regions enclosing 1~s.d. and 2~s.d. fluctuations around the median
expected value assuming only background is present, respectively.  The red dashed curve shows the median
expected value assuming a Higgs boson signal is present, separately at each $m_H$.}}
\end{centering}
\end{figure}

\vspace*{1cm}
\section{Combined Results} 

Before extracting the combined limits we study the distributions of the log-likelihood 
ratio (LLR) for different hypotheses to quantify the expected sensitivity across the mass 
range tested.  Figure~\ref{fig:comboLLR} and Table~\ref{tab:llrVals} display the LLR distributions for the combined 
analyses as functions of $m_{H}$. Included are the median of the LLR distributions for the
background-only hypothesis (LLR$_{b}$), the signal-plus-background hypothesis (LLR$_{s+b}$), 
and the observed value for the data (LLR$_{\rm{obs}}$).  The shaded bands represent the 
one and two s.d. departures for LLR$_{b}$ centered on the median.  
The separation between the medians of the LLR$_{b}$ and LLR$_{s+b}$ distributions
provides a measure of the discriminating power of the search.  The sizes
of the one- and two-s.d. LLR$_{b}$ bands indicate the width of the LLR$_{b}$
distribution, assuming no signal is truly present and only statistical fluctuations
and systematic effects are present.  The value of LLR$_{\rm{obs}}$ relative to
LLR$_{s+b}$ and LLR$_{b}$ indicates whether the data distribution appears to resemble
what we expect if a signal is present (i.e. closer to the LLR$_{s+b}$ distribution,
which is negative by construction) or whether it resembles the background expectation
more closely; the significance of departures of LLR$_{\rm{obs}}$ from LLR$_{b}$
can be evaluated by the width of the LLR$_{b}$ bands.  
  The data are consistent with the prediction of the background-only hypothesis (the black
  dashed line) above $\sim$ 145 GeV/$c^2$.  For $m_{H}$ from 110 to 140~GeV/$c^2$, 
  an excess in the data has an amplitude consistent with the expectation for
a standard model Higgs boson in this mass range (dashed red line). In this region our 
ability to distinguish the signal-plus-background and background-only hypotheses is, 
as indicated by the separation of the LLR$_{s+b}$ and LLR$_{b}$ values, at the 2~s.d. level.

Using the combination procedures outlined in Section III, we extract
limits on the SM Higgs boson production $\sigma \times B(H\rightarrow X)$
in \pp~collisions at $\sqrt{s}=1.96$~TeV for $100\leq m_H \leq 200$ GeV/$c^2$.
To facilitate comparisons with the standard model and to accommodate
analyses with different degrees of sensitivity and acceptance for more than one
signal production mechanism, we present our resulting limit
divided by the SM Higgs boson production
cross section, as a function of Higgs boson mass, for test masses for
which both experiments have performed dedicated searches in different
channels.  A value of the combined limit ratio which is less than or
equal to one indicates that that particular Higgs boson mass is excluded
at the 95\% C.L.

The combinations of results~\cite{CDFHiggs,DZHiggs} of each single
experiment, as used in this Tevatron combination, yield the following
ratios of 95\% C.L. observed (expected) limits to the SM cross section:
2.37~(1.16) for CDF and 2.17~(1.58) for D0 at $m_{H}=115$~GeV/$c^2$,
2.90~(1.41) for CDF and 2.53~(1.85) for D0 at $m_{H}=125$~GeV/$c^2$, and
0.42~(0.69) for CDF and 0.94~(0.76) for D0 at $m_{H}=165$~GeV/$c^2$.

\begin{figure}[hb]
\begin{centering}
\includegraphics[width=16.5cm]{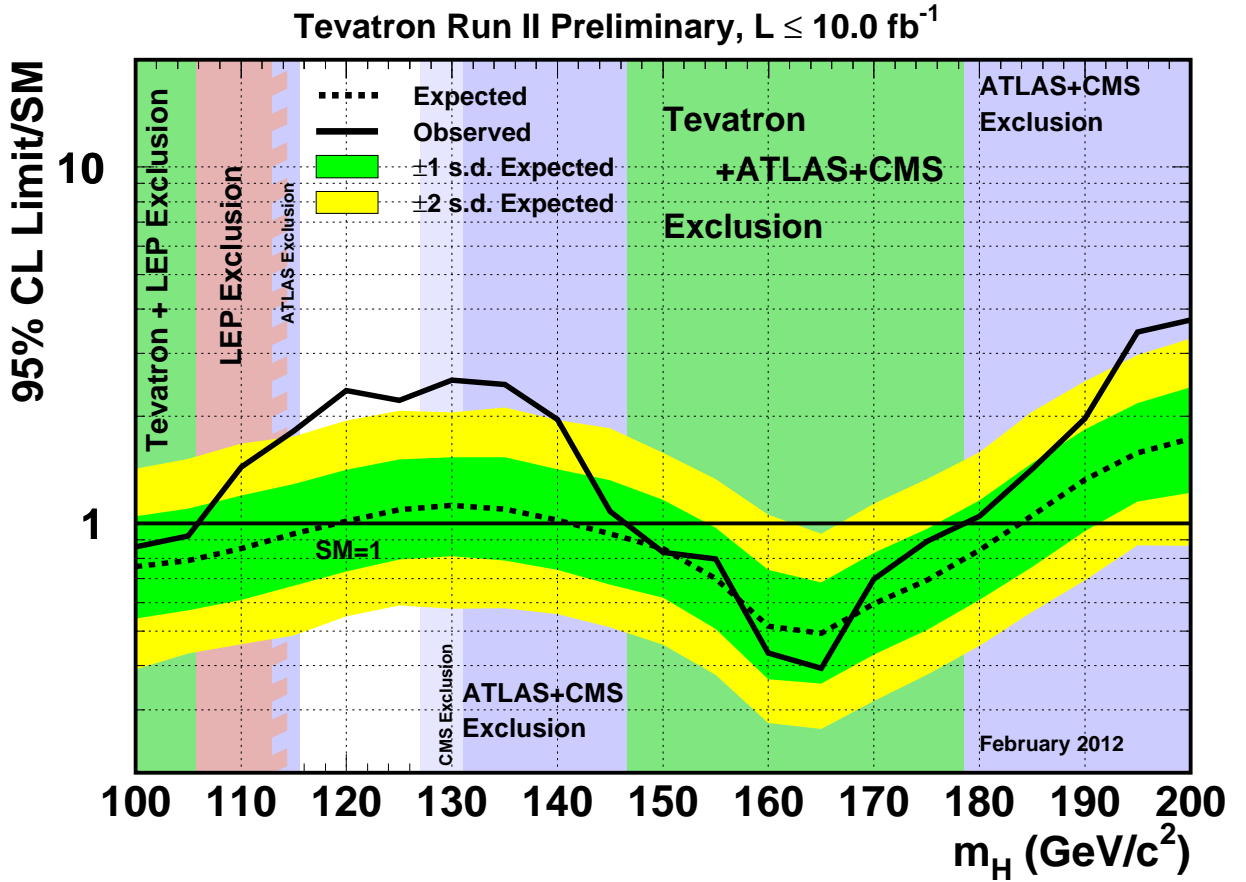}
\caption{
\label{fig:comboRatio}
Observed and expected (median, for the background-only hypothesis)
95\% C.L. upper limits on the ratios to the SM cross section, as
functions of the Higgs boson mass
for the combined CDF and D0 analyses.
The limits are expressed as a multiple of the SM prediction
for test masses (every 5 GeV/$c^2$)
for which both experiments have performed dedicated
searches in different channels.
The points are joined by straight lines
for better readability.
  The bands indicate the
68\% and 95\% probability regions where the limits can
fluctuate, in the absence of signal.
The limits displayed in this figure
are obtained with the Bayesian calculation.
}
\end{centering}
\end{figure}

\begin{table}[ht]
\caption{\label{tab:ratios} Ratios of median expected and observed 95\% C.L.
limit to the SM cross section for the combined CDF and D0 analyses as a function
of the Higgs boson mass in GeV/$c^2$, obtained with the Bayesian and with the ${\rm CL}_{\rm s}$ method.}
\begin{ruledtabular}
\begin{tabular}{lccccccccccc}\\
Bayesian             &  100 &  105 &  110 &  115 &  120 &  125 &  130 &  135 &  140 &  145 &  150 \\ \hline
Expected             & 0.76 & 0.79 & 0.85 & 0.94 & 1.01 & 1.10 & 1.12 & 1.10 & 1.02 & 0.93 & 0.85 \\
Observed             & 0.86 & 0.92 & 1.44 & 1.82 & 2.36 & 2.22 & 2.52 & 2.46 & 1.96 & 1.08 & 0.83 \\

\hline
\hline\\
${\rm CL}_{\rm s}$   &  100 &  105 &  110 &  115 &  120 &  125 &  130 &  135 &  140 &  145 &  150 \\ \hline
Expected             & 0.76 & 0.80 & 0.86 & 0.92 & 1.02 & 1.11 & 1.13 & 1.12 & 1.05 & 0.95 & 0.84 \\
Observed             & 0.84 & 0.97 & 1.52 & 1.88 & 2.20 & 2.23 & 2.65 & 2.62 & 1.93 & 1.07 & 0.83 \\
\end{tabular}
\end{ruledtabular}
\end{table}

\begin{table}[ht]
\caption{\label{tab:ratios-3}
Ratios of median expected and observed 95\% C.L.
limit to the SM cross section for the combined CDF and D0 analyses as a function
of the Higgs boson mass in GeV/$c^2$, obtained with the Bayesian and with the ${\rm CL}_{\rm s}$ method.}
\begin{ruledtabular}
\begin{tabular}{lccccccccccc}
Bayesian             &  155 &  160 &  165 &  170 &  175 &  180 &  185 &  190 &  195 &  200 \\ \hline
Expected             & 0.70 & 0.52 & 0.49 & 0.60 & 0.69 & 0.84 & 1.05 & 1.33 & 1.58 & 1.73 \\
Observed             & 0.80 & 0.43 & 0.39 & 0.70 & 0.89 & 1.05 & 1.42 & 1.97 & 3.45 & 3.73 \\
\hline
\hline\\
${\rm CL}_{\rm s}$   &  155 &  160 &  165 &  170 &  175 &  180 &  185 &  190 &  195 &  200 \\ \hline
Expected             & 0.74 & 0.53 & 0.50 & 0.62 & 0.73 & 0.87 & 1.10 & 1.38 & 1.61 & 1.84 \\
Observed             & 0.74 & 0.43 & 0.38 & 0.68 & 0.89 & 1.04 & 1.47 & 2.09 & 3.56 & 4.06 \\
\end{tabular}
\end{ruledtabular}
\end{table}

The ratios of the 95\% C.L. expected and observed limit to the SM cross
section are shown in Figure~\ref{fig:comboRatio} for the combined CDF
and D0 analyses.  The observed and median expected ratios are listed
for the tested Higgs boson masses in Table~\ref{tab:ratios} for $m_{H}
\leq 150$~GeV/$c^2$, and in Table~\ref{tab:ratios-3} for $m_{H} \geq
155$~GeV/$c^2$, as obtained by the Bayesian and the ${\rm CL}_{\rm s}$
methods.  In the following summary we quote only the limits obtained
with the Bayesian method, which was decided upon {\it a priori}.
The corresponding limits and expected limits obtained using the
${\rm CL}_{\rm s}$ method are shown alongside the Bayesian limits in
the tables.  We obtain the observed (expected) values of
0.92~(0.79) at $m_{H}=105$~GeV/$c^2$, 
1.82~(0.94) at $m_{H}=115$~GeV/$c^2$,
2.22~(1.10) at $m_{H}=125$~GeV/$c^2$,
1.08~(0.93) at $m_{H}=145$~GeV/$c^2$, 
0.39~(0.49) at $m_{H}=165$~GeV/$c^2$,
and 1.42~(1.05) at $m_{H}=185$~GeV/$c^2$. 

We choose to use the intersections of piecewise linear interpolations of our observed and expected
rate limits in order to quote ranges of Higgs boson masses that are excluded and that are expected
to be excluded.  The sensitivities of our searches to Higgs bosons are smooth functions of the Higgs
boson mass and depend most strongly on the predicted cross sections and the decay branching ratios
(the decay $H\rightarrow W^+W^-$ is the dominant decay for the region of highest sensitivity).
We therefore use the linear interpolations to extend the results from the 5~GeV/$c^2$ mass grid 
investigated to points in between.  The regions of Higgs boson masses excluded at the 95\% C.L. 
thus obtained are $100<m_H<106$~GeV/$c^{2}$ and $147<m_{H}<179$~GeV/$c^{2}$.  The expected exclusion 
regions are, given the current sensitivity, $100<m_H<119$~GeV/$c^{2}$ and $141<m_{H}<184$~GeV/$c^{2}$.
Higgs boson masses 
below 100~GeV/$c^{2}$ were not studied. We also show in Figure~\ref{fig:comboCLS}, and list in Table~\ref{tab:clsVals},
the observed values of 1-${\rm CL}_{\rm s}$ and their expected distributions for the background-only hypothesis as functions 
of the Higgs boson mass. The excluded regions obtained by finding 
the intersections of the linear interpolations of the observed $1-{\rm CL}_{\rm s}$ curve are
 nearly identical to those obtained with the Bayesian calculation.  

\begin{figure}[t]
\begin{centering}
\includegraphics[width=0.73\textwidth]{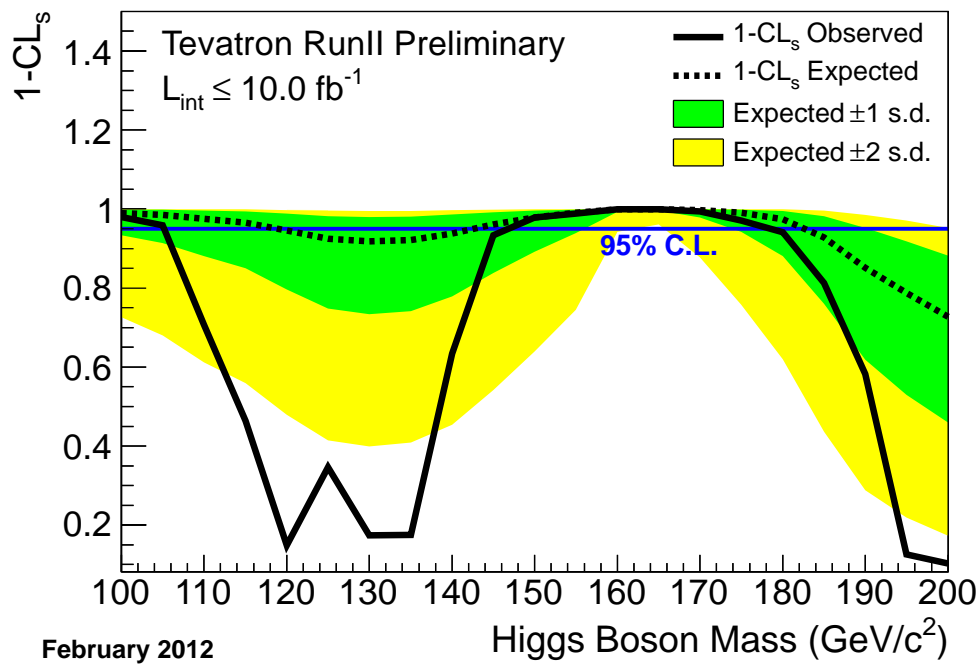} \\
\caption{
\label{fig:comboCLS}
The exclusion strength 1-${\rm CL}_{\rm s}$ as a function of the Higgs boson mass (in steps 
of 5 GeV/$c^2$), for the combination of the CDF and D0 analyses. The green and yellow bands 
correspond to the regions enclosing 1~s.d. and 2~s.d. fluctuations around the median predicted
value in the background-only hypothesis, 
respectively.}
\end{centering}
\end{figure}

Figure~\ref{fig:comboCLSB} shows the $p$-value ${\rm CL}_{\rm s+b}$ as a function of $m_H$ as well as the expected 
distributions in the absence of a Higgs boson signal.   Figure~\ref{fig:comboCLB} shows 
the $p$-value 1-${\rm CL}_{\rm b}$ as a function of $m_H$, i.e., the 
probability that an upward fluctuation of the background can give an outcome as signal-like as the data or more.
In the absence of a Higgs boson signal, the observed $p$-value is expected to be uniformly distributed between 0 and 1.
A small $p$-value indicates that the data are not easily explained by the background-only hypothesis,
and that the data prefer the signal-plus-background prediction.   Our sensitivity to a Higgs boson
with a mass of 165 GeV/$c^{2}$ is such that
we would expect to see a $p$-value corresponding to $\sim 4$~s.d. in half of the experimental outcomes.
The smallest observed $p$-value corresponds to a Higgs boson mass of 120 GeV/$c^{2}$.  
The fluctuations seen in the observed $p$-value as a function of the tested $m_H$ result from excesses seen 
in different search channels, as well as from point-to-point fluctuations due to the separate discriminants 
at each $m_H$, and are discussed in more detail below.  The width of the dip in the $p$-values from 115 to 
135 GeV/$c^{2}$ is consistent with the resolution of the combination of the $H \to b\bar{b}$ and $H \to W^+W^-$ channels.   The
effective resolution of this search comes from two independent sources of information.  The reconstructed candidate
masses help constrain $m_H$, but more importantly, the expected cross sections times the relevant branching ratios for the
$H \to b\bar{b}$ and $H \to W^+W^-$ channels are strong functions of $m_H$ in the SM.  The observed excesses in
the $H \to b\bar{b}$ channels coupled with a more background-like outcome in the $H \to W^+W^-$ channels
determines the shape of the observed $p$-value as a function of $m_H$.

\begin{figure}[t]
\begin{centering}
\includegraphics[width=0.7\textwidth]{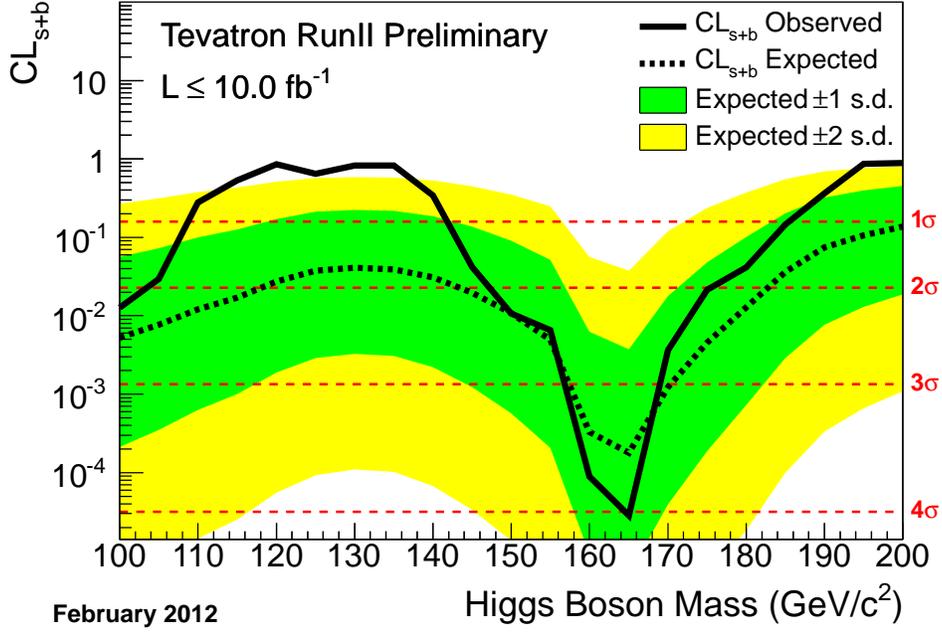} \\
\caption{
\label{fig:comboCLSB}
The signal $p$-values ${\rm CL}_{\rm s+b}$ as a function of the Higgs boson mass (in steps 
of 5 GeV/$c^2$), for the combination of the CDF and D0 analyses. The green and yellow bands 
correspond to the regions enclosing 1~s.d. and 2~s.d. fluctuations around the median predicted
value in the background-only hypothesis, 
respectively.}
\end{centering}
\end{figure}

\begin{figure}[t]
\begin{centering}
\includegraphics[width=0.7\textwidth]{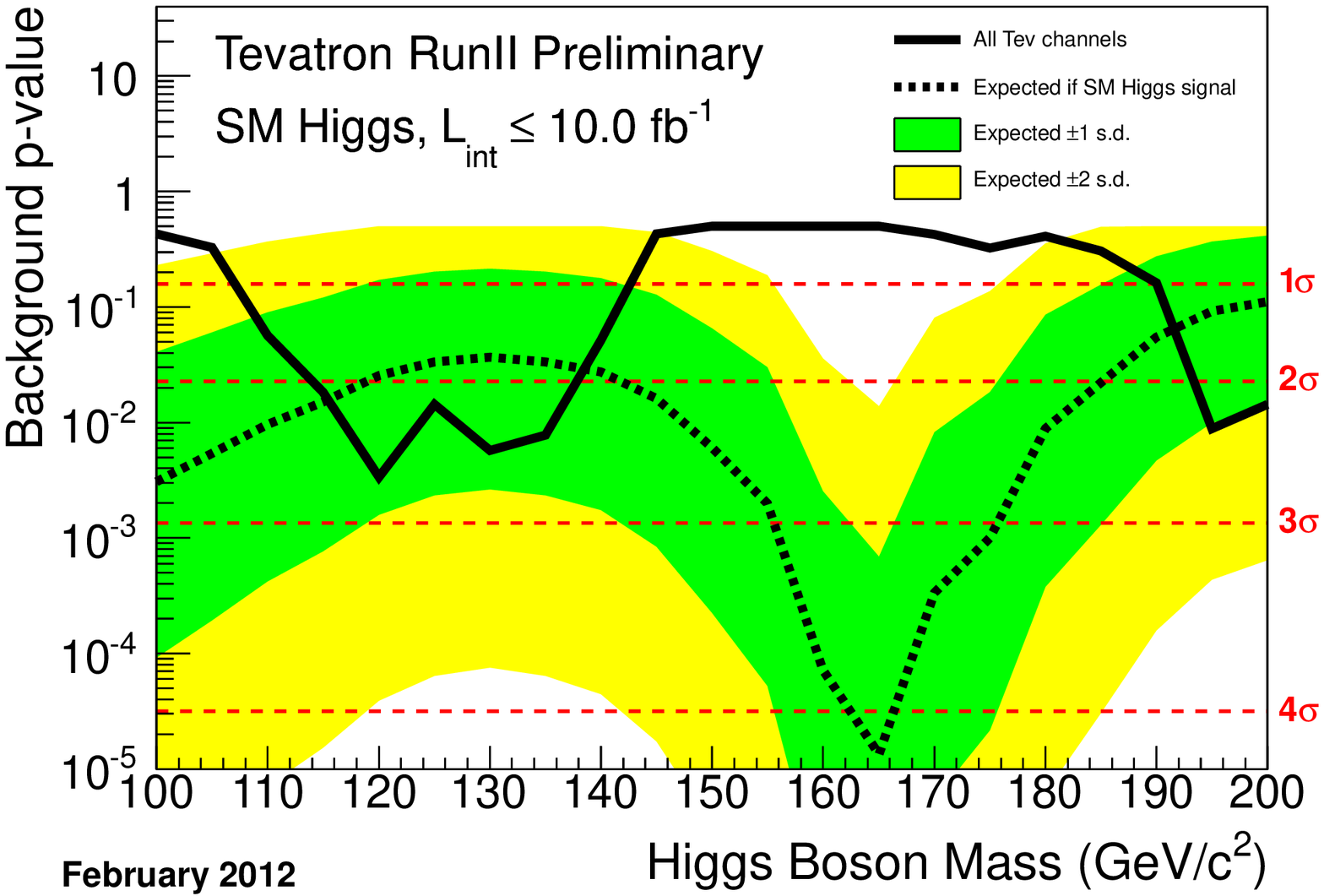} \\
\caption{
\label{fig:comboCLB}
The background $p$-values 1-${\rm CL}_{\rm b}$ as a function of the Higgs boson mass (in steps 
of 5 GeV/$c^2$), for the combination of the CDF and D0 analyses. The green and yellow bands 
correspond respectively to the regions enclosing 1~s.d. and 2~s.d. fluctuations around the median prediction
in the signal plus background hypothesis at each value of $m_H$.}
\end{centering}
\end{figure}

We perform a fit of the signal-plus-background hypothesis to the 
observed data, allowing the signal strength to vary as a function of $m_H$.  The resulting best-fit signal
strength is shown in Figure~\ref{fig:combBestFit}, normalized to the SM prediction.  
The signal strength is within 1~s.d. of the SM expectation with a Higgs boson signal in the range 
$110<m_H<140$~GeV$/c^2$.   
The largest signal fit in this range, normalized to the SM prediction, is obtained at 130 GeV$/c^2$.  
The reason the highest signal strength is at 130 GeV/$c^{2}$ while 
the smallest $p$-value from Figure~\ref{fig:comboCLB} is at 120 GeV$/c^2$ is because a signal 
at 120 GeV$/c^2$ would have a higher cross section than a signal at 130 GeV$/c^2$, and since the 
resolution of the discriminants cannot distinguish very well such a small 
mass difference, a signal at 120 GeV$/c^2$ would be similar to a signal at 130 GeV$/c^2$ with a larger 
scale factor for the predicted cross section.

\begin{figure}[ht]
\begin{centering}
\includegraphics[width=0.7\textwidth]{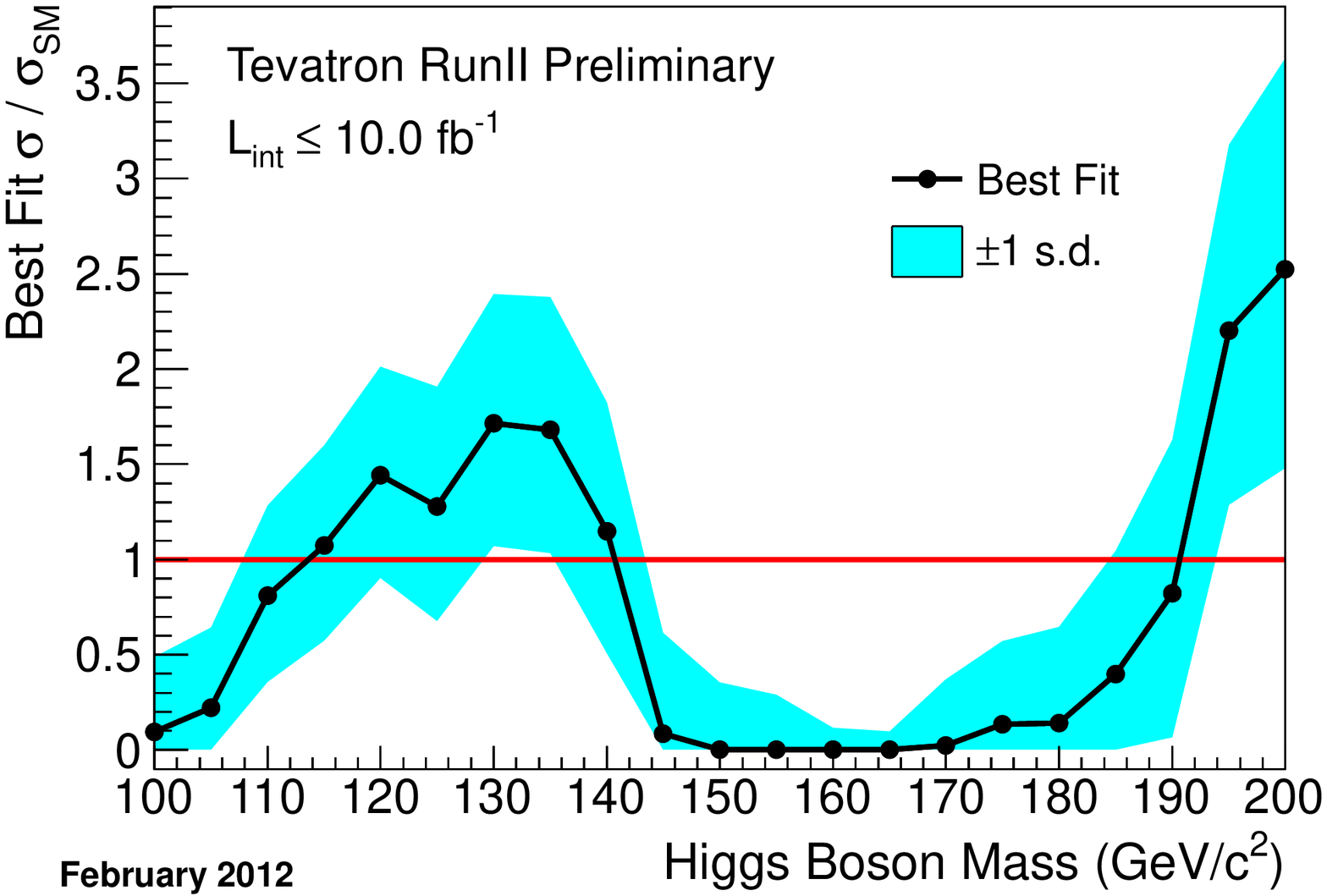}
\caption{
\label{fig:combBestFit}
The best fit signal cross section of all CDF and D0 search channels combined shown as a ratio to the 
standard model cross section as a function of the tested Higgs boson mass.  The horizontal line at 1 represents the signal 
strength expected for a standard model Higgs boson hypothesis.  The blue band shows the 1~s.d. 
uncertainty on the signal fit. 
}
\end{centering}
\end{figure}

Figure~\ref{fig:tevcombochi2} shows $\Delta\chi^2 = {\rm{LLR}_{\rm{obs}}}-{\rm{LLR}_b}$,
which is an estimate of how discrepant the observed data are with the median expectation from
the prediction of the background-only hypothesis,
as a function of $m_H$.  Significantly negative values of $\Delta\chi^2$ indicate a preference in the data for the 
signature of Higgs boson production.

\begin{figure}[ht]
\begin{centering}
\includegraphics[width=0.7\textwidth]{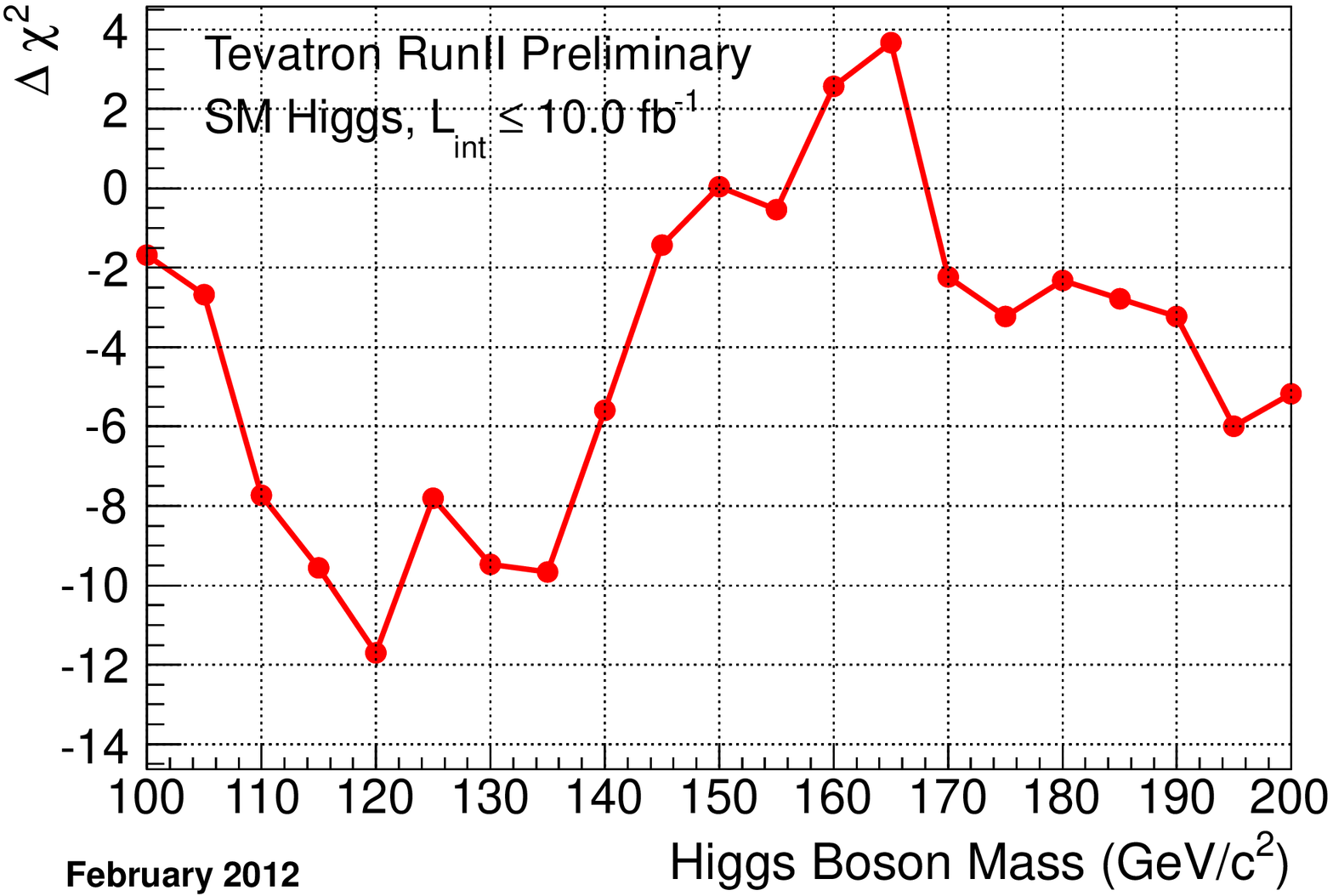}
\caption{
\label{fig:tevcombochi2}
The curve shows $\Delta\chi^2 = {\rm{LLR}_{\rm{obs}}}-{\rm{LLR}_b}$, an 
estimate of how discrepant the observed data are with the median expectation from
the prediction of the background-only hypothesis, as a function of $m_H$.
Significantly negative values of  $\Delta\chi^2$ indicate
a preference in the data for the signature of Higgs boson production.  
}
\end{centering}
\end{figure}
  

We also investigate combinations of CDF and \DZ searches based on the $H\rightarrow b{\bar{b}}$ 
and $H\rightarrow W^+W^-$ decay modes.   Below 125~GeV/$c^2$, the $H\rightarrow b{\bar{b}}$ searches 
contribute the majority of our sensitivity.  The $WH\rightarrow \ell\nu b\bar{b}$, $ZH\rightarrow 
\nu\bar{\nu} b\bar{b}$, and $ZH\rightarrow \ell^+\ell^- b\bar{b}$ channels from both experiments 
are included in this combination.  The result is shown in Figure~\ref{fig:comboRatiobb}.  The 
distribution of the LLR
demonstrates the compatibility of the observed data with both the background-only and 
signal-plus-background hypotheses, and is shown in Figure~\ref{fig:hbbLLR}.  An interesting feature
of this graph is that as $m_H$ increases towards the high end of the range shown, $Br(H\rightarrow b{\bar{b}})$
falls rapidly, and the expected signal yield becomes small.  Thus LLR approaches zero as $m_H$ gets larger,
independent of the experimental outcome.   This feature can also be seen with the shaded bands which also
converge on zero at high $m_H$.  If there is a broad excess in the $H\rightarrow b{\bar{b}}$ searches,
then LLR will fall to a minimum value and rise again.  

Figure~\ref{fig:hbbCLsb} 
shows the observed and expected values of CL$_{s+b}$  as functions of m$_H$.
Figure~\ref{fig:hbbCLb} shows the $p$-value for the background-only hypothesis 1 - CL$_b$, which 
represents the probability for the background to fluctuate to produce an outcome as signal-like as 
the observed data or more.  The smallest $p$-value within the mass range where these searches are 
performed, $100<m_H<150$~GeV/$c^2$, corresponds to a significance of approximately 2.8~s.d.

These probabilities do not include the look-elsewhere effect (LEE), and are thus local $p$-values, 
corresponding to searches at each value of $m_H$ separately.  The LEE accounts for the probability 
of observing an upwards fluctuation of the background at any of the tested values of $m_H$ in our 
search region, at least as significant as the one observed at the value of $m_H$ with the most 
significant local excess.  A simple and correct method of calculating the LEE, and thus the global 
significance of the excess, is to simulate many possible experimental outcomes assuming the absence
of a signal, and for each one, compute the LLR and the fitted cross section curves and find the deviation with the smallest 
background-only-hypothesis $p$-value.  Using this minimum $p$-value as a test statistic, another 
$p$-value is then computed, which is the probability of observing that minimum $p$-value or less.  
This method is difficult to pursue in the Tevatron Higgs boson searches due to the fact that in 
most search analyses, a distinct multivariate analysis (MVA) discriminant function is trained for 
each value of $m_H$ that is tested.  This step is an important optimization, because the kinematic 
distributions and signal branching ratios are functions of $m_H$, but it introduces the difficulty of 
running the same set of simulated events separately through many MVA functions in order to compute 
the LEE with the simple method.  The use of a separate MVA function at each $m_H$ also introduces 
additional point-to-point randomness as individual events are reclassified from bins with lower
$s/b$ to higher $s/b$ and vice versa.  Even though the discriminants are nearly optimal and are 
thus highly similar from one $m_H$ value to the next, small variations are amplified by the discrete 
nature of the data which are processed through these MVAs.  One may see this in the variations 
of observed limits, LLR values and $p$-values from one mass point to the next which show more rapid 
variation than can be explained from mass resolution effects alone.

Gross and Vitells~\cite{grossvitells} provide a technique that extrapolates from a smaller sample of 
background-only Monte Carlo simulations fully propagated through the MVA discriminant functions.  We lack the ability to 
perform this propagation through all of our channels, as we rely on exchanged histograms of distributions 
of selected events.  We therefore estimate the LEE effect in a simplified manner.  In the mass range 
100--130~GeV$/c^2$, where the low-mass $H\rightarrow b{\bar{b}}$ searches dominate, the reconstructed 
mass resolution is approximately 10-15\%, or about 15~GeV$/c^2$.  We therefore estimate a LEE factor 
of $\sim 2$ for the low-mass region.  The $H\rightarrow\gamma\gamma$ searches have a much better mass 
resolution, of order 3\%, but their contribution to the final LLR is small due to the much smaller 
$s/b$ in those searches.  They introduce more rapid oscillations of LLR as a function of $m_H$, but the 
magnitude of these oscillations is much smaller than those induced by the $H\rightarrow b{\bar{b}}$ 
searches.  The $H\rightarrow \tau^+\tau^-$ searches have both worse reconstructed mass resolution and 
lower $s/b$ than the $H\rightarrow b{\bar{b}}$ searches and therefore similarly do not play a significant 
role in the estimation of the LEE.  Applying the LEE of 2 to the most significant local $p$-value obtained 
from our $H\rightarrow b{\bar{b}}$ combination, we obtain a global significance of approximately 2.6~s.d.

We perform a fit of the signal-plus-background hypothesis to the observed data, allowing the signal strength 
to vary as a function of $m_H$. The resulting best-fit signal strength is shown in Figure~\ref{fig:hbbBest},
normalized to the SM prediction. The $H \to b\bar{b}$ excess comes mainly from the CDF channels, which have 
combined $>2$~s.d. excesses, with the most signal-like candidates coming from CDF's $ZH \to \ell\ell 
b\bar{b}$ channel.  The $WH\rightarrow \ell\nu b\bar{b}$, $ZH \rightarrow \nu\bar{\nu} b\bar{b}$, and 
$ZH \rightarrow \ell\ell b\bar{b}$ search channels all contribute to the increase in significance of the 
CDF excess with respect to previous combinations.  The larger excesses found in each individual channel are 
consistent with the large numbers of new events being added to the searches through the analysis of new data 
and use of the improved neural-network $b$-tagging algorithm.  For the $ZH \to \ell\ell b\bar{b}$ channel,
which sees the largest change in the significance of its observed excess, more than half of the currently 
analyzed data events were not contained within previous analyses of this channel.  The \Dzero $H\to b\bar{b}$ 
channels see a $\sim$~1~s.d. excess, consistent with the signal-plus-background hypothesis. 
 
Above 125 GeV/$c^2$, the $H\rightarrow W^+W^-$ channels contribute the majority of our search sensitivity. 
We combine all $H \to W^+W^-$ searches from CDF and D0, incorporating potential signal contributions 
from $gg \to H$, $WH$, $ZH$, and VBF production.   The result of this combination is shown in 
Figure~\ref{fig:comboRatioWW}.  The distribution of the LLR
is shown in Figure~\ref{fig:hwwLLR}, which shows good agreement overall with the background-only 
hypothesis.  Where the sensitivity is low, for $m_H=115$~GeV/$c^2$ and $m_H \ge$~190~GeV/$c^2$, the data 
are slightly more compatible with the signal-plus-background hypothesis.  Figure~\ref{fig:hwwCLsb} shows 
the observed and expected 
CL$_{s+b}$ distribution 
as a function of m$_H$.  Figure~\ref{fig:hwwCLb} shows the $p$-value for the background-only hypothesis.
We perform a fit of the observed data to the signal-plus-background hypothesis, allowing the signal 
strength to vary in the fit as a function of $m_H$ as shown in Figure~\ref{fig:hwwBest}.  Consistent 
with Figure~\ref{fig:hwwLLR} the combined observed data do not indicate any significant excesses, though the 
\Dzero $H \to W^+W^-$ analysis has a slight excess ($\sim 1.5$~s.d.) from 130 to 140 GeV/$c^2$ consistent 
with the signal-plus-background hypothesis.

The $H\rightarrow W^+W^-$ analyses which dominate the sensitivity of our high mass searches have   
poor resolution for reconstructing $m_H$ due to the presence of two neutrinos in the final 
states of the most sensitive channels, and we thus expect the outcomes in these searches at each $m_H$ in the
high-mass range to be highly correlated with each other.
Above $m_H=2M_W$, the $W$ bosons are on shell, and the kinematic 
variables take on different weights in the training of the MVAs than they do at masses below 
$2M_W$.  At very high masses, the discriminating variable 
$\Delta R_{\rm{leptons}}=\sqrt{\Delta\phi_{\rm{leptons}}^2 + \Delta\eta_{\rm{leptons}}}$~\cite{cdfHWW,dzHWW} plays 
less of a role than it does near the $W^+W^-$ threshold.  We therefore expect a LEE factor 
of approximately two for our high-mass searches in the mass range 130~$< m_H <$~200~GeV/$c^2$.  Over 
the entire mass range of our Higgs searches, 100~$< m_H <$~200~GeV/$c^2$, we therefore expect that 
there are roughly four possible independent locations for uncorrelated excesses to appear in our 
analysis.  The global $p$-value associated with our entire suite of Higgs searches is therefore 
$1-(1-p_{\rm{min}})^4$, using the Dunn-\^Sid\'ak correction~\cite{dunn}.  Based on this approach, if   
we simply chose to consider the region not currently excluded by other experiments, our resulting LEE
factor would be one, making the global significance equivalent to the local significance.  The smallest 
local $p$-value obtained from the full combination of CDF and \DZ SM Higgs searches has a significance 
of approximately 2.7~s.d.  Applying a LEE of 4 to this value, we obtain a global significance of 
approximately 2.2~s.d.  

As a final step, we separately combine CDF and D0 searches for $H\rightarrow\gamma\gamma$, and display 
the resulting limits on the production cross section times the decay branching ratio normalized to the 
SM prediction in Figure~\ref{fig:comboRatiogamgam}.

In summary, we combine all available CDF and D0 results on SM Higgs boson searches, based on 
luminosities ranging from 4.3 to 10.0 fb$^{-1}$. Compared to our previous combination, more 
data have been added to the existing channels, additional channels have been included, and 
analyses have been further optimized to gain sensitivity.  The results presented here 
significantly extend the individual limits of each collaboration and those obtained in our 
previous combination.  The sensitivity of our combined search is sufficient to exclude a Higgs 
boson at high mass and is, in the absence of signal, expected to grow in the future as further 
improvements are made to our analysis techniques.  
There is an excess of data events with respect 
to the background estimation in the mass range $115<m_{H}<135$~GeV/$c^{2}$ which 
causes our limits to not be as stringent as expected.  At $m_H=120$~GeV/$c^{2}$, 
the $p$-value for a background fluctuation to produce this excess is $\sim$3.5$\times$10$^{-3}$, 
corresponding to a local significance of 2.7 standard deviations.  The global significance for 
such an excess anywhere in the full mass range is approximately 2.2 standard deviations, after accounting
for the look-elsewhere effect. 

In addition, we separate the CDF and D0 searches into combinations focusing on the $H\to b\bar{b}$ 
and $H \to W^+W^-$ channels. The largest excess is observed in the $H\to b\bar{b}$ channels, 
corresponding to a local significance of $\approx 2.8$~s.d. prior to accounting for the look 
elsewhere effect of $\sim$2, which, when included, yields a global significance of $\approx 2.6$~s.d.

\begin{figure}[hb]
\begin{centering}
\includegraphics[width=0.7\textwidth]{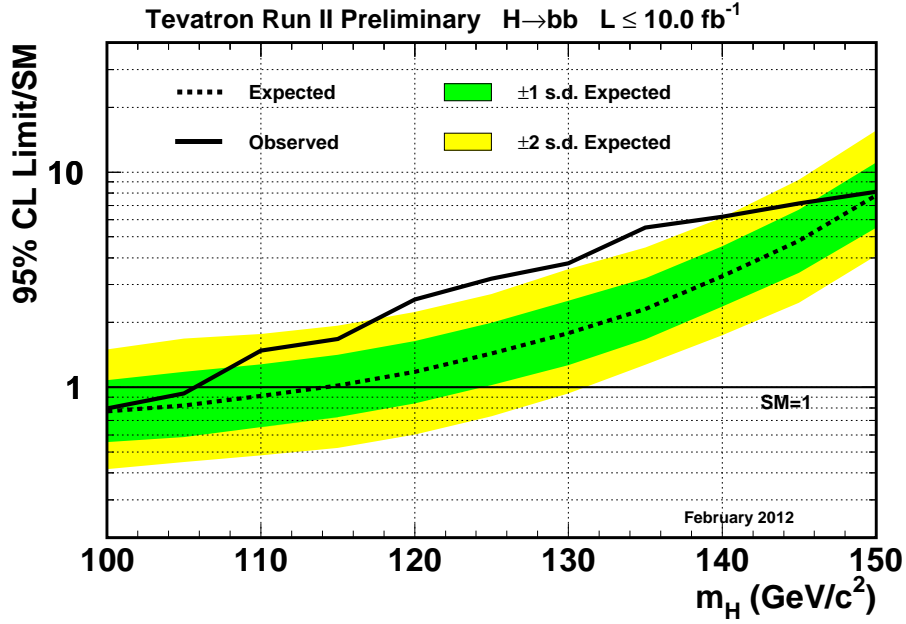}
\caption{
\label{fig:comboRatiobb}
Observed and expected (median, for the background-only hypothesis)
95\% C.L. upper limits on the ratios to the SM cross section,
as functions of the Higgs boson mass
for the combination of CDF and D0 analyses focusing on the $H
\rightarrow b\bar{b}$ decay channel.  The limits are expressed
as a multiple of the SM prediction for test masses (every 5
GeV/$c^2$) for which both experiments have performed dedicated
searches in different channels.
The points are joined by straight lines for better readability.
The bands indicate the 68\% and 95\% probability regions where
the limits can fluctuate, in the absence of signal.  The limits
displayed in this figure are obtained with the Bayesian calculation.
}
\end{centering}
\end{figure}

\begin{figure}[hb]
\begin{centering}
\includegraphics[width=0.7\textwidth]{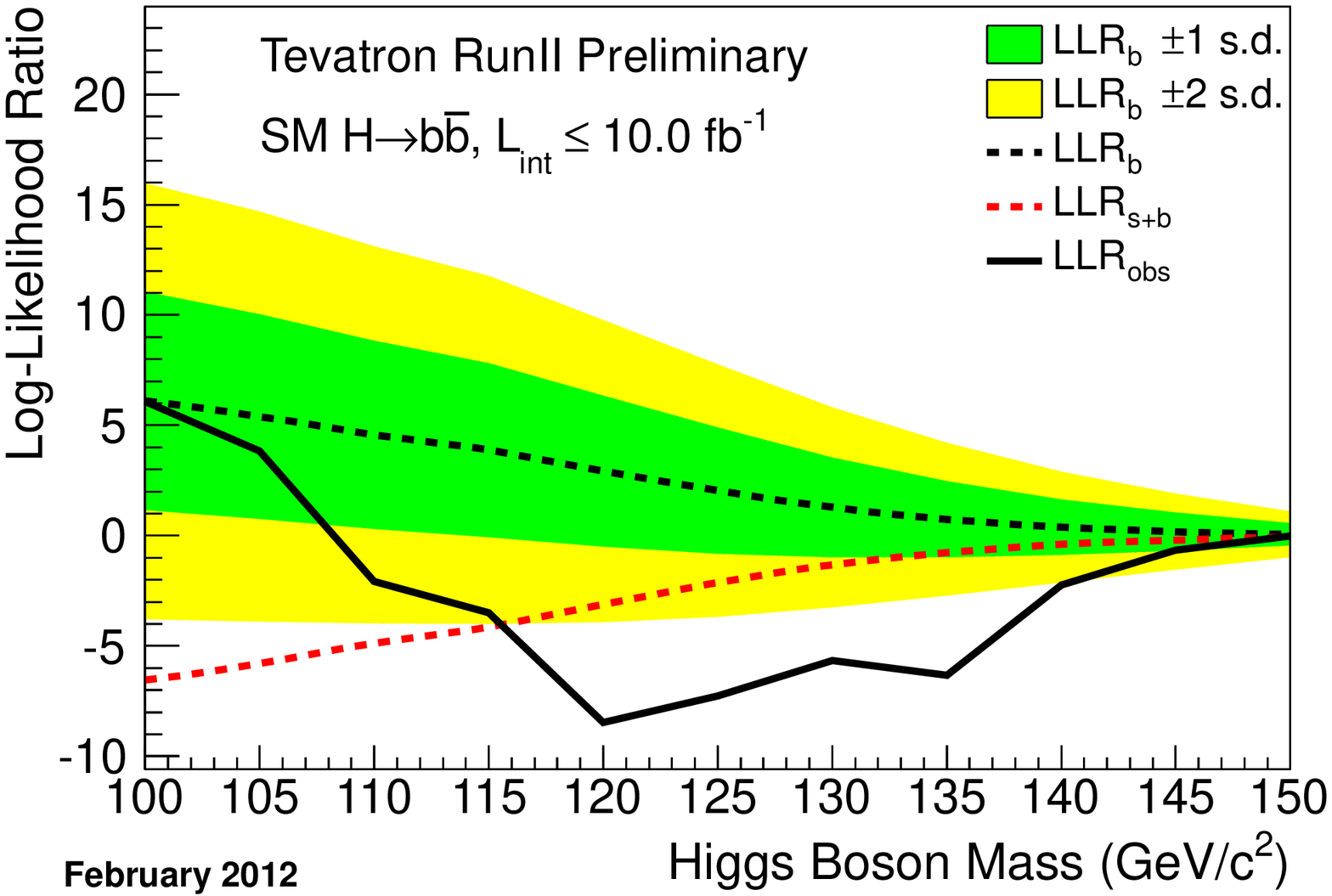}
\caption{
\label{fig:hbbLLR}
Distributions of the log-likelihood ratio (LLR) as a function of Higgs boson mass obtained with
the ${\rm CL}_{\rm s}$ method for the combination of all CDF and D0 analyses in the $H \to b\bar{b}$ channels. 
The green and yellow bands correspond to the regions enclosing 1~s.d. and 2~s.d. fluctuations 
of the background, respectively.}
\end{centering}
\end{figure}

\newpage
\begin{figure}[t]
\begin{centering}
\includegraphics[width=0.7\textwidth]{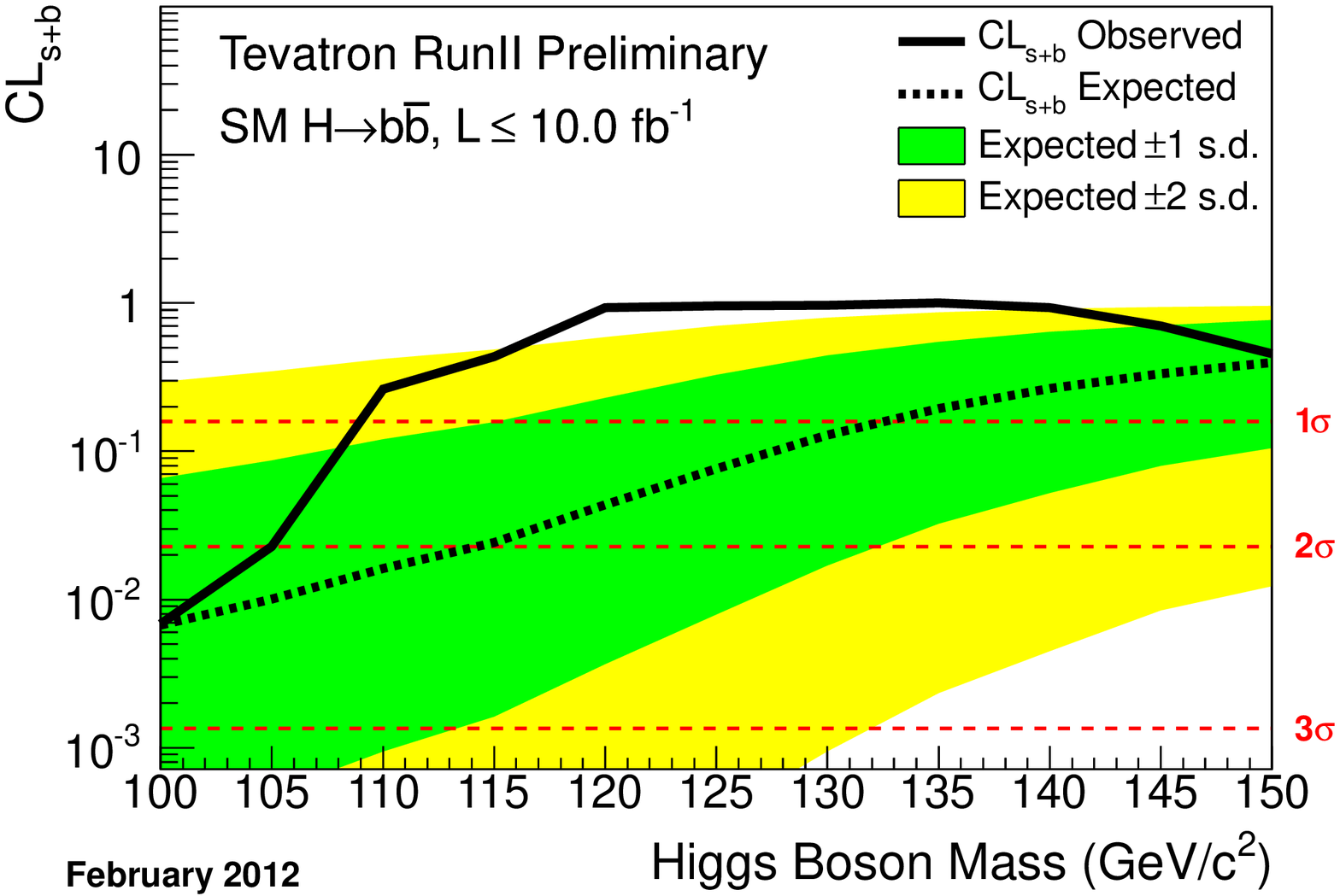} \\
\caption{
\label{fig:hbbCLsb}
The signal $p$-values ${\rm CL}_{\rm s+b}$ for the signal plus background hypothesis 
as a function of the Higgs boson mass (in steps 
of 5 GeV/$c^2$), for the combination of all CDF and D0 analyses in the $H \to b\bar{b}$ channels. 
 The green and yellow bands 
correspond to the regions enclosing 1~s.d. and 2~s.d. fluctuations of the background, 
respectively.}
\end{centering}
\end{figure}

\begin{figure}[t]
\begin{centering}
\includegraphics[width=0.7\textwidth]{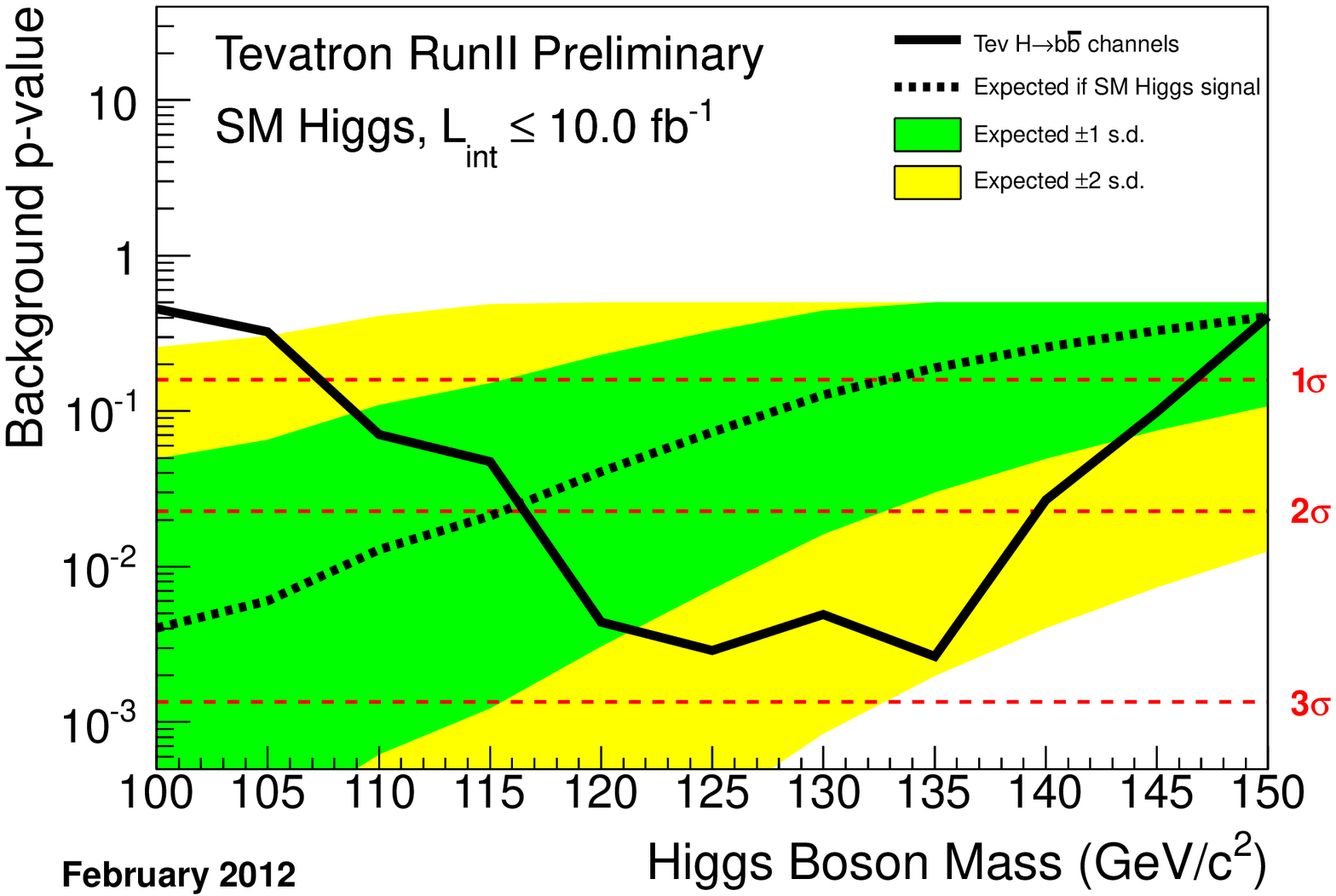} \\
\caption{
\label{fig:hbbCLb}
The background $p$-values 1-${\rm CL}_{\rm b}$ for the null hypothesis as a function of the Higgs boson mass (in steps 
of 5 GeV/$c^2$), for the combination of all CDF and D0 analyses in the $H \to b\bar{b}$ channels. 
The green and yellow bands correspond to the regions enclosing 1~s.d. and 2~s.d. fluctuations of the background, 
respectively.}
\end{centering}
\end{figure}

\newpage
 \begin{figure}[t]
 \begin{centering}
\includegraphics[width=0.7\textwidth]{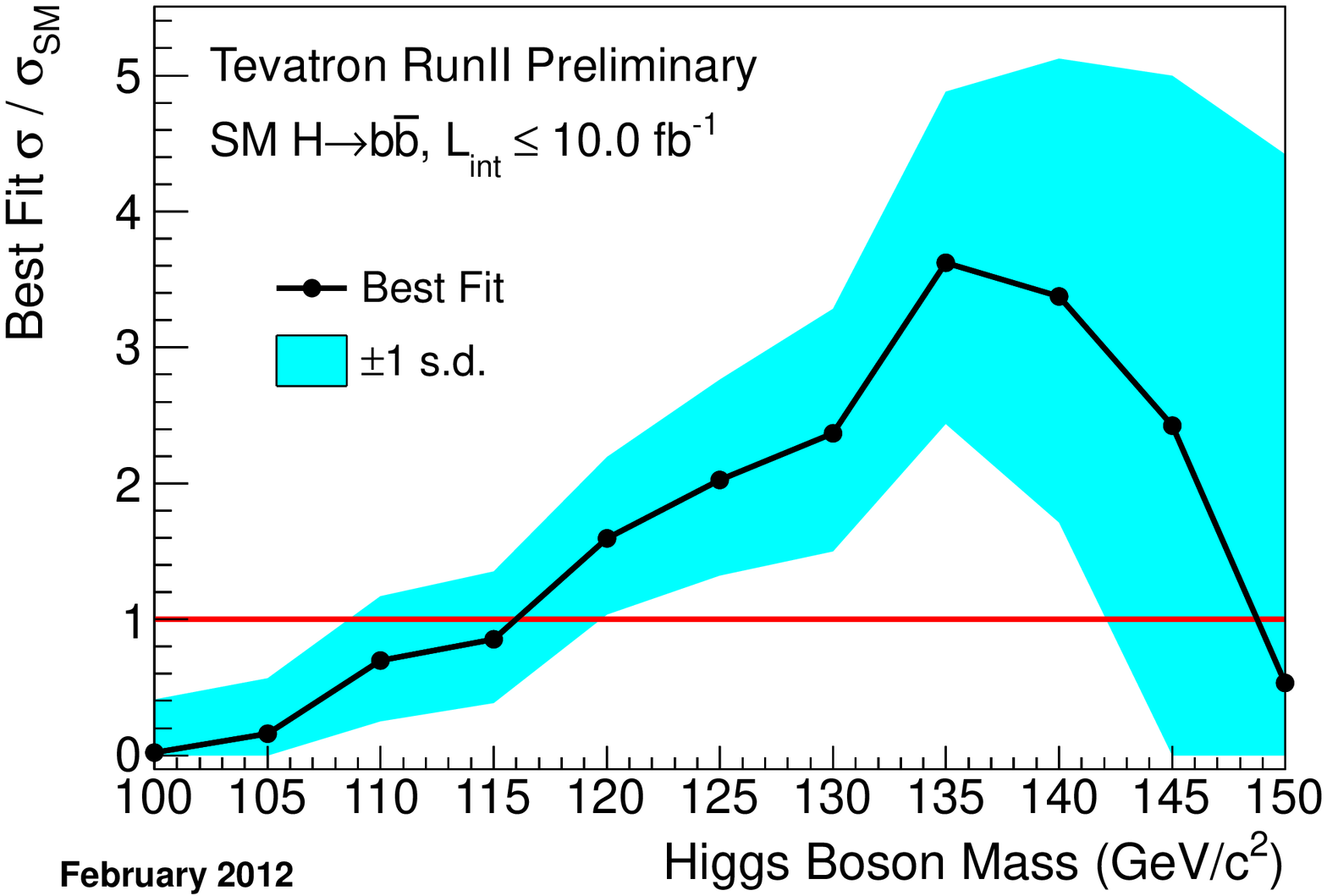} \\
\caption{
 \label{fig:hbbBest}
 The best fit of the signal cross section as a function of the Higgs boson mass
(in steps of 5 GeV/$c^2$), for the combination of all 
 CDF and D0 analyses in the $H\to b\bar{b}$ channels. The blue band shows the 1~s.d. uncertainty on the signal fit.}
 \end{centering}
 \end{figure}


\begin{figure}[hb]
\begin{centering}
\includegraphics[width=0.7\textwidth]{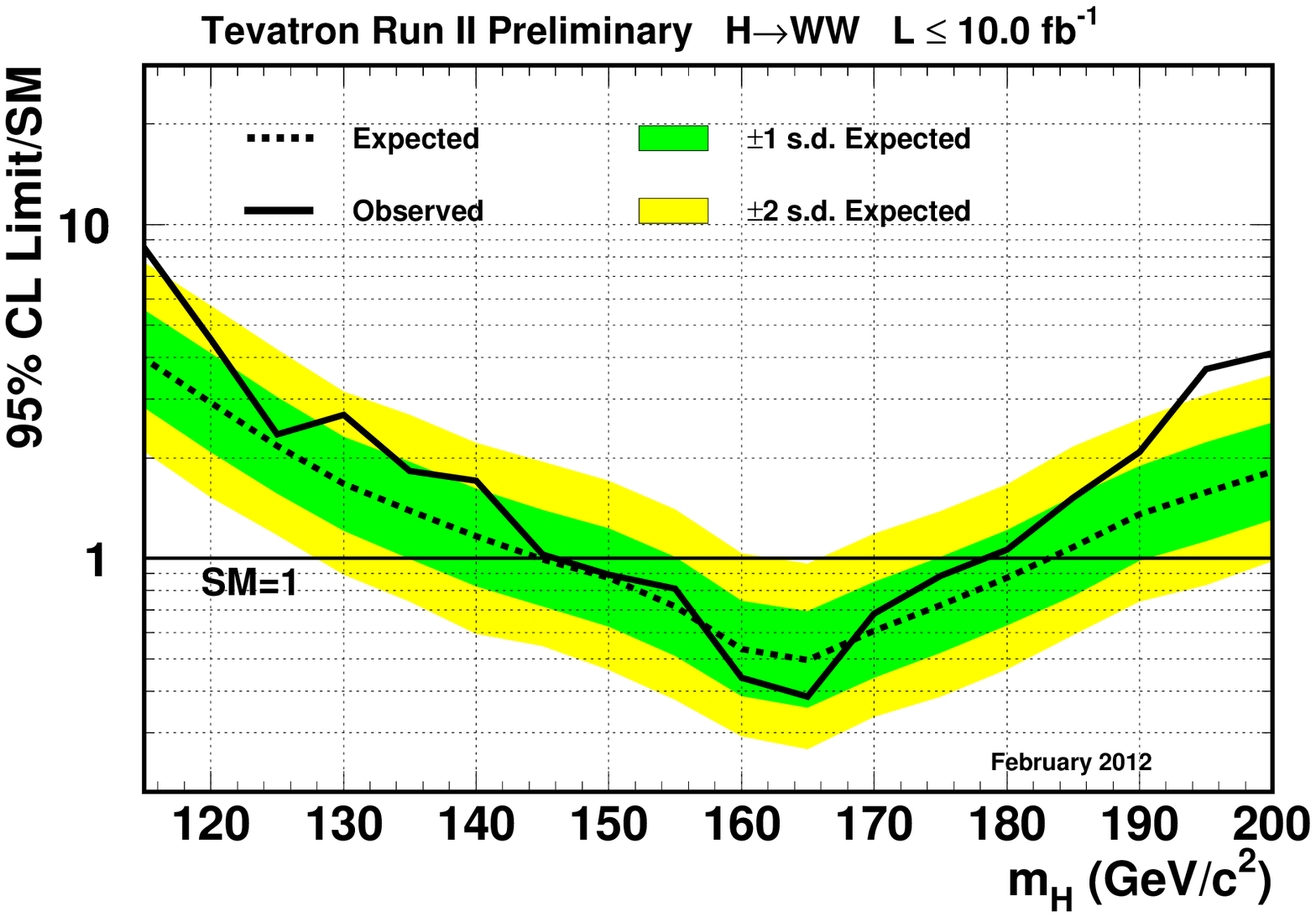}
\caption{
\label{fig:comboRatioWW}
Observed and expected (median, for the background-only hypothesis)
95\% C.L. upper limits on the ratios to the SM cross section,
as functions of the Higgs boson mass
for the combination of CDF and D0 analyses focusing on the $H
\to W^+W^-$ decay channel.  The limits are expressed
as a multiple of the SM prediction for test masses (every 5
GeV/$c^2$) for which both experiments have performed dedicated
searches in different channels.
The points are joined by straight lines for better readability.
The bands indicate the 68\% and 95\% probability regions where
the limits can fluctuate, in the absence of signal.  The limits
displayed in this figure are obtained with the Bayesian calculation.
}
\end{centering}
\end{figure}

\begin{figure}[hb]
\begin{centering}
\includegraphics[width=0.7\textwidth]{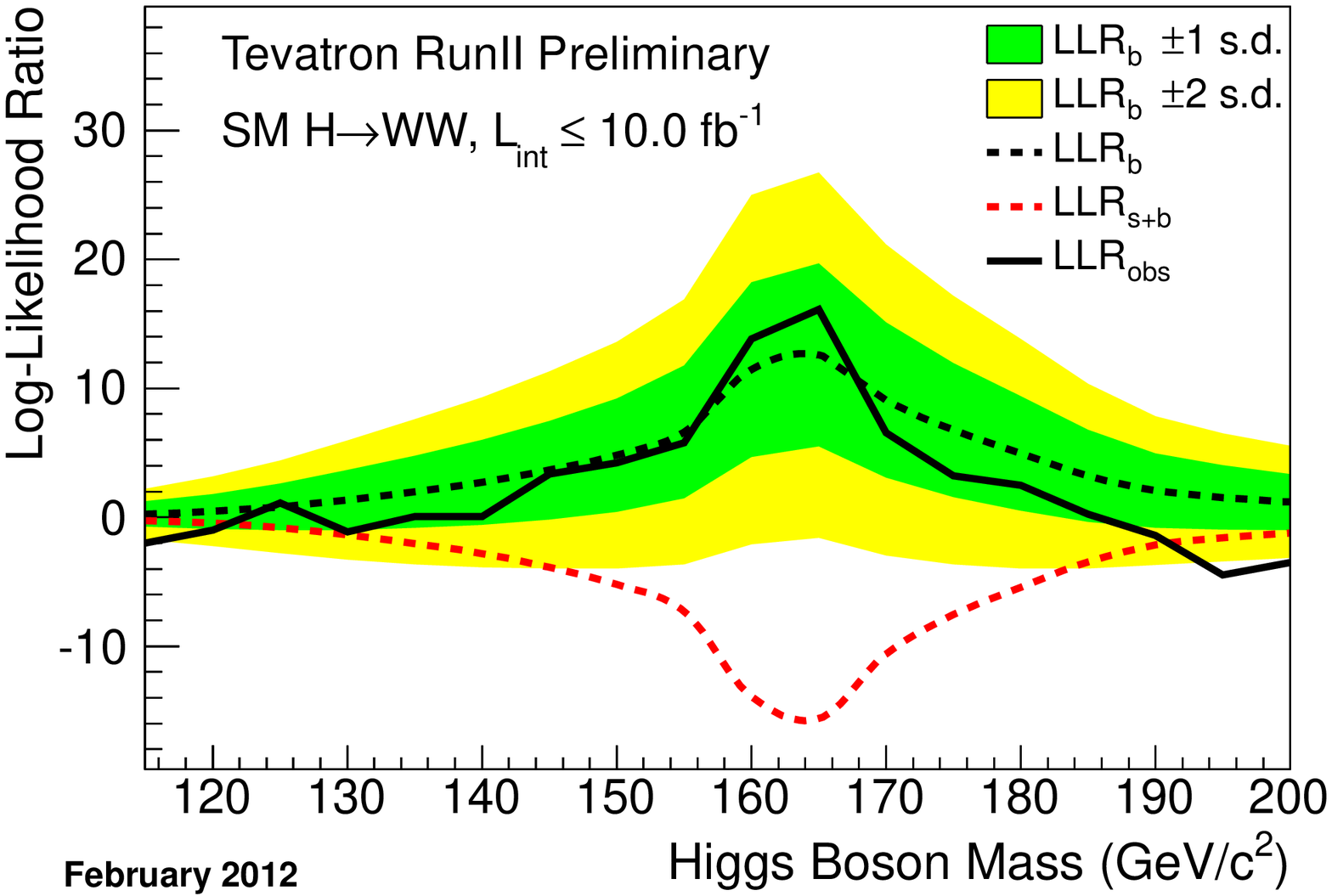}
\caption{
\label{fig:hwwLLR}
Distributions of the log-likelihood ratio (LLR) as a function of Higgs boson mass obtained with
the ${\rm CL}_{\rm s}$ method for the combination of all CDF and D0 analyses in the $H \to W^+W^-$ channels. 
The green and yellow bands correspond to the regions enclosing 1~s.d. and 2~s.d. fluctuations 
of the background, respectively.}
\end{centering}
\end{figure}

\newpage
\begin{figure}[t]
\begin{centering}
\includegraphics[width=0.7\textwidth]{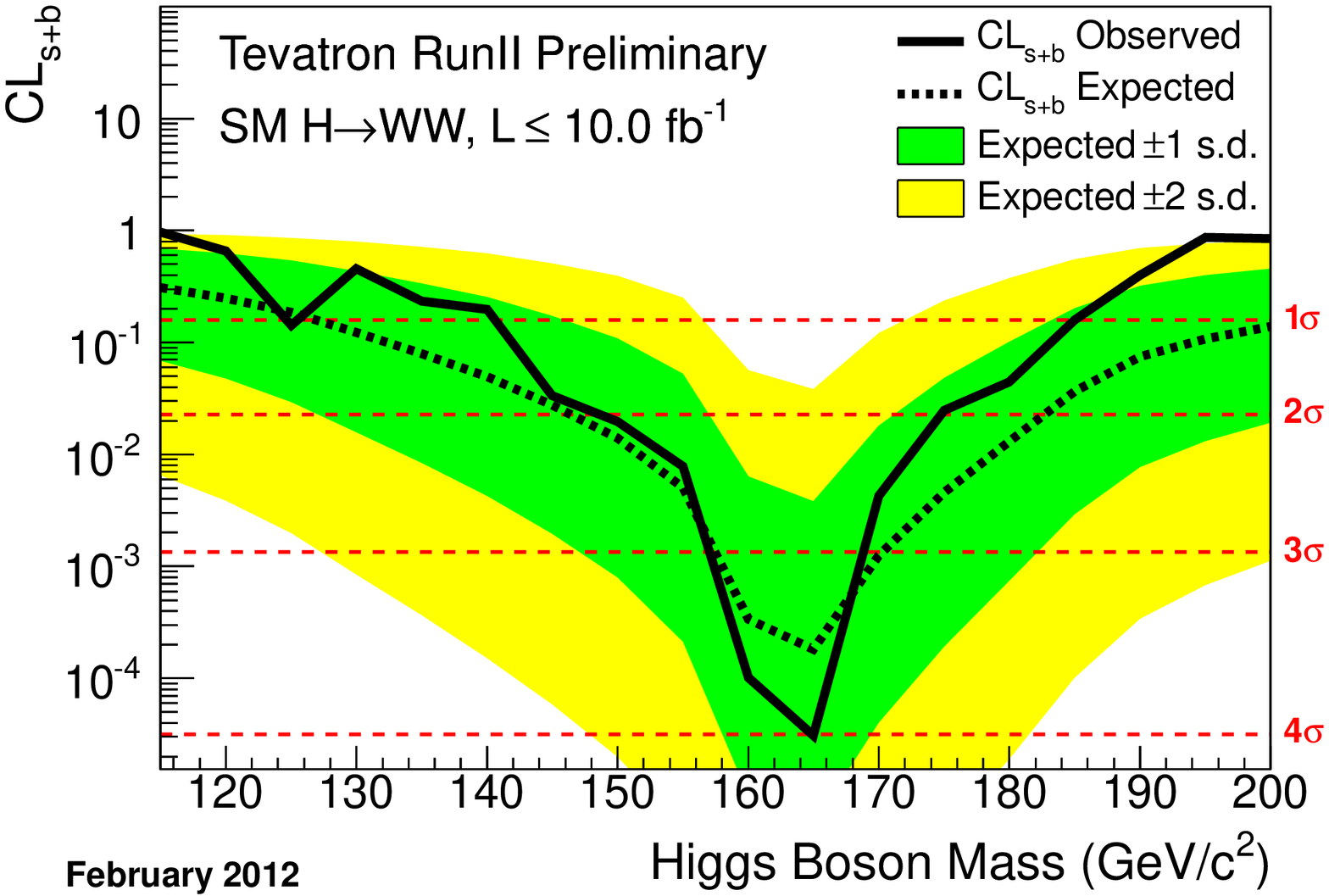} \\
\caption{
\label{fig:hwwCLsb}
The signal $p$-values ${\rm CL}_{\rm s+b}$ for the signal plus background hypothesis 
as a function of the Higgs boson mass (in steps 
of 5 GeV/$c^2$), for the combination of all CDF and D0 analyses in the $H \to W^+W^-$ channels. 
 The green and yellow bands 
correspond to the regions enclosing 1~s.d. and 2~s.d. fluctuations of the background, 
respectively.}
\end{centering}
\end{figure}

\begin{figure}[t]
\begin{centering}
\includegraphics[width=0.7\textwidth]{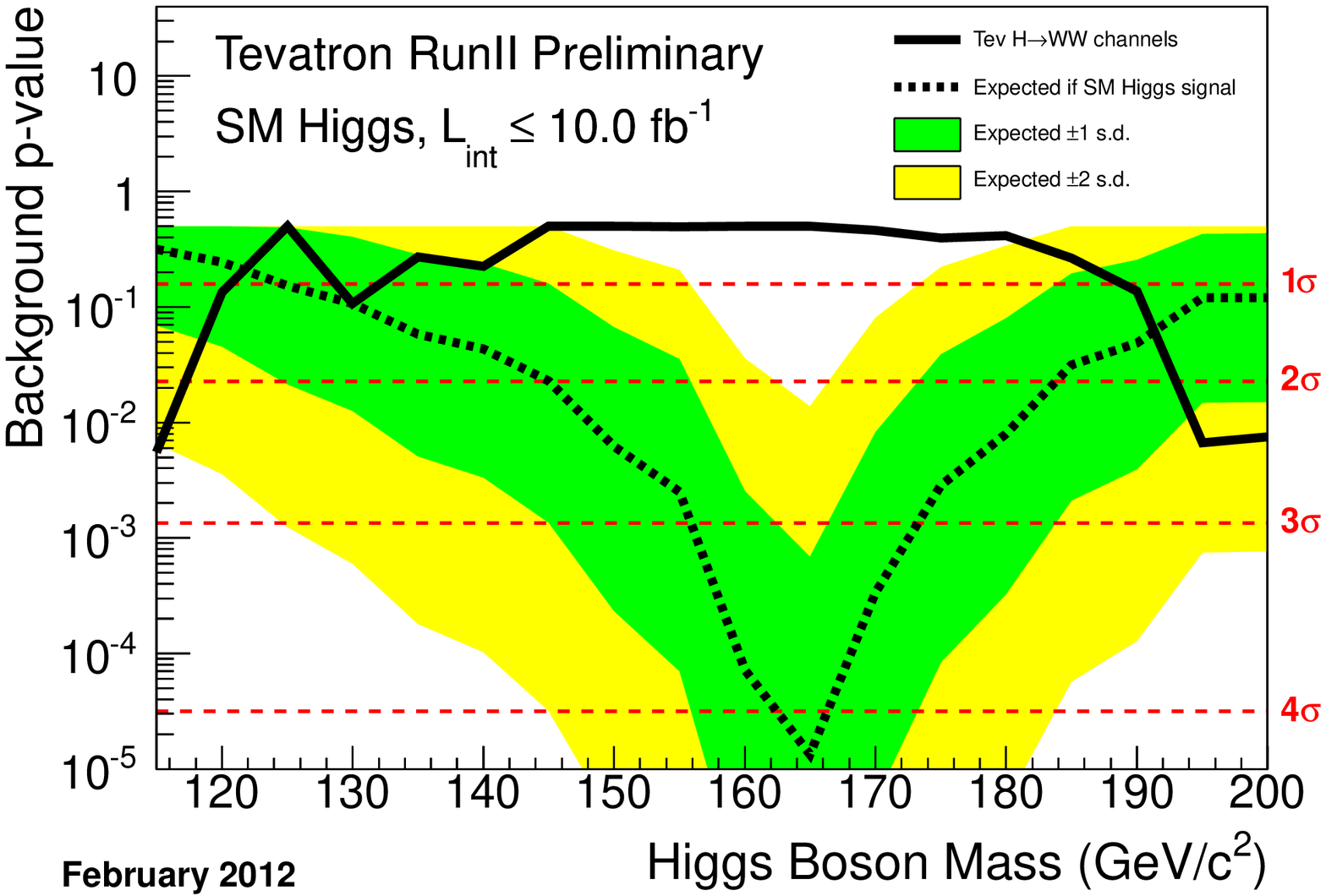} \\
\caption{
\label{fig:hwwCLb}
The background $p$-values 1-${\rm CL}_{\rm b}$ for the null hypothesis as a function of the Higgs boson mass (in steps 
of 5 GeV/$c^2$), for the combination of all CDF and D0 analyses in the $H \to W^+W^-$ channels. 
The green and yellow bands 
correspond to the regions enclosing 1~s.d. and 2~s.d. fluctuations of the background, 
respectively.}
\end{centering}
\end{figure}

\newpage
 \begin{figure}[t]
 \begin{centering}
\includegraphics[width=0.7\textwidth]{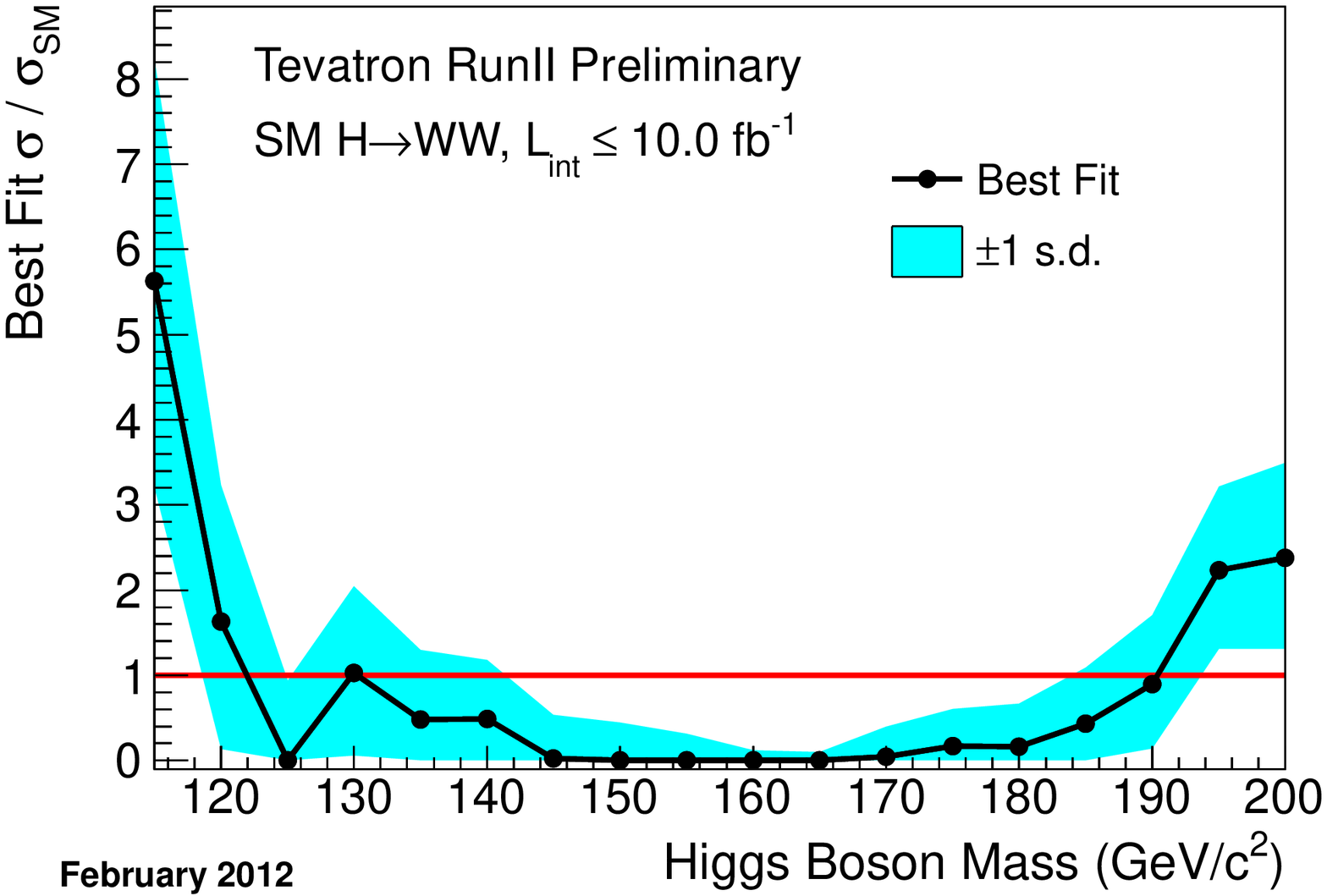} \\
\caption{
 \label{fig:hwwBest}
 The best fit of the signal cross section as a function of the Higgs boson mass
(in steps of 5 GeV/$c^2$), for the combination of all 
 CDF and D0 analyses in the $H\to W^+W^-$ channels.  The blue band shows the 1~s.d. uncertainty on the signal fit. }
 \end{centering}
 \end{figure}

\clearpage\newpage


\begin{figure}[hb]
\begin{centering}
\includegraphics[width=0.7\textwidth]{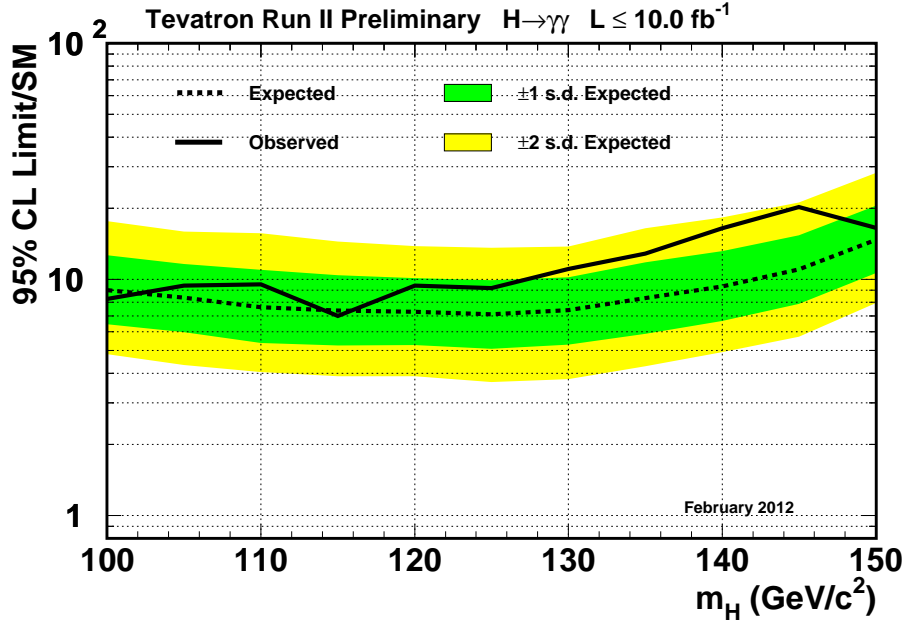}
\caption{
\label{fig:comboRatiogamgam}
Observed and expected (median, for the background-only hypothesis)
95\% C.L. upper limits on the ratios to the SM cross section,
as functions of the Higgs boson mass
for the combination of CDF and D0 analyses focusing on the 
$H \to \gamma\gamma$ decay channel.  The limits are expressed
as a multiple of the SM prediction for test masses (every 5
GeV/$c^2$).  The points are joined by straight lines for better 
readability.  The bands indicate the 68\% and 95\% probability 
regions where the limits can fluctuate, in the absence of signal.  
The limits displayed in this figure are obtained with the Bayesian 
calculation.
}
\end{centering}
\end{figure}

\begin{table}[htpb]
\caption{\label{tab:llrVals} Log-likelihood ratio (LLR) values for the combined CDF + \Dzero Higgs boson search obtained using the {\rm CL}$_{S}$ method.}
\begin{ruledtabular}
\begin{tabular}{lccccccc}
$m_{H}$ (GeV/$c^2$ &  LLR$_{\rm{obs}}$ & LLR$_{S+B}^{\rm{med}}$ &
LLR$_{B}^{-2{\rm s.d.}}$ & LLR$_{B}^{-1{\rm s.d.}}$ & LLR$_{B}^{\rm{med}}$ &  LLR$_{B}^{+1{\rm s.d.}}$ & LLR$_{B}^{+2{\rm s.d.}}$ \\
\hline
100 & 4.84 & -6.96 & 16.75 & 11.64 & 6.53 & 1.42 & -3.69 \\ 
105 & 3.18 & -6.27 & 15.56 & 10.71 & 5.87 & 1.02 & -3.82 \\ 
110 & -2.66 & -5.36 & 14.08 & 9.57 & 5.07 & 0.57 & -3.94 \\ 
115 & -5.07 & -4.74 & 12.96 & 8.72 & 4.49 & 0.25 & -3.99 \\ 
120 & -8.01 & -3.90 & 11.38 & 7.54 & 3.69 & -0.15 & -3.99 \\ 
125 & -4.63 & -3.29 & 10.28 & 6.73 & 3.17 & -0.39 & -3.95 \\ 
130 & -6.45 & -3.15 & 9.98 & 6.50 & 3.02 & -0.45 & -3.93 \\ 
135 & -6.56 & -3.23 & 10.14 & 6.62 & 3.10 & -0.42 & -3.94 \\ 
140 & -2.10 & -3.66 & 10.97 & 7.23 & 3.49 & -0.25 & -3.98 \\ 
145 & 2.82 & -4.55 & 12.48 & 8.36 & 4.24 & 0.12 & -4.00 \\ 
150 & 5.29 & -5.72 & 14.43 & 9.84 & 5.26 & 0.67 & -3.91 \\ 
155 & 6.11 & -7.33 & 16.95 & 11.80 & 6.64 & 1.49 & -3.67 \\ 
160 & 14.06 & -13.94 & 25.05 & 18.27 & 11.49 & 4.71 & -2.07 \\ 
165 & 16.27 & -15.66 & 26.81 & 19.71 & 12.61 & 5.51 & -1.60 \\ 
170 & 6.88 & -10.62 & 21.18 & 15.15 & 9.11 & 3.07 & -2.96 \\ 
175 & 3.57 & -7.58 & 17.22 & 12.00 & 6.79 & 1.58 & -3.63 \\ 
180 & 2.64 & -5.47 & 13.88 & 9.43 & 4.97 & 0.51 & -3.95 \\ 
185 & 0.45 & -3.48 & 10.41 & 6.82 & 3.23 & -0.37 & -3.96 \\ 
190 & -1.14 & -2.18 & 7.87 & 4.98 & 2.09 & -0.80 & -3.69 \\ 
195 & -4.44 & -1.61 & 6.54 & 4.05 & 1.55 & -0.94 & -3.43 \\ 
200 & -3.97 & -1.24 & 5.59 & 3.39 & 1.20 & -0.99 & -3.18 \\ 
\hline
\end{tabular}
\end{ruledtabular}
\end{table}


\begin{table}[htpb]
\caption{\label{tab:clsVals} The observed and expected 1-{\rm CL}$_{\rm s}$ values as functions of $m_H$, for the combined
CDF and \Dzero Higgs boson searches.}
\begin{ruledtabular}
\begin{tabular}{lcccccc}
$m_H$ (GeV/$c^2$) & 1-{\rm CL}$_{\rm s}^{\rm{obs}}$ &
1-{\rm CL}$_{\rm s}^{-2{\rm s.d.}}$ &
1-{\rm CL}$_{\rm s}^{-1{\rm s.d.}}$ &
1-{\rm CL}$_{\rm s}^{\rm{median}}$ &
1-{\rm CL}$_{\rm s}^{+1{\rm s.d.}}$ &
1-{\rm CL}$_{\rm s}^{+2{\rm s.d.}}$ \\ \hline
100 & 0.980 & 1.000 & 0.999 & 0.989 & 0.933 & 0.726 \\ 
105 & 0.958 & 1.000 & 0.998 & 0.985 & 0.914 & 0.680 \\ 
110 & 0.707 & 0.999 & 0.996 & 0.976 & 0.881 & 0.612 \\ 
115 & 0.463 & 0.999 & 0.994 & 0.966 & 0.850 & 0.559 \\ 
120 & 0.148 & 0.998 & 0.988 & 0.945 & 0.796 & 0.479 \\ 
125 & 0.347 & 0.996 & 0.982 & 0.925 & 0.748 & 0.415 \\ 
130 & 0.174 & 0.995 & 0.979 & 0.918 & 0.734 & 0.400 \\ 
135 & 0.175 & 0.996 & 0.981 & 0.922 & 0.742 & 0.409 \\ 
140 & 0.634 & 0.997 & 0.986 & 0.938 & 0.779 & 0.454 \\ 
145 & 0.934 & 0.999 & 0.992 & 0.961 & 0.838 & 0.541 \\ 
150 & 0.979 & 0.999 & 0.996 & 0.978 & 0.892 & 0.639 \\ 
155 & 0.988 & 1.000 & 0.999 & 0.990 & 0.939 & 0.745 \\ 
160 & 1.000 & 1.000 & 1.000 & 0.999 & 0.993 & 0.943 \\ 
165 & 1.000 & 1.000 & 1.000 & 1.000 & 0.996 & 0.961 \\ 
170 & 0.994 & 1.000 & 1.000 & 0.998 & 0.979 & 0.877 \\ 
175 & 0.971 & 1.000 & 0.999 & 0.991 & 0.943 & 0.758 \\ 
180 & 0.941 & 0.999 & 0.995 & 0.974 & 0.881 & 0.619 \\ 
185 & 0.813 & 0.996 & 0.982 & 0.928 & 0.760 & 0.436 \\ 
190 & 0.582 & 0.985 & 0.952 & 0.852 & 0.619 & 0.288 \\ 
195 & 0.125 & 0.971 & 0.918 & 0.787 & 0.530 & 0.219 \\ 
200 & 0.103 & 0.952 & 0.882 & 0.727 & 0.459 & 0.173 \\ 
\end{tabular}
\end{ruledtabular}
\end{table}

\begin{center}
{\bf Acknowledgments}  
\end{center}

We thank the Fermilab staff and the technical staffs of the
participating institutions for their vital contributions, and we
acknowledge support from the
DOE and NSF (USA);
CONICET and UBACyT (Argentina);
ARC (Australia);
CNPq, FAPERJ, FAPESP and FUNDUNESP (Brazil);
CRC Program and NSERC (Canada);
CAS, CNSF, and NSC (China);
Colciencias (Colombia);
MSMT and GACR (Czech Republic);
Academy of Finland (Finland);
CEA and CNRS/IN2P3 (France);
BMBF and DFG (Germany);
INFN (Italy);
DAE and DST (India);
SFI (Ireland);
Ministry of Education, Culture, Sports, Science and Technology (Japan);
KRF, KOSEF and World Class University Program (Korea);
CONACyT (Mexico);
FOM (The Netherlands);
FASI, Rosatom and RFBR (Russia);
Slovak R\&D Agency (Slovakia);
Ministerio de Ciencia e Innovaci\'{o}n, and Programa Consolider-Ingenio 2010 (Spain);
The Swedish Research Council (Sweden);
Swiss National Science Foundation (Switzerland);
STFC and the Royal Society (United Kingdom);
and
the A.P. Sloan Foundation (USA).


\clearpage
\newpage


\appendix
\appendixpage
\addappheadtotoc
\section{Systematic Uncertainties}

\begin{table}[h]
\begin{center}
\caption{\label{tab:cdfsystwh3jet} Systematic uncertainties on the signal and background
contributions for CDF's $WH\rightarrow\ell\nu b{\bar{b}}$ single tight $b$-tag (Tx) 
and single loose $b$-tag (Lx) categories.  Systematic uncertainties are listed by name; 
see the original references for a detailed explanation of their meaning and on how they 
are derived.  Systematic uncertainties for $WH$ shown in this table are obtained for 
$m_H=115$~GeV/$c^2$.  Uncertainties are relative, in percent, and are symmetric unless 
otherwise indicated. Shape uncertainties are labeled with an "S". }

\vskip 0.1cm
{\centerline{CDF: single tight $b$-tag (Tx) $WH\rightarrow\ell\nu b{\bar{b}}$ channel relative uncertainties (\%)}}
\vskip 0.099cm
\begin{ruledtabular}
\begin{tabular}{lcccccc}\\
Contribution              & $W$+HF & Mistags & Top & Diboson & Non-$W$ & $WH$  \\ \hline
Luminosity ($\sigma_{\mathrm{inel}}(p{\bar{p}})$)
                          & 3.8      & 0       & 3.8 & 3.8     & 0       &    3.8   \\
Luminosity Monitor        & 4.4      & 0       & 4.4 & 4.4     & 0       &    4.4   \\
Lepton ID                 & 2.0-4.5      & 0       & 2.0-4.5   & 2.0-4.5       & 0       &    2.0-4.5   \\
Jet Energy Scale          & 3.2-6.9(S)      & 0.9-1.8(S)       & 0.8-9.7(S)   & 3.6-13.2(S)       & 0       &    3.0-5.0(S)   \\
Mistag Rate (tight)               & 0      & 19    & 0   & 0       & 0       &    0   \\
Mistag Rate (loose)               & 0      & 0    & 0   & 0       & 0       &    0   \\
$B$-Tag Efficiency (tight)         & 0      & 0       & 3.9 & 3.9     & 0       &    3.9   \\
$B$-Tag Efficiency (loose)        & 0      & 0       & 0 & 0     & 0       &    0   \\
$t{\bar{t}}$ Cross Section  & 0    & 0       & 10  & 0       & 0       &    0   \\
Diboson Rate                & 0      & 0       & 0   & 6.0    & 0       &    0   \\
Signal Cross Section        & 0      & 0       & 0   & 0       & 0       &    5 \\
HF Fraction in W+jets       &    30  & 0       & 0   & 0       & 0       &    0   \\
ISR+FSR+PDF                 & 0      & 0       & 0   & 0       & 0       &    3.8-6.8 \\
$Q^2$                       &  3.2-6.9(S)           &    0.9-1.8(S)     &    0       &   0            & 0        &    0        \\
QCD Rate                    & 0      & 0       & 0   & 0       & 40      &    0   \\
\end{tabular}
\end{ruledtabular}

\vskip 0.3cm
{\centerline{CDF: single loose $b$-tag (Lx) $WH\rightarrow\ell\nu b{\bar{b}}$ channel relative uncertainties (\%)}}
\vskip 0.099cm
\begin{ruledtabular}
\begin{tabular}{lcccccc}\\
Contribution              & $W$+HF & Mistags & Top & Diboson & Non-$W$ & $WH$  \\ \hline
Luminosity ($\sigma_{\mathrm{inel}}(p{\bar{p}})$)
                          & 3.8      & 0       & 3.8 & 3.8     & 0       &    3.8   \\
Luminosity Monitor        & 4.4      & 0       & 4.4 & 4.4     & 0       &    4.4   \\
Lepton ID                 & 2      & 0       & 2   & 2       & 0       &    2   \\
Jet Energy Scale          & 2.2-6.0(S)      & 0.9-1.8(S)       & 1.6-8.6(S)   & 4.6-9.6(S)       & 0       &    3.1-4.8(S)   \\
Mistag Rate (tight)               & 0      & 0    & 0   & 0       & 0       &    0   \\
Mistag Rate (loose)               & 0      & 10    & 0   & 0       & 0       &    0   \\
$B$-Tag Efficiency (tight)         & 0      & 0       & 0 & 0     & 0       &    0   \\
$B$-Tag Efficiency (loose)        & 0      & 0       & 3.2 & 3.2     & 0       &    3.2   \\
$t{\bar{t}}$ Cross Section  & 0    & 0       & 10  & 0       & 0       &    0   \\
Diboson Rate              & 0      & 0       & 0   & 6.0    & 0       &    0   \\
Signal Cross Section      & 0      & 0       & 0   & 0       & 0       &    10 \\
HF Fraction in W+jets     &    30  & 0       & 0   & 0       & 0       &    0   \\
ISR+FSR+PDF               & 0      & 0       & 0   & 0       & 0       &    2.4-4.9 \\
QCD Rate                  & 2.1-6.0(S)      & 0.9-1.8(S)       & 0   & 0       & 40      &    0   \\
\end{tabular}
\end{ruledtabular}

\end{center}
\end{table}

\begin{table}[h]
\begin{center}
\caption{\label{tab:cdfsystwh2jet} Systematic uncertainties on the signal and
background contributions for CDF's $WH\rightarrow\ell\nu b{\bar{b}}$ two tight
$b$-tag (TT), one tight $b$-tag and one loose $b$-tag (TL), and two loose 
$b$-tag (LL) channels. Systematic uncertainties are listed by name; see the 
original references for a detailed explanation of their meaning and on how
they are derived.  Systematic uncertainties for $WH$ shown in this table are
obtained for $m_H=115$ GeV/$c^2$.  Uncertainties are relative, in percent, and
are symmetric unless otherwise indicated. Shape uncertainties are labeled with 
an "S".}
\vskip 0.1cm
{\centerline{CDF: two tight $b$-tag (TT) $WH\rightarrow\ell\nu b{\bar{b}}$ channel relative uncertainties (\%)}}
\vskip 0.099cm
\begin{ruledtabular}
\begin{tabular}{lcccccc}\\
Contribution              & $W$+HF & Mistags & Top & Diboson & Non-$W$ & $WH$  \\ \hline
Luminosity ($\sigma_{\mathrm{inel}}(p{\bar{p}})$)
                          & 3.8      & 0       & 3.8 & 3.8     & 0       &    3.8   \\
Luminosity Monitor        & 4.4      & 0       & 4.4 & 4.4     & 0       &    4.4   \\
Lepton ID                 & 2.0-4.5      & 0       & 2.0-4.5   & 2.0-4.5       & 0       &    2.0-4.5   \\
Jet Energy Scale          & 4.0-16.6(S)      & 0.9-3.3(S)       & 0.9-10.4(S)   & 4.7-19.7(S)       & 0       &    2.3-13.6(S)   \\
Mistag Rate (tight)               & 0      & 40     & 0   & 0       & 0       &    0   \\
Mistag Rate (loose)               & 0      & 0     & 0   & 0       & 0       &    0   \\
$B$-Tag Efficiency (tight)         & 0      & 0       & 7.8 & 7.8     & 0       &    7.8   \\
$B$-Tag Efficiency (loose)        & 0      & 0       & 0 & 0     & 0       &    0   \\
$t{\bar{t}}$ Cross Section  & 0    & 0       & 10  & 0       & 0       &    0   \\
Diboson Rate              & 0      & 0       & 0   & 6.0   & 0       &    0   \\
Signal Cross Section      & 0      & 0       & 0   & 0       & 0       &    5 \\
HF Fraction in W+jets     &    30  & 0       & 0   & 0       & 0       &    0   \\
ISR+FSR+PDF               & 0      & 0       & 0   & 0       & 0       &    6.4-12.6 \\
$Q^2$                     & 4.0-8.8(S)          & 0.9-1.8(S)        & 0    &   0      &   0      &  0   \\
QCD Rate                  & 0      & 0       & 0   & 0       & 40      &    0   \\
\end{tabular}
\end{ruledtabular}

\vskip 0.3cm
{\centerline{CDF: one tight and one loose $b$-tag (TL) $WH\rightarrow\ell\nu b{\bar{b}}$ channel relative uncertainties (\%)}}
\vskip 0.099cm
\begin{ruledtabular}
\begin{tabular}{lcccccc}\\
Contribution              & $W$+HF & Mistags & Top & Diboson & Non-$W$ & $WH$  \\ \hline
Luminosity ($\sigma_{\mathrm{inel}}(p{\bar{p}})$)
                          & 3.8      & 0       & 3.8 & 3.8     & 0       &    3.8   \\
Luminosity Monitor        & 4.4      & 0       & 4.4 & 4.4     & 0       &    4.4   \\
Lepton ID                 & 2.0-4.5      & 0       & 2.0-4.5   & 2.0-4.5       & 0       &    2.0-4.5   \\
Jet Energy Scale          & 3.9-12.4(S)      &  0.9-3.3(S)      & 1.4-11.5(S)   & 5.0-16.0(S)       &        &    2.5-16.1(S)   \\
Mistag Rate (tight)               & 0      & 19     & 0   & 0       & 0       &    0   \\
Mistag Rate (loose)               & 0      & 10     & 0   & 0       & 0       &    0   \\
$B$-Tag Efficiency (tight)         & 0      & 0       & 3.9 & 3.9     & 0       &    3.9   \\
$B$-Tag Efficiency (loose)        & 0      & 0       & 3.2 & 3.2     & 0       &    3.2   \\
$t{\bar{t}}$ Cross Section  & 0    & 0       & 10  & 0       & 0       &    0   \\
Diboson Rate              & 0      & 0       & 0   & 6.0    & 0       &    0   \\
Signal Cross Section      & 0      & 0       & 0   & 0       & 0       &    5 \\
HF Fraction in W+jets     &    30  & 0       & 0   & 0       & 0       &    0   \\
ISR+FSR+PDF               & 0      & 0       & 0   & 0       & 0       &    3.3-10.3 \\
$Q^2$                     & 3.9-7.7(S)             &  0.9-1.9(S)       &     0       &   0             & 0        & 0 \\
QCD Rate                  & 0      & 0       & 0   & 0       & 40      &    0   \\
\end{tabular}
\end{ruledtabular}

\vskip 0.3cm
{\centerline{CDF: two loose $b$-tag (LL) $WH\rightarrow\ell\nu b{\bar{b}}$ channel relative uncertainties (\%)}}
\vskip 0.099cm
\begin{ruledtabular}
\begin{tabular}{lcccccc}\\
Contribution              & $W$+HF & Mistags & Top & Diboson & Non-$W$ & $WH$  \\ \hline
Luminosity ($\sigma_{\mathrm{inel}}(p{\bar{p}})$)
                          & 3.8      & 0       & 3.8 & 3.8     & 0       &    3.8   \\
Luminosity Monitor        & 4.4      & 0       & 4.4 & 4.4     & 0       &    4.4   \\
Lepton ID                 & 2      & 0       & 2   & 2       & 0       &    2   \\
Jet Energy Scale          & 3.6-6.9(S)      & 0.9-1.8(S)       & 1.7-7.9(S)   & 1.2-8.5       & 0       &    2.7-5.4(S)   \\
Mistag Rate (tight)               & 0      & 0     & 0   & 0       & 0       &    0   \\
Mistag Rate (loose)               & 0      & 20     & 0   & 0       & 0       &    0   \\
$B$-Tag Efficiency (tight)         & 0      & 0       & 0 & 0     & 0       &    0   \\
$B$-Tag Efficiency (loose)        & 0      & 0       & 6.3 & 6.3     & 0       &    6.3   \\
$t{\bar{t}}$ Cross Section  & 0    & 0       & 10  & 0       & 0       &    0   \\
Diboson Rate              & 0      & 0       & 0   & 6.0    & 0       &    0   \\
Signal Cross Section      & 0      & 0       & 0   & 0       & 0       &    10 \\
HF Fraction in W+jets     &    30  & 0       & 0   & 0       & 0       &    0   \\
ISR+FSR+PDF               & 0      & 0       & 0   & 0       & 0       &    2.0-13.6 \\
QCD Rate                  & 3.6-6.9(S)      & 0.9-1.8(S)       & 0   & 0       & 40      &    0   \\
\end{tabular}
\end{ruledtabular}

\end{center}
\end{table}

\begin{table}[h]
\begin{center}
\caption{\label{tab:d0systwh1} Systematic uncertainties on the signal and background
contributions for D0's $WH\rightarrow\ell\nu b{\bar{b}}$ single and double tag channels.
Systematic uncertainties are listed by name, see the original
references for a detailed explanation of their meaning and on how they are derived.
Systematic uncertainties for $WH$ shown in this table are obtained for $m_H=115$ GeV/$c^2$. Uncertainties are
relative, in percent, and are symmetric unless otherwise indicated.   Shape uncertainties are labeled with an ``(S)'',
and ``SH'' represents shape only uncertainty.
}
\vskip 0.2cm
{\centerline{$WH \rightarrow\ell\nu b\bar{b}$ Single Tag (TST) channels relative uncertainties (\%)}}
\vskip 0.099cm
\begin{ruledtabular}
\begin{tabular}{l c c c c c c c }\\
Contribution             &~Dibosons~ & $W+b\bar{b}/c\bar{c}$& $W$+l.f. & $~~~t\bar{t}~~~$ &single top&Multijet& ~~~$WH$~~~\\
\hline
Luminosity                &  6.1  &  6.1  &  6.1  &  6.1  &  6.1  &   --    &  6.1  \\ 
Electron ID/Trigger eff. (S)& 1--5  & 2--4  &  2--4 & 1--2  & 1--2  &   --    &  2--3 \\       
Muon Trigger eff. (S)     & 1    &  1    &  1    &  1    &  1    &   --    &  1 \\       
Muon ID/Reco eff./resol.  & 4.1  &   4.1 &   4.1 &   4.1 &   4.1 &   --    &   4.1 \\        
Jet ID/Reco eff.          & 2    &  2    &  2    &  2    &  2    &   --    &  2    \\ 
Jet Resolution    (S)     & 1--2 &  2--4 &  2--3 &  2--5 &  1--2 &   --    &  2 \\       
Jet Energy Scale  (S)     & 4--7 &  1--5 &  2--5 &  2--7 &  1--2 &   --    &  2--6 \\       
Vertex Conf. Jet  (S)     & 4--6 & 3--4 & 2--3 & 6--10 & 2--4 &   --    &  3--7 \\       
$b$-tag/taggability (S)   & 1--3 &  1--4 & 7--10  &  1--6 &  1--2 &   --    &  2--9 \\ 
Heavy-Flavor K-factor     & --   &    20 &      -- &   --    &   --    &   --    &   --    \\       
Inst.-WH $e\nu b\bar{b}$ (S) & 1--2 & 2--4 & 1--3 & 1--2  &  1--3 &  15  &  1--2 \\ 
Inst.-WH $\mu\nu b\bar{b} $  &  --  &   2.4 &   2.4 &   --    &   --    &  20  &   --    \\ 
Cross Section             & 6    &     9 &     6 &    7 &    7 &   --    &     6.1 \\ 
Signal Branching Fraction & --   &   --  &  --  &  --  &  --  &  --  & 1-9 \\      
ALPGEN MLM pos/neg(S)     & --   &   --  &    SH  &     --    &   --    &   --    &   --    \\       
ALPGEN Scale (S)          & --   &  SH   &    SH  &     --    &   --    &   --    &   --    \\       
Underlying Event (S)      & --   &  SH   &    SH  &     --    &   --    &   --    &   --    \\       
PDF, reweighting          &  2   &  2    & 2     & 2     &  2    &   --    &  2    \\
\end{tabular}
\end{ruledtabular}

\vskip 0.5cm
{\centerline{$WH \rightarrow\ell\nu b\bar{b}$ Loose Double Tag (LDT) channels relative uncertainties (\%)}}
\vskip 0.099cm
\begin{ruledtabular}
\begin{tabular}{ l c c c c c c c }   \\
Contribution  &~Dibosons~&$W+b\bar{b}/c\bar{c}$&$W$+l.f.&$~~~t\bar{t}~~~$&single top&Multijet& ~~~$WH$~~~\\
\hline
Luminosity                &  6.1  &  6.1  &  6.1  &  6.1  &  6.1  &   --    &  6.1  \\ 
Electron ID/Trigger eff. (S)  & 2--5  & 2--3  &  2--3 & 1--2  & 1--2  &   --    &  1--2 \\       
Muon Trigger eff. (S)     &  1    &  1    &  1    &  1    &  1    &   --    &  1 \\       
Muon ID/Reco eff./resol.  &   4.1 &   4.1 &   4.1 &   4.1 &   4.1 &   --    &   4.1 \\        
Jet ID/Reco eff.          &  2    &  2    &  2    &  2    &  2    &   --    &  2    \\ 
Jet Resolution    (S)     &  1--7 &  2--7 &  2--3 &  2--7 &  2--4 &   --    &  1--5 \\       
Jet Energy Scale  (S)     &  2--11 & 2--5 &  2--7 &  2--7 &  2--5 &   --    &  2--8 \\       
Vertex Conf. Jet  (S)     &  2--11 & 2--12 & 2--3 & 4--15 & 2--3 &   --    &  3--7 \\       
$b$-tag/taggability (S)   & 2--15  &  2--6 & 6--10 & 2--5 & 2--3 &   --    &  1--5 \\ 
Heavy-Flavor K-factor     &   --    &    20 &      -- &   --    &   --    &   --    &   --    \\       
Inst.-WH $e\nu b\bar{b}$ (S) & 1--2 & 2--4 & 1--3 & 1--2  &  1--3 &  15  &  1--2 \\ 
Inst.-WH $\mu\nu b\bar{b} $  &  --  &   2.4 &   2.4 &   --    &   --    &  20  &   --    \\ 
Cross Section             &     6 &     9 &     6 &    7 &    7 &   --    &     6.1 \\
Signal Branching Fraction &   --  &   --  &  --  &  --  &  --  &  --  & 1-9 \\      
ALPGEN MLM pos/neg(S)     &   --    &   --  &     SH &   --    &   --    &   --    &   --    \\       
ALPGEN Scale (S)          &   --    &   SH  &    SH &   --    &   --    &   --    &   --    \\       
Underlying Event (S)      &   --    &   SH  &    SH &   --    &   --    &   --    &   --    \\       
PDF, reweighting          &  2    &  2    & 2     & 2     &  2    &   --    &  2    \\
\end{tabular}
\end{ruledtabular}

\end{center}
\end{table}

\begin{table}
\begin{center}
\vskip 0.5cm
{\centerline{$WH \rightarrow\ell\nu b\bar{b}$ Tight Double Tag (TDT) channels relative uncertainties (\%)}}
\vskip 0.099cm
\begin{ruledtabular}
\begin{tabular}{ l c c c c c c c }   \\
Contribution  &~Dibosons~&$W+b\bar{b}/c\bar{c}$&$W$+l.f.&$~~~t\bar{t}~~~$&single top&Multijet& ~~~$WH$~~~\\
\hline
Luminosity                &  6.1  &  6.1  &  6.1  &  6.1  &  6.1  &   --    &  6.1  \\ 
Electron ID/Trigger eff. (S)  & 2--5  & 2--3  &  2--3 & 1--2  & 1--2  &   --    &  1--2 \\       
Muon Trigger eff. (S)     &  1    &  1    &  1    &  1    &  1    &   --    &  1 \\            
Muon ID/Reco eff./resol.  &   4.1 &   4.1 &   4.1 &   4.1 &   4.1 &   --    &   4.1 \\        
Jet ID/Reco eff.          &  2    &  2    &  2    &  2    &  2    &   --    &  2    \\    
Jet Resolution    (S)     &  2--5 &  2--4 &  2--6 &  2--7 &  1--2 &   --    &  4--6 \\       
Jet Energy Scale  (S)     &  3--8 &  2--5 &  1--8 &  2--9 &  2--4 &   --    &  2--6 \\       
Vertex Conf. Jet  (S)     & 2--3  &  2--4 & 2--5  & 5--7  & 2--3  &   --    &  2--4 \\       
$b$-tag/taggability (S)   & 3--15 &  4--15& 10--15 & 5--10 & 5--9 &   --    &  4--12 \\ 
Heavy-Flavor K-factor     &   --    &    20 &      -- &   --    &   --    &   --    &   --    \\       
Inst.-WH $e\nu b\bar{b}$ (S) & 1--2 & 2--4 & 1--3 & 1--2  &  1--3 &  15  &  1--2 \\ 
Inst.-WH $\mu\nu b\bar{b} $  &  --  &   2.4 &   2.4 &   --    &   --    &  20  &   --    \\ 
Cross Section             &     6 &     9 &     6 &    7 &    7 &   --    &     6.1 \\
Signal Branching Fraction &   --  &   --  &  --  &  --  &  --  &  --  & 1-9 \\      
ALPGEN MLM pos/neg(S)     &   --    &   --  &    SH &   --    &   --    &   --    &   --    \\       
ALPGEN Scale (S)          &   --    &   SH  &    SH &   --    &   --    &   --    &   --    \\       
Underlying Event (S)      &   --    &   SH  &    SH &   --    &   --    &   --    &   --    \\       
PDF, reweighting          &  2    &  2    & 2     & 2     &  2    &   --    &  2    \\
\end{tabular}
\end{ruledtabular}
\end{center}
\end{table}

\begin{table}
\begin{center}
\caption{\label{tab:d0sysVHtau}  Systematic uncertainties on the signal and background contributions for 
D0's
$\tau \tau jj$ Run IIb channel. 
Systematic uncertainties for the 
Higgs signal shown in this table are obtained for $m_H=135$ GeV/$c^2$.  Systematic uncertainties are listed 
by name; see the original references for a detailed explanation of their meaning and on how they are derived.  
Uncertainties are relative, in percent, and are symmetric unless otherwise indicated. Shape
      uncertainties are labeled with an ``(S).''}
\vskip 0.1cm
{\centerline{$\mu \tau_{had} jj$ Run~IIb channel relative uncertainties (\%)}}
\vskip 0.099cm
\begin{ruledtabular}
\begin{tabular}{lccccccccc}\\
Contribution &$VH$  &$VBF$& $ggH$  &$W$+jets & $Z$+jets & Top& Dibosons& Multijet\\ \hline

Luminosity   & 6.1 & 6.1 & 6.1 & 6.1 & 6.1 & 6.1 &6.1 &-- \\ 
$\mu$ ID & 2.9 & 2.9 & 2.9 & 2.9 & 2.9 & 2.9 &2.9 &-- \\ 
$Single \mu$ trigger & 5.0 & 5.0 & 5.0 & 5.0 & 5.0 & 5.0 &5.0 &-- \\ 
inclusive trigger relative & 7.0 & 7.0 & 7.0 & 7.0 & 7.0 & 7.0 &7.0 &-- \\ 
$\tau$ energy correction & 9.8 & 9.8 & 9.8 & 9.8 & 9.8 & 9.8 &9.8 &-- \\ 
$\tau$ track efficiency & 1.4 & 1.4 & 1.4 & 1.4 & 1.4 & 1.4 &1.4 &-- \\ 
$\tau$ selection by type & 10,4,5 & 10,4,5 & 10,4,5 & 10,4,5 & 10,4,5 & 10,4,5 &10,4,5 &-- \\ 
Cross section & 6.2 &4.9& 33 &6.0&6.0&10.0&7.0&--\\
GGF Signal PDF &--&--&29&--&--&--&--&--\\
GGF $H p_T$ Reweighting (S)& -- & -- $\sim$5.0 & -- &  -- &  -- & -- &-- &--\\
Signal Branching Fraction  & 0-7.3 & 0-7.3  & 0-7.3 &  -- &  -- &  -- &  -- &  --\\
Vertex confirmation for jets(S)  & $\sim$ 5.0 &  $\sim$5.0 &  $\sim$5.0 &  $\sim$5.0 &  $\sim$5.0 &  $\sim$5.0 & $\sim$5.0 &-- \\

Jet ID(S)  & $\sim$5 & $\sim$5 & $\sim$5 & $\sim$5& $\sim$5 & $\sim$5 &$\sim$5 &-- \\
Jet Energy Resolution (S)  & $\sim$5 & $\sim$5 & $\sim$5 & $\sim$5& $\sim$5 & $\sim$5 &$\sim$5 &-- \\
Jet energy Scale (S)  & $\sim$5 & $\sim$5 & $\sim$5 & $\sim$5& $\sim$5 & $\sim$5 &$\sim$5 &-- \\

Jet pT  & 5.5 & 5.5 & 5.5 & 5.5 & 5.5 & 5.5 &5.5 &-- \\
PDF reweighting   & 1.6 & 1.6 & 1.6 & 1.6 & 2 & 2 &2 &-- \\
Multijet Normalization  & -- & -- & -- & -- & -- & -- &-- &5.3 \\
Multijet Shape   & -- & -- & -- & -- & -- & -- &-- &$\sim$15 \\

\end{tabular}
\end{ruledtabular}

\vskip 0.3cm
{\centerline{$e \tau_{had} jj$ Run~IIb relative uncertainties (\%)}}
\vskip 0.099cm
\begin{ruledtabular}
\begin{tabular}{lccccccccc}\\
Contribution &$VH$  &$VBF$& $ggH$  &$W$+jets & $Z$+jets & Top& Dibosons& Multijet\\ \hline
\hline
Luminosity  & 6.1 & 6.1 & 6.1 & 6.1 & 6.1 & 6.1 &6.1 &-- \\ 
Electron ID & 4 & 4 & 4 & 4 & 4 & 4 &4 &-- \\ 
Electron trigger & 2 & 2 & 2 & 2 & 2 & 2 &2 &-- \\ 
$\tau$ energy correction & 9.8 & 9.8 & 9.8 & 9.8 & 9.8 & 9.8 &9.8 &-- \\ 
$\tau$ track efficiency & 1.4 & 1.4 & 1.4 & 1.4 & 1.4 & 1.4 &1.4 &-- \\ 
$\tau$ selection by type & 10,4,5 & 10,4,5 & 10,4,5 & 10,4,5 & 10,4,5 & 10,4,5 &10,4,5 &-- \\ 
Cross section & 6.1 &4.9& 33 &6.0&6.0&10.0&7.0&--\\
GGF Signal PDF &--&--&29&--&--&--&--&--\\
GGF $H p_T$ Reweighting (S)& -- & -- & $\sim$ 5.0 & -- & -- & -- &-- &--\\
Signal Branching Fraction  & 0-7.3 & 0-7.3  & 0-7.3 &  -- &  -- &  -- &  -- &  --\\
Vertex confirmation for jets(S)  & $\sim$ 5.0 &  $\sim$5.0 &  $\sim$5.0 &  $\sim$5.0 &  $\sim$5.0 &  $\sim$5.0 & $\sim$5.0 &-- \\ 

Jet ID(S)  & $\sim$10 & $\sim$5 & $\sim$5 & $\sim$5& $\sim$5 & $\sim$5 &$\sim$5 &-- \\
Jet Energy Resolution (S)  & $\sim$10 & $\sim$10 & $\sim$10 & $\sim$10& $\sim$10 & $\sim$10 &$\sim$10 &-- \\
Jet energy Scale (S)  & $\sim$10 & $\sim$10 & $\sim$10 & $\sim$10& $\sim$10 & $\sim$10 &$\sim$10 &-- \\

Jet pT  & 5.5 & 5.5 & 5.5 & 5.5 & 5.5 & 5.5 &5.5 &-- \\
PDF reweighting  & 1.6 & 1.6 & 1.6 & 1.6 & 2 & 2 &2 &-- \\
Multijet Normalization  & -- & -- & -- & -- & -- & -- &-- &4.7 \\
Multijet Shape   & -- & -- & -- & -- & -- & -- &-- & $\sim$15 \\

\end{tabular}
\end{ruledtabular}

\end{center}
\end{table}

\begin{table}
\begin{center}
\caption{\label{tab:cdfvvbb1} Systematic uncertainties on the signal and background contributions for CDF's
$WH,ZH\rightarrow\MET b{\bar{b}}$ tight double tag (SS), loose double tag (SJ), and single tag (1S) channels.
Systematic uncertainties are listed by name; see the original references for a detailed explanation of their
meaning and on how they are derived.  Systematic uncertainties for $ZH$ and $WH$ shown in this table are
obtained for $m_H=120$~GeV/$c^2$.  Uncertainties are relative, in percent, and are symmetric unless otherwise
indicated. Shape uncertainties are labeled with an "S".}
\vskip 0.1cm
{\centerline{CDF: tight double-tag (SS) $WH,ZH\rightarrow\MET b{\bar{b}}$ channel relative uncertainties (\%)}}
\vskip 0.099cm
\begin{ruledtabular}
      \begin{tabular}{lccccccccc}\\
        Contribution & ZH & WH & Multijet & Mistags & Top Pair & S. Top  & Diboson  & W + HF  & Z + HF \\\hline
        Luminosity       & 3.8 & 3.8 &     &  & 3.8 & 3.8 & 3.8     & 3.8     & 3.8     \\
        Lumi Monitor      & 4.4 & 4.4 &     &  & 4.4 & 4.4 & 4.4     & 4.4     & 4.4     \\
        Tagging SF        & 10.4& 10.4&      & & 10.4& 10.4& 10.4    & 10.4    & 10.4    \\
      Trigger Eff. (S)& 0.9 & 1.4 & 0.9 & & 0.9 & 1.6 & 2.0     & 1.8     & 1.2     \\
        Lepton Veto       & 2.0 & 2.0 &      & & 2.0 & 2.0 &2.0      & 2.0     & 2.0     \\
        PDF Acceptance    & 3.0 & 3.0 &    &   & 3.0 & 3.0 &3.0      & 3.0     & 3.0     \\
        JES (S)       & $^{+1.7}_{-1.8}$
                                  & $^{+2.4}_{-2.3}$
                                          & &
                                                  & $^{+0.0}_{-0.1}$
                                                          & $^{+2.5}_{-2.4}$
                                                                  & $^{+4.1}_{-4.5}$
                                                                             & $^{+4.3}_{-4.6}$
                                                                                          & $^{+8.8}_{-3.2}$    \\
        ISR/FSR               & \multicolumn{2}{c}{$^{+3.0}_{+3.0}$} &       &       &       &           &           &      \\
        Cross-Section     &  5  & 5 &   &    & 10 & 10 & 6    & 30      & 30      \\
        Multijet Norm.  (shape)   &   &    & 2.5 &  &      & &          &           &           \\
        Mistag (S) & & & & $^{+36.7}_{-30}$ & & & & &\\
      \end{tabular}
\end{ruledtabular}

\vskip 0.3cm
{\centerline{CDF: loose double-tag (SJ) $WH,ZH\rightarrow\MET b{\bar{b}}$ channel relative uncertainties (\%)}}
\vskip 0.099cm
 \begin{ruledtabular}
     \begin{tabular}{lccccccccc}\\
        Contribution & ZH & WH & Multijet & Mistags & Top Pair & S. Top  & Diboson  & W + HF  & Z + HF \\\hline
        Luminosity       & 3.8  & 3.8  &   &  & 3.8  & 3.8  & 3.8      & 3.8      & 3.8     \\
        Lumi Monitor      & 4.4  & 4.4  &   &  & 4.4  & 4.4  & 4.4      & 4.4      & 4.4     \\
        Tagging SF        & 8.3 & 8.3 &   &  & 8.3 & 8.3 & 8.3     & 8.3     & 8.3     \\
      Trigger Eff. (S)& 1.2 & 1.7 & 1.6 & & 0.9 & 1.8 & 2.0     & 2.5     & 1.9     \\
        Lepton Veto       & 2.0  & 2.0  &    & & 2.0  & 2.0  &2.0       & 2.0      & 2.0     \\
        PDF Acceptance    & 3.0  & 3.0  &  &   & 3.0  & 3.0  & 3.0       & 3.0      & 3.0     \\
        JES (S)       & $^{+1.9}_{-1.9}$
                                   & $^{+2.4}_{-2.4}$
                                          & &
                                                          & $^{+3.0}_{-2.8}$
                                                                        & $^{-0.6}_{0.2}$
                                                                    & $^{+4.2}_{-4.2}$
                                                                                 & $^{+6.8}_{-5.9}$
                                                                                              & $^{+8.3}_{-3.1}$    \\
        ISR/FSR               & \multicolumn{2}{c}{$^{+2.4}_{-2.4}$} &    &   &       &       &           &           &      \\
        Cross-Section     &  5.0   & 5.0 &   &   & 10 & 10 & 6    & 30      & 30      \\
        Multijet Norm.  &       & & 1.6 &       & &          &           &           \\
        Mistag (S) & & & & $^{+65.2}_{-38.5}$ & & & & &\\
      \end{tabular}
\end{ruledtabular}

\vskip 0.3cm
{\centerline{CDF: single-tag (1S) $WH,ZH\rightarrow\MET b{\bar{b}}$ channel relative uncertainties (\%)}}
\vskip 0.099cm
\begin{ruledtabular}
      \begin{tabular}{lccccccccc}\\
        Contribution & ZH & WH & Multijet & Mistags & Top Pair & S. Top  & Diboson  & W + HF  & Z + HF \\\hline
        Luminosity       & 3.8  & 3.8  &    & & 3.8  & 3.8  & 3.8      & 3.8      & 3.8     \\
        Lumi Monitor      & 4.4  & 4.4  &   & & 4.4  & 4.4  & 4.4      & 4.4      & 4.4     \\
        Tagging SF        & 5.2  & 5.2  &     & & 5.2  & 5.2  & 5.2      & 5.2      & 5.2     \\
      Trigger Eff. (S)& 1.2 & 1.7 & 1.6 & & 0.9 & 1.8 & 2.0     & 2.5     & 1.9     \\
        Lepton Veto       & 2.0  & 2.0  &  &   & 2.0  & 2.0  &2.0       & 2.0      & 2.0     \\
        PDF Acceptance    & 3.0  & 3.0  &   &  & 3.0  & 3.0  & 3.0       & 3.0      & 3.0     \\
        JES (S)       & $^{+2.6}_{-2.6}$
                                  & $^{+3.3}_{-3.1}$
                                          & &
                                                                & $^{-0.8}_{+0.6}$
                                                                        & $^{+2.7}_{-2.8}$
                                                                                & $^{+5.1}_{-5.1}$
                                                                                          & $^{+8.2}_{-6.8}$
                                                                                                         & $^{+10.8}_{-3.4}$    \\
        ISR/FSR               & \multicolumn{2}{c}{$^{+2.0}_{-2.0}$} &       &       &       &           &           &      \\
        Cross-Section     &  5.0   & 5.0 &   &   & 10 & 10 & 6    & 30      & 30      \\
        Multijet Norm.  &       & & 0.7 &       & &          &           &           \\
        Mistag (S) & & & & $^{+17.9}_{-17.4}$ & & & & &\\
      \end{tabular}
\end{ruledtabular}

\end{center}
\end{table}

\begin{table}[h]
\caption{\label{tab:d0vvbb} Systematic uncertainty ranges on the signal and background 
contributions and the error on the total background
for D0's $ZH\rightarrow\nu\nu b{\bar{b}}$ medium-tag and tight-tag channels.
Systematic uncertainties are listed by name, see the original
references for a detailed explanation of their meaning and on how they
are derived.  Systematic uncertainties for $VH$ ($WH$+$ZH$) shown in
this table are obtained for $m_H=115$ GeV/$c^2$. Uncertainties are
relative, in percent, and are symmetric unless otherwise indicated.
Shape uncertainties are labeled with an ``(S)'', and ``SH'' represents
shape only uncertainty.}
\vskip 0.2cm
{\centerline{$ZH \rightarrow\nu\nu b\bar{b}$ medium-tag channel relative uncertainties (\%)}}
\vskip 0.099cm
\begin{ruledtabular}
\begin{tabular}{l c c c c c c }\\
Contribution            & Top  & $V+b\bar{b}/c\bar{c}$ & $V$+l.f. & Dibosons  & Total Bkgd & $VH$  \\
\hline
Jet ID/Reco Eff (S)             &   2.0 &   2.0  & 2.0    &  2.0     &  1.9 & 2.0  \\
Jet Energy Scale (S)            &   1.3 &   1.5  & 2.8    &  1.5    &1.9  & 0.3  \\
Jet Resolution (S)              &   0.5 &   0.4  & 0.5    &  0.8     & 0.5  & 0.9  \\
Vertex Conf. / Taggability (S)  &   3.4 &   2.2  & 2.0    &  2.3     & 2.2  & 2.1  \\
b Tagging (S)                   &   1.5 &   2.6  & 8.0    &  3.6     & 3.7  & 0.6  \\
Lepton Identification           &   1.5 &   0.9  & 0.8    &  0.9     & 0.9  & 0.9  \\
Trigger                         &   2.0 &   2.0  & 2.0    &  2.0     & 1.9  & 2.0  \\
Heavy Flavor Fractions          & --   &  20.0  &  --    &  --      & 8.4  & --   \\
Cross Sections                  &  10.0 & 10.2   & 10.2   &  7.0     &  9.8 & 7.0  \\
Signal Branching Fraction       & --   &  --    &  --    &  --      &  --  & 1-9  \\
Luminosity                      &   6.1 &  6.1   & 6.1    &  6.1     & 5.8  & 6.1  \\
Multijet Normalilzation         & --   &  --    &  --    &  --      & 1.1  & --   \\
ALPGEN MLM (S)                  & --   &  --    &  SH    &  --      & --   & --   \\
ALPGEN Scale (S)                & --   &  SH    &  SH    &  --      & --   & --   \\
Underlying Event (S)            & --   &  SH    &  SH    &  --      & --   & --   \\
PDF, reweighting (S)            & SH   &  SH    &  SH    &  SH      & SH   & SH   \\
Total uncertainty     &  12.8 & 23.8   &  15.1  &  10.8    & 14.2 & 10.0  \\
\end{tabular}
\vskip 0.5cm
{\centerline{$ZH \rightarrow\nu\nu b\bar{b}$ tight-tag channel relative uncertainties (\%)}}
\vskip 0.099cm
\begin{tabular}{ l c c c c c c  }   \\
Contribution            & Top  & $V+b\bar{b}/c\bar{c}$ & $V$+l.f. & Dibosons  & Total Bkgd & $VH$  \\
\hline
Jet ID/Reco Eff (S)             & 2.0 & 2.0    & 2.0    & 2.0      &  2.0 & 2.0  \\
Jet Energy Scale (S)            & 1.0 & 1.6    & 3.9    & 1.6      &  1.6 & 0.5  \\
Jet Resolution (S)              & 0.7 & 0.6    & 2.6    & 1.4      &  0.8 & 1.3  \\
Vertex Conf. / Taggability (S)  & 3.0  & 1.9    & 2.4    & 2.0      & 2.3  & 1.9  \\
b Tagging (S)                   & 8.9  & 7.3    & 12.5    & 6.4      & 7.4  & 7.8  \\
Lepton Identification           & 1.9  & 0.8    & 0.3    &  0.7     & 1.1  & 0.8  \\
Trigger                         & 2.0  & 2.0    & 2.0    &  2.0     & 2.0  & 2.0  \\
Heavy Flavor Fractions          & --   &  20.0  &  --    &  --      & 11.0  & --   \\
Cross Sections                  &  10.0 & 10.2   & 10.2   & 7.0      & 10.0  & 7.0  \\
Signal Branching Fraction       & --   &  --    &  --    &  --      &  --  & 1-9  \\
Luminosity                      & 6.1  &  6.1   &  6.1   &  6.1     & 6.1  & 6.1  \\
Multijet Normalilzation         & --   &  --    &  --    &  --      & 0.2  & --   \\
ALPGEN MLM (S)                  & --   &  --    &  SH    &  --      & --   & --   \\
ALPGEN Scale (S)                & --   &  SH    &  SH    &  --      & --   & --   \\
Underlying Event (S)            & --   &  SH    &  SH    &  --      & --   & --   \\
PDF, reweighting (S)            & SH   &  SH    &  SH    &  SH      & SH   & SH   \\
Total uncertainty     &  15.5 & 24.7   & 18.3   &  12.0    & 16.8 & 12.7  \\
\end{tabular}
\end{ruledtabular}

\end{table}

\begin{table}
\begin{center}
\caption{\label{tab:cdfllbb1} Systematic uncertainties on the signal and background 
    contributions for CDF's $ZH\rightarrow \ell^+\ell^-b{\bar{b}}$ tight double tag 
    (TT) and one tight tag and one loose tag (TL) channels. Systematic uncertainties 
    are listed by name; see the original references for a detailed explanation of 
    their meaning and on how they are derived.  Uncertainties are relative, in percent 
    on the event yield. Shape uncertainties are labeled with an ``(S)''.}
\vskip 0.2cm
{\centerline{CDF: tight double tag (TT) $\ell\ell b \bar{b}$ channels relative uncertainties (\%)}}
\vskip 0.099cm
\begin{ruledtabular}
\begin{tabular}{lccccccccc} \\
Contribution   & ~Fakes~ & ~~~$t\bar{t}$~~~  & ~~$WW$~~ & ~~$WZ$~~ & ~~$ZZ$~~  & ~$Z+c{\bar{c}}$~ & ~$Z+b{\bar{b}}$~& ~Mistags~ & ~~~$ZH$~~~ \\ \hline
Luminosity ($\sigma_{\mathrm{inel}}(p{\bar{p}})$)          &     &    3.8 &     3.8&3.8&3.8 &    3.8           &    3.8          &        &    3.8  \\
Luminosity Monitor        &     &    4.4 & 4.4 & 4.4 &      4.4 &    4.4           &    4.4          &       &    4.4  \\
Lepton ID    &     &    1 &    1& 1& 1 &      1           &    1          &       &    1  \\
Lepton Energy Scale    &     &    1.5 &      1.5 & 1.5& 1.5 &    1.5           &    1.5          &        &    1.5  \\
Fake $Z\rightarrow e^+ e^-$       & 50    &   & &   &    &              &             &        &     \\
Fake $Z\rightarrow \mu^+ \mu^-$       & 5    &   & &   &    &              &             &        &     \\
Tight Mistag Rate  &   &   & & &  &  &  & 40 &  \\
Loose Mistag Rate  &   &   & & &  &  &  &  &  \\
JES  [$e^+ e^-$, 2 jet]     &     &
$^{+0.8}_{-0.7}$   &   
$^{+14.4}_{-13.2}$   &   
$^{+6.2}_{-6.2}$   &   
$^{+8.2}_{-8.3}$   &   
$^{+5.6}_{-5.6}$   &   
$^{+8.1}_{-7.9}$   &   
$^{+10.4}_{-10.4}$   &   
$^{+3.6}_{-4.2}$   \\  
JES [$e^+ e^-$, 3 jet]        &     &
$^{+8.3}_{-8.2}$   &   
$^{-0.7}_{+1.7}$   &   
$^{-4.2}_{+4.3}$   &   
$^{+14.4}_{-13.3}$   &   
$^{+10.6}_{-10.5}$   &   
$^{+13.2}_{-13.2}$   &   
$^{+12.4}_{-12.4}$   &   
$^{+15.1}_{-14.9}$   \\   
JES  [$\mu^+ \mu^-$, 2 jet]     &     &
$^{+1.0}_{-0.9}$   &   
$^{+5.4}_{+2.1}$   &   
$^{+13.4}_{-13.4}$   &   
$^{+7.7}_{-7.7}$   &   
$^{-1.5}_{+1.5}$   &   
$^{+8.2}_{-8.2}$   &   
$^{+5.7}_{-5.8}$   &   
$^{+3.1}_{-3.5}$   \\   
JES  [$\mu^+ \mu^-$, 3 jet]     &     &
$^{+9.3}_{-9.1}$   &   
$^{+3.9}_{-3.0}$   &   
$^{+4.8}_{-5.7}$   &   
$^{+15.5}_{-15.5}$   &   
$^{+7.3}_{-7.3}$   &   
$^{+14.2}_{-14.5}$   &   
$^{+20.5}_{-18.0}$   &   
$^{+12.5}_{-13.3}$   \\   
Tight $b$-tag Rate       &     &    7.8 &      7.8 & 7.8 & 7.8 &    7.8           &   7.8         &      &    7.8 \\
Loose $b$-tag Rate       &     &     &       &  &  &              &           &      &     \\
$t{\bar{t}}$ Cross Section         &     &   10&  &&      &              &             &        &     \\
Diboson Cross Section        &     &    & 6 & 6& 6    &              &             &        &     \\
$Z+$HF Cross Section      &        &  &&  &    &  40            & 40           &        &     \\
$ZH$ Cross Section    &     &    &   &&   &              &             &        &    5 \\
ISR/FSR           &     &    &     &&  &              &             &        &   5.5--7.6 \\
Electron Trigger Eff.  &        & 1    & 1 & 1& 1   & 1              & 1             &      &   1     \\
Muon Trigger Eff.  &        & 5   & 5 & 5& 5   & 5              & 5             &      &   5    \\
\end{tabular}
\end{ruledtabular}
\vskip 0.5cm
{\centerline{CDF: one tight and one loose tag (TL) $\ell\ell b \bar{b}$ channels relative uncertainties (\%)}}
\vskip 0.099cm
\begin{ruledtabular}
\begin{tabular}{lccccccccc} \\
Contribution   & ~Fakes~ & ~~~$t\bar{t}$~~~  & ~~$WW$~~ & ~~$WZ$~~ & ~~$ZZ$~~  & ~$Z+c{\bar{c}}$~ & ~$Z+b{\bar{b}}$~& ~Mistags~ & ~~~$ZH$~~~ \\ \hline
Luminosity ($\sigma_{\mathrm{inel}}(p{\bar{p}})$)          &     &    3.8 &     3.8&3.8&3.8 &    3.8           &    3.8          &        &    3.8  \\
Luminosity Monitor        &     &    4.4 & 4.4 & 4.4 &      4.4 &    4.4           &    4.4          &       &    4.4  \\
Lepton ID    &     &    1 &    1& 1& 1 &      1           &    1          &       &    1  \\
Lepton Energy Scale    &     &    1.5 &      1.5 & 1.5& 1.5 &    1.5           &    1.5          &        &    1.5  \\
Fake $Z\rightarrow e^+ e^-$       & 50    &   & &   &    &              &             &        &     \\
Fake $Z\rightarrow \mu^+ \mu^-$       & 5    &   & &   &    &              &             &        &     \\
Tight Mistag Rate  &   &   & & &  &  &  & 19 &  \\
Loose Mistag Rate  &   &   & & &  &  &  & 10 &  \\
JES  [$e^+ e^-$, 2 jet]     &     &
$^{+0.9}_{-1.0}$   &   
$^{+13.0}_{-12.6}$   &   
$^{+9.3}_{-9.4}$   &   
$^{+10.3}_{-10.2}$   &   
$^{+10.3}_{-10.3}$   &   
$^{+8.9}_{-9.3}$   &   
$^{+10.4}_{-10.4}$   &   
$^{+4.0}_{-4.2}$   \\   
JES [$e^+ e^-$, 3 jet]        &     &
$^{+6.9}_{-7.0}$   &   
$^{+10.3}_{-8.3}$   &   
$^{+16.2}_{-16.0}$   &   
$^{+14.6}_{-14.5}$   &   
$^{+22.8}_{-23.4}$   &   
$^{+15.1}_{-15.2}$   &   
$^{+18.5}_{-18.5}$   &   
$^{+14.3}_{-14.4}$   \\   
JES  [$\mu^+ \mu^-$, 2 jet]     &     &
$^{+1.1}_{-1.1}$   &   
$^{+3.7}_{1.8}$   &   
$^{+6.5}_{-6.5}$   &   
$^{+7.5}_{-7.5}$   &   
$^{+12.5}_{-12.4}$   &   
$^{+10.1}_{-10.1}$   &   
$^{+11.0}_{-11.0}$   &   
$^{+4.0}_{-4.1}$   \\   
JES  [$\mu^+ \mu^-$, 3 jet]     &     &
$^{+8.0}_{-8.0}$   &   
$^{+2.0}_{-1.6}$   &   
$^{+14.4}_{-14.5}$   &   
$^{+24.1}_{-24.1}$   &   
$^{+16.0}_{-14.7}$   &   
$^{+17.5}_{-17.6}$   &   
$^{+14.3}_{-14.2}$   &   
$^{+13.1}_{-14.0}$   \\   
Tight $b$-tag Rate       &     &    3.9 &      3.9 & 3.9 & 3.9 &    3.9           &   3.9         &      &    3.9 \\
Loose $b$-tag Rate     &     &    3.2 &      3.2 & 3.2 & 3.2 &    3.2           &   3.2        &      &    3.2 \\
$t{\bar{t}}$ Cross Section         &     &   10&  &&      &              &             &        &     \\
Diboson Cross Section        &     &    & 6 & 6& 6    &              &             &        &     \\
$Z+$HF Cross Section      &        &  &&  &    &  40            & 40           &        &     \\
$ZH$ Cross Section    &     &    &   &&   &              &             &        &    5 \\
ISR/FSR           &     &    &     &&  &              &             &        &   3.4--7.0 \\
Electron Trigger Eff.  &        & 1    & 1 & 1& 1   & 1              & 1             &      &   1     \\
Muon Trigger Eff.  &        & 5   & 5 & 5& 5   & 5              & 5             &      &   5    \\
\end{tabular}
\end{ruledtabular}

\end{center}
\end{table}

\begin{table}
\begin{center}
\caption{\label{tab:cdfllbb2}  Systematic uncertainties on the signal and 
    background contributions for CDF's $ZH\rightarrow \mu^+\mu^-b{\bar{b}}$ 
    single tight tag (Tx) and double loose tag (LL) channels.  Systematic 
    uncertainties are listed by name; see the original references for a 
    detailed explanation of their meaning and on how they are derived.  
    Uncertainties are relative, in percent on the event yield. Shape 
    uncertainties are labeled with an ``(S)''.}
\vskip 0.2cm
{\centerline{CDF: single tight tag (TT) $\ell\ell b \bar{b}$ channels relative uncertainties (\%)}}
\vskip 0.099cm
\begin{ruledtabular}
\begin{tabular}{lccccccccc} \\
Contribution   & ~Fakes~ & ~~~$t\bar{t}$~~~  & ~~$WW$~~ & ~~$WZ$~~ & ~~$ZZ$~~  & ~$Z+c{\bar{c}}$~ & ~$Z+b{\bar{b}}$~& ~Mistags~ & ~~~$ZH$~~~ \\ \hline
Luminosity ($\sigma_{\mathrm{inel}}(p{\bar{p}})$)          &     &    3.8 &     3.8&3.8&3.8 &    3.8           &    3.8          &        &    3.8  \\
Luminosity Monitor        &     &    4.4 & 4.4 & 4.4 &      4.4 &    4.4           &    4.4          &       &    4.4  \\
Lepton ID    &     &    1 &    1& 1& 1 &      1           &    1          &       &    1  \\
Lepton Energy Scale    &     &    1.5 &      1.5 & 1.5& 1.5 &    1.5           &    1.5          &        &    1.5  \\
Fake $Z\rightarrow e^+ e^-$       & 50    &   & &   &    &              &             &        &     \\
Fake $Z\rightarrow \mu^+ \mu^-$       & 5    &   & &   &    &              &             &        &     \\
Tight Mistag Rate  &   &   & & &  &  &  & 19 &  \\
Loose Mistag Rate  &   &   & & &  &  &  &  &  \\
JES  [$e^+ e^-$, 2 jet]     &     &
$^{-0.3}_{+0.3}$   &   
$^{+13.7}_{-13.5}$   &   
$^{+8.5}_{-8.5}$   &   
$^{+6.5}_{-6.3}$   &   
$^{+13.2}_{-13.2}$   &   
$^{+11.0}_{-11.1}$   &   
$^{+12.0}_{-12.0}$   &   
$^{+3.5}_{-3.8}$   \\   
JES [$e^+ e^-$, 3 jet]        &     &
$^{+7.1}_{-7.1}$   &   
$^{+8.9}_{-8.2}$   &   
$^{+17.0}_{-17.0}$   &   
$^{+15.4}_{-15.4}$   &   
$^{+16.4}_{-16.4}$   &   
$^{+15.8}_{-15.9}$   &   
$^{+18.6}_{-18.5}$   &   
$^{+15.4}_{-15.7}$   \\   
JES  [$\mu^+ \mu^-$, 2 jet]     &     &
$^{+0.6}_{-0.7}$   &   
$^{+3.9}_{-3.3}$   &   
$^{+8.6}_{-8.6}$   &   
$^{+7.6}_{-7.7}$   &   
$^{+10.2}_{-10.5}$   &   
$^{+9.3}_{-9.3}$   &   
$^{+11.1}_{-11.1}$   &   
$^{+3.4}_{-3.7}$   \\  
JES  [$\mu^+ \mu^-$, 3 jet]     &     &
$^{+5.5}_{-5.5}$   &   
$^{+5.7}_{-1.9}$   &   
$^{+16.6}_{-16.6}$   &   
$^{+16.8}_{-16.8}$   &   
$^{+16.1}_{-16.2}$   &   
$^{+16.1}_{-16.2}$   &   
$^{+17.5}_{-17.5}$   &   
$^{+13.8}_{-13.9}$   \\   
Tight $b$-tag Rate       &     &    3.9 &      3.9 & 3.9 & 3.9 &    3.9           &   3.9         &      &    3.9 \\
Loose $b$-tag Rate     &     &     &      &  &  &               &           &      &     \\
$t{\bar{t}}$ Cross Section         &     &   10&  &&      &              &             &        &     \\
Diboson Cross Section        &     &    & 6 & 6& 6    &              &             &        &     \\
$Z+$HF Cross Section      &        &  &&  &    &  40            & 40           &        &     \\
$ZH$ Cross Section    &     &    &   &&   &              &             &        &    5 \\
ISR/FSR           &     &    &     &&  &              &             &        &   0.9--12.8\\
Electron Trigger Eff.  &        & 1    & 1 & 1& 1   & 1              & 1             &      &   1     \\
Muon Trigger Eff.  &        & 5   & 5 & 5& 5   & 5              & 5             &      &   5    \\
\end{tabular}
\end{ruledtabular}
\vskip 0.5cm
{\centerline{CDF: double loose tag (LL) $\ell\ell b \bar{b}$ channels relative uncertainties (\%)}}
\vskip 0.099cm
\begin{ruledtabular}
\begin{tabular}{lccccccccc} \\
Contribution   & ~Fakes~ & ~~~$t\bar{t}$~~~  & ~~$WW$~~ & ~~$WZ$~~ & ~~$ZZ$~~  & ~$Z+c{\bar{c}}$~ & ~$Z+b{\bar{b}}$~& ~Mistags~ & ~~~$ZH$~~~ \\ \hline
Luminosity ($\sigma_{\mathrm{inel}}(p{\bar{p}})$)          &     &    3.8 &     3.8&3.8&3.8 &    3.8           &    3.8          &        &    3.8  \\
Luminosity Monitor        &     &    4.4 & 4.4 & 4.4 &      4.4 &    4.4           &    4.4          &       &    4.4  \\
Lepton ID    &     &    1 &    1& 1& 1 &      1           &    1          &       &    1  \\
Lepton Energy Scale    &     &    1.5 &      1.5 & 1.5& 1.5 &    1.5           &    1.5          &        &    1.5  \\
Fake $Z\rightarrow e^+ e^-$       & 50    &   & &   &    &              &             &        &     \\
Fake $Z\rightarrow \mu^+ \mu^-$       & 5    &   & &   &    &              &             &        &     \\
Tight Mistag Rate  &   &   & & &  &  &  &  &  \\
Loose Mistag Rate  &   &   & & &  &  &  & 20 &  \\
JES  [$e^+ e^-$, 2 jet]     &     &
$^{+0.5}_{-0.5}$   &   
$^{+7.5}_{-4.8}$   &   
$^{+8.6}_{-8.7}$   &   
$^{+9.0}_{-8.9}$   &   
$^{+10.0}_{-9.3}$   &   
$^{+11.3}_{-11.0}$   &   
$^{+12.5}_{-12.5}$   &   
$^{+4.0}_{-4.4}$   \\   
JES [$e^+ e^-$, 3 jet]        &     &
$^{+8.6}_{-8.6}$   &   
$^{+32.9}_{-29.5}$   &   
$^{+14.6}_{-14.9}$   &   
$^{+16.5}_{-15.2}$   &   
$^{+20.8}_{-20.8}$   &   
$^{+17.8}_{-17.9}$   &   
$^{+18.9}_{-19.0}$   &   
$^{+14.6}_{-15.4}$   \\   
JES  [$\mu^+ \mu^-$, 2 jet]     &     &
$^{+2.5}_{-2.5}$   &   
$^{+4.5}_{-3.0}$   &   
$^{+6.7}_{-6.7}$   &   
$^{+10.2}_{-9.9}$   &   
$^{+9.2}_{-9.3}$   &   
$^{+7.7}_{-7.6}$   &   
$^{+11.5}_{-11.5}$   &   
$^{+3.9}_{-4.3}$   \\   
JES  [$\mu^+ \mu^-$, 3 jet]     &     &
$^{+9.2}_{-9.2}$   &   
$^{+13.4}_{-10.4}$   &   
$^{+14.1}_{-14.1}$   &   
$^{+16.6}_{-16.6}$   &   
$^{+14.7}_{-14.7}$   &   
$^{+16.8}_{-16.9}$   &   
$^{+17.5}_{-17.5}$   &   
$^{+11.6}_{-12.2}$   \\   
Tight $b$-tag Rate     &     &     &      &  &  &               &           &      &     \\
Loose $b$-tag Rate       &     &    6.3 &      6.3 & 6.3 & 6.3 &    6.3          &   6.3         &      &    6.3 \\
$t{\bar{t}}$ Cross Section         &     &   10&  &&      &              &             &        &     \\
Diboson Cross Section        &     &    & 6 & 6& 6    &              &             &        &     \\
$Z+$HF Cross Section      &        &  &&  &    &  40            & 40           &        &     \\
$ZH$ Cross Section    &     &    &   &&   &              &             &        &    5 \\
ISR/FSR           &     &    &     &&  &              &             &        &   3.1--15.2 \\
Electron Trigger Eff.  &        & 1    & 1 & 1& 1   & 1              & 1             &      &   1     \\
Muon Trigger Eff.  &        & 5   & 5 & 5& 5   & 5              & 5             &      &   5    \\
\end{tabular}
\end{ruledtabular}

\end{center}
\end{table}

\begin{table}
\caption{\label{tab:d0llbb1}Systematic uncertainties on the contributions for 
D0's $ZH\rightarrow \ell^+\ell^-b{\bar{b}}$ channels.
Systematic uncertainties are listed by name; see the original references for a 
detailed explanation of their meaning and on how they
are derived.
Systematic uncertainties for $ZH$  shown in this table are obtained for $m_H=115$ GeV/$c^2$.
Uncertainties are relative, in percent, and are symmetric unless otherwise indicated. Shape uncertainties are 
labeled with an ``(S)''. }
\vskip 0.8cm
{\centerline{$ZH \rightarrow \ell\ell b \bar{b}$ Single Tag (ST) channels relative uncertainties (\%)}}
\vskip 0.099cm
\begin{ruledtabular}
\begin{tabular}{  l  c  c  c  c  c  c  c  c }   
Contribution              & $ZH$  & Multijet& $Z$+l.f.  &  $Z+$\bb & $Z+$\cc & Dibosons & Top\\ \hline
Jet Energy Scale (S)      &  4.2  &   --    &  6.8   &  4.9   &  5.2  &  6.7   &  3.3  \\
Jet Energy Resolution (S) &  1.2  &   --    &  5.2   &  3.3   &  3.2  &  2.2   &  0.4  \\
Jet ID (S)                &  0.3  &   --    &  0.7   &  0.3   &  0.5  &  0.5   &  0.6  \\
Taggability (S)           &  1.5  &   --    &  1.0   &  1.0   &  1.3  &  1.3   &  0.8  \\
$Z p_T$ Model (S)         &   --  &   --    &  2.7   &  1.4   &  1.5  &   --   &   --  \\
HF Tagging Efficiency (S) &  0.4  &   --    &   --   &  1.1   &  4.0  &   --   &  1.3  \\
LF Tagging Efficiency (S) &   --  &   --    &   73   &   --   &   --  &  3.0   &   --  \\
$ee$ Multijet Shape (S)   &   --  &   54    &   --   &   --   &   --  &   --   &   --  \\
Multijet Normalization    &   --  &  1-70   &   --   &   --   &   --  &   --   &   --  \\
$Z$+jets Jet Angles (S)   &   --  &   --    &  1.7   &  2.9   &  3.4  &   --   &   --  \\
Alpgen MLM (S)            &   --  &   --    &  0.3   &   --   &   --  &   --   &   --  \\
Alpgen Scale (S)          &   --  &   --    &  0.4   &  0.4   &  0.4  &   --   &   --  \\
Underlying Event (S)      &   --  &   --    &  0.2   &  0.4   &  0.3  &   --   &   --  \\
Trigger (S)               & 0.4   &   --    &  0.03  &  0.2   &  0.2  &  0.2   &  0.5  \\
Cross Sections            & 6     &   --    &   --   &  20    &  20   &  7     &  10   \\
Signal Branching Fraction & 1-9   &   --    &   --   &  --    &  --   &  --    &  --   \\
Normalization             & 2.5   &   --    &  0.3   &  0.3   &  0.3  &  2.5   &  2.5  \\
PDFs                      & 0.6   &   --    &  1.0   &  2.4   &  1.1  &  0.7   &  5.9 \\

\end{tabular}
\end{ruledtabular}
\vskip 0.8cm
{\centerline{$ZH \rightarrow \ell\ell b \bar{b}$ Double Tag (DT) channels relative uncertainties (\%)}}
\vskip 0.099cm
\begin{ruledtabular}
\begin{tabular}{  l  c c  c  c  c  c  c  c }  \\
Contribution             & $ZH$ & Multijet& $Z$+l.f.  &  $Z+$\bb & $Z+$\cc & Dibosons & Top\\  \hline
Jet Energy Scale (S)     &  2.6  &   --    &  7.4   &  6.5   &  5.1   &  5.8   &  1.0  \\
Jet Energy Resolution(S) &  1.0  &   --    &  4.0   &  4.4   &  4.7   &  0.9   &  0.9  \\
JET ID (S)               &  0.8  &   --    &  0.8   &  0.1   &  0.1   &  0.8   &  0.8  \\
Taggability (S)          &  0.9  &   --    &  0.5   &  1.0   &  0.8   &  0.7   &  0.9  \\
$Z_{p_T}$ Model (S)      &   --  &   --    &  1.3   &  1.3   &  2.0   &   --   &   --  \\
HF Tagging Efficiency (S)&  5.3  &   --    &   --   &  5.7   &  5.9   &   --   &  4.0  \\
LF Tagging Efficiency (S)&   --  &   --    &  47    &   --   &   --   &  6.2   &   --  \\
$ee$ Multijet Shape (S)  &   --  &    59   &   --   &   --   &   --   &   --   &   --  \\
Multijet Normalization   &   --  &  1-70   &   --   &   --   &   --   &   --   &   --  \\
$Z$+jets Jet Angles (S)  &   --  &   --    &  1.4   &  3.7   &  3.7   &   --   &   --  \\
Alpgen MLM (S)           &   --  &   --    &  0.2   &   --   &   --   &   --   &   --  \\
Alpgen Scale (S)         &   --  &   --    &  0.3   &  0.4   &  0.4   &   --   &   --  \\
Underlying Event(S)      &   --  &   --    &  0.3   &  0.4   &  0.4   &   --   &   --  \\
Trigger (S)              &  0.4  &   --    &  0.4   &  0.3   &  0.2   &  0.3   &  0.5  \\
Cross Sections           &  6    &   --    &   --   &  20    &  20    &  7     &  10    \\
Signal Branching Fraction& 1-9   &   --    &   --   &  --    &  --    &  --    &  --   \\
Normalization            &  2.5  &   --    &  0.3   &  0.3   &  0.3   &  2.5   &  2.5     \\
PDFs                     & 0.6   &   --    &  1.0   &  2.4   &  1.1   &  0.7   &  5.9 \\
\end{tabular}
\end{ruledtabular}
\end{table}

\clearpage

\begin{table}
\begin{center}
\caption{\label{tab:cdfsystww0} Systematic uncertainties on the signal and background contributions for CDF's
$H\rightarrow W^+W^-\rightarrow\ell^{\pm}\ell^{\prime \mp}$ channels with zero, one, and two or more associated
jets.  These channels are sensitive to gluon fusion production (all channels) and $WH, ZH$ and VBF production.
Systematic uncertainties are listed by name (see the original references for a detailed explanation of their
meaning and on how they are derived).  Systematic uncertainties for $H$ shown in this table are obtained for
$m_H=160$ GeV/$c^2$.  Uncertainties are relative, in percent, and are symmetric unless otherwise indicated.
The uncertainties associated with the different background and signal processed are correlated within individual
jet categories unless otherwise noted.  Boldface and italics indicate groups of uncertainties which are correlated
with each other but not the others on the line.}

\vskip 0.1cm
{\centerline{CDF: $H\rightarrow W^+W^-\rightarrow\ell^{\pm}\ell^{\prime \mp}$ with no associated jet channel relative uncertainties (\%)}}
\vskip 0.099cm
\begin{ruledtabular}
\begin{tabular}{lccccccccccc} \\
Contribution               &   $WW$       &  $WZ$        &  $ZZ$        &  $t\bar{t}$  &  DY          &  $W\gamma$  & $W$+jet & $gg\to H$   &  $WH$        &  $ZH$        &  VBF         \\ \hline
{\bf Cross Section}        &              &              &              &              &              &             &         &             &              &              &              \\ 
ScaleInclusive             &              &              &              &              &              &             &         & 13.4      &              &              &              \\ 
Scale1+Jets                &              &              &              &              &              &             &         & $-23.0$   &              &              &              \\ 
Scale2+Jets                &              &              &              &              &              &             &         & 0.0       &              &              &              \\ 
PDF Model                  &              &              &              &              &              &             &         & 7.6       &              &              &              \\ 
Total                      & {\it 6.0}  & {\it 6.0}  & {\it 6.0}  & 7.0        &              &             &         &             & {\bf 5.0}  & {\bf 5.0}  & 10.0       \\ 
{\bf Acceptance}           &              &              &              &              &              &             &         &             &              &              &              \\ 
Scale (jets)               & {\it 0.3}s &              &              &              &              &             &         &             &              &              &              \\  
PDF Model (leptons)        &              &              &              &              &              &             &         & 2.7       &              &              &              \\  
PDF Model (jets)           & {\it 1.1}  &              &              &              &              &             &         & 5.5       &              &              &              \\  
Higher-order Diagrams      &              & {\it 10.0} & {\it 10.0} & 10.0       &              & 10.0      &         &             & {\bf 10.0} & {\bf 10.0} & {\bf 10.0} \\ 
$\MET$ Modeling            &              &              &              &              & 19.0       &             &         &             &              &              &              \\ 
Conversion Modeling        &              &              &              &              &              &  6.8      &         &             &              &              &              \\ 
Jet Fake Rates             &              &              &              &              &              &             &         &             &              &              &              \\ 
(Low S/B)                  &              &              &              &              &              &             & 15.0  &             &              &              &              \\
(High S/B)                 &              &              &              &              &              &             & 24.0  &             &              &              &              \\ 
Jet Energy Scale           & {\it 3.1}  & {\it 6.2}  & {\it 3.5}  & {\it 28.2} & {\it 18.0} & {\it 3.5} &         & {\it 5.7} & {\it 9.9}  & {\it 5.3}  & {\it 12.9} \\ 
Lepton ID Efficiencies     & {\it 3.8}  & {\it 3.8}  & {\it 3.8}  & {\it 3.8}  & {\it 3.8}  &             &         & {\it 3.8} & {\it 3.8}  & {\it 3.8}  & {\it 3.8}  \\ 
Trigger Efficiencies       & {\it 2.0}  & {\it 2.0}  & {\it 2.0}  & {\it 2.0}  & {\it 2.0}  &             &         & {\it 2.0} & {\it 2.0}  & {\it 2.0}  & {\it 2.0}  \\ 
{\bf Luminosity}           & {\it 5.9}  & {\it 5.9}  & {\it 5.9}  & {\it 5.9}  & {\it 5.9}  &             &         & {\it 5.9} & {\it 5.9}  & {\it 5.9}  & {\it 5.9}  \\ 
\end{tabular}
\end{ruledtabular}

\vskip 0.3cm
{\centerline{CDF: $H\rightarrow W^+W^-\rightarrow\ell^{\pm}\ell^{\prime \mp}$ with one associated jet channel relative uncertainties (\%)}}
\vskip 0.099cm
\begin{ruledtabular}
\begin{tabular}{lccccccccccc} \\
Contribution            &   $WW$       &  $WZ$        &  $ZZ$        &  $t\bar{t}$   &  DY          & $W\gamma$    & $W$+jet & $gg \to H$   &  $WH$        &  $ZH$        &  VBF         \\ \hline
{\bf Cross Section}     &              &              &              &               &              &              &         &              &              &              &              \\ 
ScaleInclusive          &              &              &              &               &              &              &         & 0.0        &              &              &              \\ 
Scale1+Jets             &              &              &              &               &              &              &         & 35.0       &              &              &              \\ 
Scale2+Jets             &              &              &              &               &              &              &         & $-12.7$    &              &              &              \\ 
PDF Model               &              &              &              &               &              &              &         & 17.3       &              &              &              \\ 
Total                   & {\it 6.0}  & {\it 6.0}  & {\it 6.0}  & 7.0         &              &              &         &              & {\bf 5.0}  & {\bf 5.0}  & 10.0       \\ 
{\bf Acceptance}        &              &              &              &               &              &              &         &              &              &              &              \\ 
Scale (jets)            &{\it -4.0}s &              &              &               &              &              &         &              &              &              &              \\ 
PDF Model (leptons)     &              &              &              &               &              &              &         & 3.6        &              &              &              \\ 
PDF Model (jets)        &{\it  4.7}  &              &              &               &              &              &         & -6.3       &              &              &              \\ 
Higher-order Diagrams   &              & {\it 10.0} & {\it 10.0} & 10.0        &              & 10.0       &         &              & {\bf 10.0} & {\bf 10.0} & {\bf 10.0} \\ 
$\MET$ Modeling         &              &              &              &               & 21.0       &              &         &              &              &              &              \\ 
Conversion Modeling     &              &              &              &               &              & 6.8        &         &              &              &              &              \\ 
Jet Fake Rates          &              &              &              &               &              &              &         &              &              &              &              \\ 
(Low S/B)               &              &              &              &               &              &              & 16.0  &              &              &              &              \\
(High S/B)              &              &              &              &               &              &              & 27.0  &              &              &              &              \\ 
Jet Energy Scale        & {\it -5.8} & {\it -1.1} & {\it -4.8} & {\it -13.1} & {\it -6.5} & {\it -9.5} &         & {\it -3.8} & {\it -8.5} & {\it -7.8} & {\it -6.8} \\ 
Lepton ID Efficiencies  & {\it 3.8}  & {\it 3.8}  & {\it 3.8}  & {\it 3.8}   & {\it 3.8}  &              &         & {\it 3.8}  & {\it 3.8}  & {\it 3.8}  & {\it 3.8}  \\ 
Trigger Efficiencies    & {\it 2.0}  & {\it 2.0}  & {\it 2.0}  & {\it 2.0}   & {\it 2.0}  &              &         & {\it 2.0}  & {\it 2.0}  & {\it 2.0}  & {\it 2.0}  \\ 
{\bf Luminosity}        & {\it 5.9}  & {\it 5.9}  & {\it 5.9}  & {\it 5.9}   & {\it 5.9}  &              &         & {\it 5.9}  & {\it 5.9}  & {\it 5.9}  & {\it 5.9}  \\ 
\end{tabular}
\end{ruledtabular}

\end{center}
\end{table}

\begin{table}
\begin{center}
\vskip 0.3cm
{\centerline{CDF: $H\rightarrow W^+W^-\rightarrow\ell^{\pm}\ell^{\prime \mp}$ with two or more associated jets channel relative uncertainties (\%)}}
\vskip 0.0999cm
\begin{ruledtabular}
\begin{tabular}{lccccccccccc} \\
Contribution           &  $WW$         &  $WZ$         &  $ZZ$         &  $t\bar{t}$  &  DY           &  $W\gamma$    & $W$+jet & $gg\to H$     &  $WH$        &  $ZH$        &  VBF         \\ \hline
{\bf Cross Section}    &               &               &               &              &               &               &         &               &              &              &              \\ 
ScaleInclusive         &               &               &               &              &               &               &         & 0.0         &              &              &              \\ 
Scale1+Jets            &               &               &               &              &               &               &         & 0.0         &              &              &              \\ 
Scale2+Jets            &               &               &               &              &               &               &         & 33.0        &              &              &              \\ 
PDF Model              &               &               &               &              &               &               &         & 29.7        &              &              &              \\ 
Total                  & {\it 6.0}   & {\it 6.0}   & {\it 6.0}   & 7.0        &               &               &         &               & {\bf 5.0}  & {\bf 5.0}  & 10.0       \\ 
{\bf Acceptance}       &               &               &               &              &               &               &         &               &              &              &              \\ 
Scale (jets)           & {\it -8.2}s &               &               &              &               &               &         &               &              &              &              \\ 
PDF Model (leptons)    &               &               &               &              &               &               &         & 4.8         &              &              &              \\ 
PDF Model (jets)       & {\it 4.2}   &               &               &              &               &               &         & -12.3       &              &              &              \\ 
Higher-order Diagrams  &               & {\it 10.0}  & {\it 10.0}  & 10.0       &               & 10.0        &         &               & {\bf 10.0} & {\bf 10.0} & {\bf 10.0} \\ 
$\MET$ Modeling        &               &               &               &              & 26.0        &               &         &               &              &              &              \\ 
Conversion Modeling    &               &               &               &              &               & 6.8         &         &               &              &              &              \\ 
Jet Fake Rates         &               &               &               &              &               &               & 19.0  &               &              &              &              \\ 
Jet Energy Scale       & {\it -20.5} & {\it -13.2} & {\it -13.3} & {\it -1.7} & {\it -32.7} & {\it -22.0} &         & {\it -15.1} & {\it -4.0} & {\it -2.5} & {\it -3.8} \\ 
$b$-tag Veto           &               &               &               & 3.6        &               &               &         &               &              &              &              \\ 
Lepton ID Efficiencies & {\it 3.8}   & {\it 3.8}   & {\it 3.8}   & {\it 3.8}  & {\it 3.8}   &               &         & {\it 3.8}   & {\it 3.8}  & {\it 3.8}  & {\it 3.8}  \\ 
Trigger Efficiencies   & {\it 2.0}   & {\it 2.0}   & {\it 2.0}   & {\it 2.0}  & {\it 2.0}   &               &         & {\it 2.0}   & {\it 2.0}  & {\it 2.0}  & {\it 2.0}  \\ 
{\bf Luminosity}       & {\it 5.9}   & {\it 5.9}   & {\it 5.9}   & {\it 5.9}  & {\it 5.9}   &               &         & {\it 5.9}   & {\it 5.9}  & {\it 5.9}  & {\it 5.9}  \\ 
\end{tabular}
\end{ruledtabular}
\end{center}
\end{table}

\begin{table}
\begin{center}
\caption{\label{tab:cdfsystww4} Systematic uncertainties on the signal and background contributions for CDF's low-$M_{\ell\ell}$
$H\rightarrow W^+W^-\rightarrow\ell^{\pm}\ell^{\prime \mp}$ channel with zero or one associated jets.  This channel is sensitive
to only gluon fusion production.  Systematic uncertainties are listed by name (see the original references for a detailed
explanation of their meaning and on how they are derived).  Systematic uncertainties for $H$ shown in this table are obtained
for $m_H=160$ GeV/$c^2$.  Uncertainties are relative, in percent, and are symmetric unless otherwise indicated.  The uncertainties
associated with the different background and signal processed are correlated within individual categories unless otherwise noted.
In these special cases, the correlated uncertainties are shown in either italics or bold face text.}
\vskip 0.1cm
{\centerline{CDF: low $M_{\ell\ell}$ $H\rightarrow W^+W^-\rightarrow\ell^{\pm}\ell^{\prime \mp}$ with zero or one associated jets channel relative uncertainties (\%)}}
\vskip 0.099cm
\begin{ruledtabular}
\begin{tabular}{lccccccccccc} \\
Contribution            & $WW$         & $WZ$         & $ZZ$         & $t\bar{t}$   & DY           & $W\gamma$    & $W$+jet(s) & $gg\to H$   &  $WH$        &  $ZH$        &  VBF         \\ \hline
{\bf Cross Section}     &              &              &              &              &              &              &            &             &              &              &              \\
ScaleInclusive          &              &              &              &              &              &              &            & 8.1       &              &              &              \\
Scale1+Jets             &              &              &              &              &              &              &            & 0.0       &              &              &              \\
Scale2+Jets             &              &              &              &              &              &              &            & $-5.1$    &              &              &              \\
PDF Model               &              &              &              &              &              &              &            & 10.5      &              &              &              \\
Total                   & {\it 6.0}  & {\it 6.0}  & {\it 6.0}  & 7.0        & 5.0        &              &            &             & {\bf 5.0}  & {\bf 5.0}  &  10.0      \\
{\bf Acceptance}        &              &              &              &              &              &              &            &             &              &              &              \\
Scale (jets)            & {\it -0.4}s&              &              &              &              &              &            &             &              &              &              \\
PDF Model (leptons)     &              &              &              &              &              &              &            & 1.0       &              &              &              \\  
PDF Model (jets)        & {\it 1.6}  &              &              &              &              &              &            & 2.1       &              &              &              \\
Higher-order Diagrams   &              & {\it 10.0} & {\it 10.0} & 10.0       & 10.0       &              &            &             & {\bf 10.0} & {\bf 10.0} & {\bf 10.0} \\
Conversion Modeling     &              &              &              &              &              & 8.4        &            &             &              &              &              \\
Jet Fake Rates          &              &              &              &              &              &              & 13.8     &             &              &              &              \\
Jet Energy Scale        & {\it 1.2}  & {\it 2.2}  & {\it 2.0}  & {\it 13.3} & {\it 15.4} & {\it 1.2 } &            & {\it 2.4} & {\it 9.2}  & {\it 6.5}  & {\it 7.8}  \\ 
Lepton ID Efficiencies  & {\it 3.8}  & {\it 3.8}  & {\it 3.8}  & {\it 3.8}  & {\it 3.8}  &              &            & {\it 3.8} & {\it 3.8}  & {\it 3.8}  & {\it 3.8}  \\
Trigger Efficiencies    & {\it 2.0}  & {\it 2.0}  & {\it 2.0}  & {\it 2.0}  & {\it 2.0}  &              &            & {\it 2.0} & {\it 2.0}  & {\it 2.0}  & {\it 2.0}  \\
{\bf Luminosity}        & {\it 5.9}  & {\it 5.9}  & {\it 5.9}  & {\it 5.9}  & {\it 5.9}  &              &            & {\it 5.9} & {\it 5.9}  & {\it 5.9}  & {\it 5.9}  \\
\end{tabular}
\end{ruledtabular}
\end{center}
\end{table}


\begin{table}
\begin{center}
\caption{\label{tab:cdfsystww5} Systematic uncertainties on the signal and background contributions for CDF's
$H\rightarrow W^+W^-\rightarrow e^{\pm} \tau^{\mp}$ and $H\rightarrow W^+W^-\rightarrow \mu^{\pm} \tau^{\mp}$
channels.  These channels are sensitive to gluon fusion production, $WH, ZH$ and VBF production.  Systematic
uncertainties are listed by name (see the original references for a detailed explanation of their meaning
and on how they are derived).  Systematic uncertainties for $H$ shown in this table are obtained for
$m_H=160$ GeV/$c^2$.  Uncertainties are relative, in percent, and are symmetric unless otherwise indicated.
The uncertainties associated with the different background and signal processed are correlated within individual
categories unless otherwise noted.  In these special cases, the correlated uncertainties are shown in either
italics or bold face text.}
\vskip 0.1cm
{\centerline{CDF: $H\rightarrow W^+W^-\rightarrow e^{\pm} \tau^{\mp}$ channel relative uncertainties ( )}}
\vskip 0.099cm
\begin{ruledtabular}
\begin{tabular}{lccccccccccccccc} \\
Contribution                 & $WW$  & $WZ$ & $ZZ$ & $t\bar{t}$  & $Z\rightarrow\tau\tau$  & $Z\rightarrow\ell\ell$  & $W$+jet  & $W\gamma$  & $gg\to H$  & $WH$ & $ZH$ & VBF  \\ \hline
Cross section                & 6.0   & 6.0  & 6.0  & 10.0 &  5.0 &  5.0 &       &      &  10.3  &   5   &   5   &  10  \\
Measured W cross-section     &       &      &      &      &      &      & 12    &      &        &       &       &      \\
PDF Model                    & 1.6   & 2.3  & 3.2  & 2.3  &  2.7 &  4.6 & 2.2   & 3.1  &   2.5  & 2.0   &  1.9  & 1.8  \\
Higher order diagrams        & 10    & 10   & 10   & 10   &  10  &  10  &       & 10   &        & 10    &  10   & 10   \\
Conversion modeling          &       &      &      &      &      &      &       & 10   &        &       &       &      \\
Trigger Efficiency           & 0.5   & 0.6  & 0.6  & 0.6  & 0.7  & 0.5  & 0.6   & 0.6  &   0.5  & 0.5   &  0.6  & 0.5  \\
Lepton ID Efficiency         & 0.4   & 0.5  & 0.5  & 0.4  & 0.4  &  0.4 & 0.5   & 0.4  &   0.4  & 0.4   &  0.4  & 0.4  \\
$\tau$ ID Efficiency         & 1.0   & 1.3  & 1.9  & 1.3  & 2.1  &      &       & 0.3  &   2.8  &  1.6  &  1.7  & 2.8  \\
Jet into $\tau$ Fake rate    & 5.8   & 4.8  & 2.0  & 5.1  &      &  0.1 & 8.8   &      &        &  4.2  &  4.0  & 0.4  \\
Lepton into $\tau$ Fake rate & 0.2   & 0.1  & 0.6  & 0.2  &      &  2.3 &       & 2.1  &  0.15  & 0.06  &  0.15 & 0.11 \\
W+jet scale                  &       &      &      &      &      &      & 1.6   &      &        &       &       &      \\
MC Run dependence            & 2.6   & 2.6  & 2.6  &      &      &      & 2.6   &      &        &       &       &      \\
Luminosity                   & 3.8   & 3.8  & 3.8  & 3.8  & 3.8  & 3.8  & 3.8   & 3.8  &  3.8   & 3.8   &  3.8  & 3.8  \\
Luminosity Monitor           & 4.4   & 4.4  & 4.4  & 4.4  & 4.4  & 4.4  & 4.4   & 4.4  &  4.4   & 4.4   &  4.4  & 4.4  \\
\end{tabular}
\end{ruledtabular}

\vskip 0.3cm
{\centerline{CDF: $H\rightarrow W^+W^-\rightarrow \mu^{\pm} \tau^{\mp}$ channel relative uncertainties (\%)}}
\vskip 0.099cm
\begin{ruledtabular}
\begin{tabular}{lcccccccccccccc} \\
    Contribution                 & $WW$  & $WZ$ & $ZZ$ & $t\bar{t}$  & $Z\rightarrow\tau\tau$  & $Z\rightarrow\ell\ell$  & $W$+jet  & $W\gamma$  & $gg\to H$  & $WH$ & $ZH$ & VBF  \\ \hline
    Cross section                & 6.0  & 6.0  & 6.0  & 10.0 & 5.0 &  5.0 &     &      & 10.4  &   5   &   5   &  10  \\
    Measured W cross-section     &      &      &      &      &     &      & 12  &      &       &       &       &      \\
    PDF Model                    & 1.5  & 2.1  & 2.9  & 2.1  & 2.5 & 4.3  & 2.0 & 2.9  &  2.6  & 2.2   &  2.0  & 2.2  \\
    Higher order diagrams        & 10   & 10   & 10   & 10   &     &      &     & 11   &       & 10    &  10   & 10   \\
    Trigger Efficiency           & 1.3  & 0.7  & 0.7  & 1.1  & 0.9 & 1.3  & 1.0 & 1.0  &  1.3  & 1.3   &  1.2  & 1.3  \\
    Lepton ID Efficiency         & 1.1  & 1.4  & 1.4  & 1.1  & 1.2 & 1.1  & 1.4 & 1.3  &  1.0  & 1.0   &  1.0  & 1.0  \\
    $\tau$ ID Efficiency         & 1.0  & 1.2  & 1.4  & 1.6  & 1.9 &      &     &      &  2.9  &  1.6  &  1.7  & 2.8  \\
    Jet into $\tau$ Fake rate    & 5.8  & 5.0  & 4.4  & 4.4  &     & 0.2  & 8.8 &      &       &  4.5  &  4.2  & 0.4  \\
    Lepton into $\tau$ Fake rate & 0.06 & 0.05 & 0.09 & 0.04 &     & 1.9  &     & 1.2  & 0.04  & 0.02  &  0.02 & 0.04 \\
    W+jet scale                  &      &      &      &      &     &      & 1.4 &      &       &       &       &      \\
    MC Run dependence            & 3.0  & 3.0  & 3.0  &      &     &      & 3.0 &      &       &       &       &      \\
    Luminosity                   & 3.8  & 3.8  & 3.8  & 3.8  & 3.8 & 3.8  & 3.8 & 3.8  & 3.8   & 3.8   &  3.8  & 3.8  \\
    Luminosity Monitor           & 4.4  & 4.4  & 4.4  & 4.4  & 4.4 & 4.4  & 4.4 & 4.4  & 4.4   & 4.4   &  4.4  & 4.4  \\
\end{tabular}
\end{ruledtabular}
\end{center}
\end{table}

\begin{table}
\begin{center}
\caption{\label{tab:cdfsystwww} Systematic uncertainties on the signal and background contributions for
CDF's $WH\rightarrow WWW \rightarrow\ell^{\pm}\ell^{\prime \pm}$ channel with one or more associated
jets and $WH\rightarrow WWW \rightarrow \ell^{\pm}\ell^{\prime \pm} \ell^{\prime \prime \mp}$ channel.
These channels are sensitive to only $WH$ and $ZH$ production.  Systematic uncertainties are listed
by name (see the original references for a detailed explanation of their meaning and on how they are
derived).  Systematic uncertainties for $H$ shown in this table are obtained for $m_H=160$ GeV/$c^2$.
Uncertainties are relative, in percent, and are symmetric unless otherwise indicated.  The uncertainties
associated with the different background and signal processed are correlated within individual categories
unless otherwise noted.  In these special cases, the correlated uncertainties are shown in either italics
or bold face text.}
\vskip 0.1cm
{\centerline{CDF: $WH \rightarrow WWW \rightarrow\ell^{\pm}\ell^{\prime\pm}$ channel relative uncertainties (\%)}}
\vskip 0.099cm
\begin{ruledtabular}
\begin{tabular}{lccccccccc} \\
Contribution               & $WW$         &   $WZ$        &  $ZZ$         & $t\bar{t}$   &  DY           & $W\gamma$    & $W$+jet &  $WH$         &  $ZH$         \\ \hline
{\bf Cross Section}        &              &               &               &              &               &              &         &               &               \\
Total                      & {\it 6.0}  & {\it 6.0}   & {\it 6.0}   & 7.0        & 5.0         &              &         & {\bf 5.0}   & {\bf 5.0}   \\
{\bf Acceptance}           &              &               &               &              &               &              &         &               &               \\
Scale (jets)               & -6.1       &               &               &              &               &              &         &               &               \\
PDF Model (jets)           & 5.7        &               &               &              &               &              &         &               &               \\ 
Higher-order Diagrams      &              & {\it 10.0}  & {\it 10.0}  & 10.0       & 10.0        & 10.0       &         & {\bf 10.0}  & {\bf 10.0}  \\
Conversion Modeling        &              &               &               &              &               & 6.8        &         &               &               \\
Jet Fake Rates             &              &               &               &              &               &              & 37.7  &               &               \\
Charge Mismeasurement Rate & {\it 25.0} &               &               &              & {\it 25.0}  &              &         &               &               \\
Jet Energy Scale           & {\it -4.1} & {\it -4.2}s & {\it -3.3}s & {\it -0.3} & {\it -4.9}s & {\it -9.1} &         & {\it -1.0}s & {\it -0.7}s \\
Lepton ID Efficiencies     & {\it 3.8}  & {\it 3.8}   & {\it 3.8}   & {\it 3.8}  & {\it 3.8}   &              &         & {\it 3.8}   & {\it 3.8}   \\
Trigger Efficiencies       & {\it 2.0}  & {\it 2.0}   & {\it 2.0}   & {\it 2.0}  & {\it 2.0}   &              &         & {\it 2.0}   & {\it 2.0}   \\
{\bf Luminosity}           & {\it 5.9}  & {\it 5.9}   & {\it 5.9}   & {\it 5.9}  & {\it 5.9}   &              &         & {\it 5.9}   & {\it 5.9}   \\
\end{tabular}
\end{ruledtabular}

\vskip 0.3cm
{\centerline{CDF: $WH\rightarrow WWW \rightarrow \ell^{\pm}\ell^{\prime \pm} \ell^{\prime \prime \mp}$ channel relative uncertainties (\%)}}
\vskip 0.0999cm
\begin{ruledtabular}
\begin{tabular}{lccccccc} \\
Contribution                & $WZ$         & $ZZ$         & $Z\gamma$    & $t\bar{t}$  & Fakes   & $WH$         & $ZH$         \\ \hline
{\bf Cross Section}         &              &              &              &             &         &              &              \\
Total                       & {\it 6.0}  & {\it 6.0}  & 10.0       & 7.0       &         & {\bf 5.0}  & {\bf 5.0}  \\
{\bf Acceptance}            &              &              &              &             &         &              &              \\ 
Higher-order Diagrams       & {\it 10.0} & {\it 10.0} & 15.0       & 10.0      &         & {\bf 10.0} & {\bf 10.0} \\
Jet Fake Rates              &              &              &              &             & 22.3  &              &              \\
$b$-Jet Fake Rates          &              &              &              & 27.3      &         &              &              \\
Jet Energy Scale            &              &              & {\it -3.0} &             &         &              &              \\
Lepton ID Efficiencies      & {\it 5.0}  &  {\it 5.0} & {\it 5.0}  & {\it 5.0} &         & {\it 5.0}  & {\it 5.0}  \\
Trigger Efficiencies        & {\it 2.0}  &  {\it 2.0} & {\it 2.0}  & {\it 2.0} &         & {\it 2.0}  & {\it 2.0}  \\
{\bf Luminosity}            & {\it 5.9}  &  {\it 5.9} & {\it 5.9}  & {\it 5.9} &         & {\it 5.9}  & {\it 5.9}  \\
\end{tabular}
\end{ruledtabular}

\end{center}
\end{table}

\begin{table}
\begin{center}
\caption{\label{tab:cdfsystzww} Systematic uncertainties on the signal and background contributions for
CDF's $ZH\rightarrow ZWW \rightarrow \ell^{\pm}\ell^{\mp} \ell^{\prime \pm}$ channels with 1 jet and 2
or more jets.  These channels are sensitive to only $WH$ and $ZH$ production.  Systematic uncertainties
are listed by name (see the original references for a detailed explanation of their meaning and on how
they are derived).  Systematic uncertainties for $H$ shown in this table are obtained for $m_H=160$
GeV/$c^2$.  Uncertainties are relative, in percent, and are symmetric unless otherwise indicated.  The
uncertainties associated with the different background and signal processed are correlated within
individual categories unless otherwise noted.  In these special cases, the correlated uncertainties are
shown in either italics or bold face text.}
\vskip 0.1cm
{\centerline{CDF: $ZH\rightarrow ZWW \rightarrow \ell^{\pm}\ell^{\mp} \ell^{\prime \pm}$ with one associated jet channel relative uncertainties (\%)}}
\vskip 0.0999cm
\begin{ruledtabular}
\begin{tabular}{lccccccc} \\
Contribution                & $WZ$         & $ZZ$         & $Z\gamma$    & $t\bar{t}$   & Fakes   & $WH$         & $ZH$         \\ \hline
{\bf Cross Section}         &              &              &              &              &         &              &              \\
Total                       & {\it 6.0}  & {\it 6.0}  & 10.0       & 7.0        &         & {\bf 5.0}  & {\bf 5.0}  \\
{\bf Acceptance}            &              &              &              &              &         &              &              \\ 
Higher-order Diagrams       & {\it 10.0} & {\it 10.0} & 15.0       & 10.0       &         & {\bf 10.0} & {\bf 10.0} \\
Jet Fake Rates              &              &              &              &              & 23.6  &              &              \\
$b$-Jet Fake Rates          &              &              &              & 42.0       &         &              &              \\
Jet Energy Scale            & {\it -7.8} & {\it -2.4} & {\it -6.4} & {\it  2.2} &         & {\it -7.0} & {\it 7.1}  \\
Lepton ID Efficiencies      & {\it 5.0}  & {\it 5.0}  & {\it 5.0}  & {\it 5.0}  &         & {\it 5.0}  & {\it 5.0}  \\
Trigger Efficiencies        & {\it 2.0}  & {\it 2.0}  & {\it 2.0}  & {\it 2.0}  &         & {\it 2.0}  & {\it 2.0}  \\
{\bf Luminosity}            & {\it 5.9}  & {\it 5.9}  & {\it 5.9}  & {\it 5.9}  &         & {\it 5.9}  & {\it 5.9}  \\
\end{tabular}
\end{ruledtabular}

\vskip 0.3cm
{\centerline{CDF: $ZH\rightarrow ZWW \rightarrow \ell^{\pm}\ell^{\mp} \ell^{\prime \pm}$ with two or more associated jets channel relative uncertainties (\%)}}
\vskip 0.0999cm
\begin{ruledtabular}
\begin{tabular}{lccccccc} \\
Contribution                & $WZ$          & $ZZ$          & $Z\gamma$     & $t\bar{t}$   & Fakes   & $WH$          & $ZH$         \\ \hline
{\bf Cross Section}         &               &               &               &              &         &               &              \\
Total                       & {\it 6.0}   & {\it 6.0}   & 10.0        & 7.0        &         & {\bf 5.0}   & {\bf 5.0}  \\
{\bf Acceptance}            &               &               &               &              &         &               &              \\ 
Higher-order Diagrams       & {\it 10.0}  & {\it 10.0}  & 15.0        & 10.0       &         & {\bf 10.0}  & {\bf 10.0} \\
Jet Fake Rates              &               &               &               &              & 18.4  &               &              \\
$b$-Jet Fake Rates          &               &               &               & 22.2       &         &               &              \\
Jet Energy Scale            & {\it -18.0} & {\it -15.4} & {\it -16.8} & {\it -2.3} &         & {\it -20.1} & {\it -5.5} \\
Lepton ID Efficiencies      & {\it 5.0}   & {\it 5.0}   & {\it 5.0}   & {\it 5.0}  &         & {\it 5.0}   & {\it 5.0}  \\
Trigger Efficiencies        & {\it 2.0}   & {\it 2.0}   & {\it 2.0}   & {\it 2.0}  &         & {\it 2.0}   & {\it 2.0}  \\
{\bf Luminosity}            & {\it 5.9}   & {\it 5.9}   & {\it 5.9}   & {\it 5.9}  &         & {\it 5.9}   & {\it 5.9}  \\
\end{tabular}
\end{ruledtabular}

\end{center}
\end{table}

\begin{table}
\begin{center}
\caption{\label{tab:d0systww}
Systematic uncertainties on the signal and background contributions
for D0's $H\rightarrow W^+W^- \rightarrow\ell^{\pm}\ell^{\mp}$
channels.  Systematic uncertainties are listed by name; see the
original references for a detailed explanation of their meaning and on
how they are derived.  Shape uncertainties are labeled with the ``s''
designation. Systematic uncertainties given in this table are obtained
for the $m_H=165$ GeV/c$^2$ Higgs selection.
  Cross section uncertainties on
the $gg\to H$ signal depend on the jet multiplicity, as described in
the main text. Uncertainties are relative, in percent, and are
symmetric unless otherwise indicated.
}\vskip 0.1cm
{\centerline{$H\rightarrow W^+W^- \rightarrow\ell^{\pm}\ell^{\mp}$ channels relative uncertainties (\%)}}
\vskip 0.099cm
\begin{ruledtabular}
\begin{tabular}{ l  c  c  c  c  c  c  c  c  }  \\
Contribution & Dibosons & ~~$Z/\gamma^* \rightarrow \ell\ell$~~&$~~W+$jet$/\gamma$~~ &~~~~$t\bar{t}~~~~$ & ~~Multijet~~ & $gg\to H$ & $qq\to qqH$ & $VH$ \\
\hline
Luminosity/Normalization                   & 4   & --  & 4   & 4   & 4   & 4          & 4     & 4      \\
Cross Section (Scale/PDF)                  & 5-7 & --  & --  & 7   & --  & 13-33/8-30 & 5     & 6      \\
$Z/\gamma^*\rightarrow\ell\ell$ n-jet norm & --  & 2-15& --  & --  & --  & --         & --    & --     \\
$Z/\gamma^*\rightarrow\ell\ell$ MET model  & --  & 5-19& --  & --  & --  & --         & --    & --     \\
$W+$jet$/\gamma$ norm                      & --  & --  & 6-30& --  & --  & --         & --    & --     \\
$W+$jet$/\gamma$ ISR/FSR model (s)         & --  & --  & 2-20& --  & --  & --         & --    & --     \\
Vertex Confirmation (s)                    & 1-5 & 1-5 & 1-5 & 5-6 & --  & 1-5        & 1-5   & 1-5    \\
Jet identification (s)                     & 1   & 1  & 1    & 1   & --  & 1          & 1     & 1      \\
Jet Energy Scale (s)                       & 1-5 & 1-5& 1-5  & 1-4 & --  & 1-5        & 1-5   & 1-4    \\
Jet Energy Resolution(s)                   & 1-4 & 1-4& 1-4  & 1-4 & --  & 1-3        & 1-4   & 1-3    \\
B-tagging (s)                              & --  & -- & --   & 1-5 & --  & --         & --    & --     \\
\end{tabular}
\end{ruledtabular}
\end{center}
\end{table}

\begin{table}
\begin{center}
\caption{\label{tab:d0systwwtau} Systematic uncertainties on the signal and background contributions for D0's
$H\rightarrow W^+ W^- \rightarrow \mu\nu \tau_{\rm{had}}\nu $ channel.  Systematic uncertainties are listed by
name; see the original references for a detailed explanation of their meaning and on how they are derived.
Shape uncertainties are labeled with the shape designation (S). Systematic uncertainties shown in this table are obtained for the $m_H=165$ GeV/c$^2$ Higgs selection.
Uncertainties are relative, in percent, and are symmetric unless otherwise indicated.}
\vskip 0.1cm
{\centerline{D0: $H\rightarrow W^+ W^- \rightarrow \mu\nu \tau_{\rm{had}}\nu $ channel relative uncertainties (\%)}}
\vskip 0.099cm
\begin{ruledtabular}
%
\begin{tabular}{ l  c  c  c  c  c  c  c  c  c}  \\
Contribution       & Diboson    & ~~$Z/\gamma^* \rightarrow \ell\ell$~~ & ~~$W$+$\rm{jets}$~~ &
~~~~$t\bar{t}$~~~~ & ~~Multijet~~ & ~~~~$gg \rightarrow H$~~~~ &
~~~~$qq \rightarrow qqH$~~~~ & ~~~~$VH$~~~~\\
\hline
Luminosity ($\sigma_{\rm{inel}}(p\bar{p})$) &4.6   & 4.6   & -  & 4.6    &    -   &   4.6 &   4.6 &   4.6    \\
Luminosity Monitor    &  4.1  & 4.1           & -          & 4.1         &-        &   4.1 &   4.1 &   4.1    \\
Trigger     &  5.0            &5.0             & -          & 5.0        &-        &   5.0  &   5.0 &   5.0  \\
Lepton ID    &  3.7   &3.7           & -           & 3.7              &-        &   3.7 &   3.7 &   3.7    \\
EM veto        &  5.0         &-         & -         & 5.0        &-        &   5.0  &   5.0 &   5.0   \\
Tau Energy Scale (S)    &  1.0        &1.1          & -        & $<$1     &-        &   $<$1 &   $<$1 &   $<$1   \\
Jet Energy Scale (S)    &  8.0     &  $<$1    & -       & 1.8     &-        &   2.5 &   2.5 &   2.5     \\
Jet identification (S)  &  $<$1       & $<$1       & -         & 7.5        &-         &   5.0 &   5.0 &   5.0    \\
Multijet  (S)  ~~~~~    &  -      & -    & -     & -       &20-50    &   -    &   - &   -  \\
Cross Section (scale/PDF)     &  7.0       & 4.0      & -      & 10       & -        &   7/8 & 4.9 & 6.1    \\
Signal Branching Fraction & -  & -        & -      &-         &-         &0-7.3  &0-7.3 &0-7.3 \\
Modeling    ~~~~~       &  1.0     &-      & 10    & -      &-        &   3.0  &   3.0 &   3.0   \\

\end{tabular}
\end{ruledtabular}
\end{center}
\end{table}

\begin{table}[h]
\begin{center}
\caption{\label{tab:d0systwww-em}
Systematic uncertainties on the signal and background contributions for D0's
$VH\rightarrow e^\pm \nu_e \mu^\pm \nu_\mu$($V=W,Z$) channels. Systematic uncertainties are
listed by name; see the original references for a detailed explanation of their meaning and on how
they are derived. Shape uncertainties are labeled with the ``shape'' designation. Systematic uncertainties
shown in this table are obtained 
for the $m_H=165$ GeV/c$^2$ Higgs selection. Uncertainties are relative,
in percent, and are symmetric unless otherwise indicated. }
\vskip 0.2cm
{\centerline{$VH \rightarrow e^\pm \nu_e \mu^\pm \nu_\mu$ like charge electron muon pair channel relative uncertainties (\%)}}
\begin{ruledtabular}
\begin{tabular}{l c c c c c c}  \\
\hline
Contribution 				& VH		& $Z+jet/\gamma$ 	& $W+jet/\gamma$ 	& $t\bar{t}$	& Diboson 	& Multijet 	 \\
\hline
Cross section				& 6.2		& -- 				& -- 				& 6 			& 7 			& --		 \\
Luminosity/Normalization		& 4   		& -- 				& 4  				& 4  			& 4  			& --		 \\
Multijet              				& --  		& -- 				& -- 				& -- 			& -- 			& 30		\\
Trigger            				& 2 		& 2 				& 2 				& 2 			& 2 			& 2		 \\
Charge flip           			& --  		& 50 				& -- 				& 50  		& 50  		& --		 \\
W+jets/$\gamma$ 			& --  		& --  				& 10  			& --  			& --  			& --		 \\
$W-p_T$ model 			& --  		& --  				& shape			& --  			& --  			& --		\\
$Z-p_T$ model  			& --  		& shape 			& --   				& --  			& --  			& --		 \\
W+jets/$\gamma$ ISR/FSR model	& --  	& --  				& shape			& --  			& --  			& --		 \\
\hline
\end{tabular}
\end{ruledtabular}
\end{center}
\end{table}

\begin{table}
\begin{center}
\caption{\label{tab:d0systlll} 
Systematic uncertainties on the signal and background contributions
for D0's $VH\rightarrow VWW \rightarrow ee\mu, \mu\mu e$ channels.  
Systematic uncertainties are listed by name; see the
original references for a detailed explanation of their meaning and on
how they are derived.  Shape uncertainties are labeled with the ``s''
designation. Systematic uncertainties given in this table are obtained
for the $m_H=145$~GeV~Higgs selection. Uncertainties are relative, 
in percent, and are symmetric unless otherwise indicated.
}\vskip 0.1cm
{\centerline{$VH\rightarrow VWW \rightarrow$ Trilepton channels relative uncertainties (\%)}}
\vskip 0.099cm
\begin{ruledtabular}
\begin{tabular}{ l  c  c  c  c  c  c  c  c  }  \\
Contribution & Dibosons & ~~$Z/\gamma^* \rightarrow \ell\ell$~~&$~~W+$jet$/\gamma$~~ &~~~~$t\bar{t}~~~~$    & ~~$Z\gamma$~~  & $VH$ & $gg\to H$ & $qq\to qqH$  \\ 
\hline
Luminosity    						& 6.1  	& 6.1  	& 6.1  	& 6.1 	& --  	& 6.1 	& 6.1  	& 6.1  \\
Cross Section (Scale/PDF)   	& 6     	& 6     	& 6     	& 7    	& --  	& 6.2    & 7	 	& 4.9  \\
PDF              						& 2.5  	& 2.5  	& 2.5  	& 2.5 	& --  	& 2.5 	& 2.5   	& 2.5  \\
Electron Identification       	& 2.5  	& 2.5  	& 2.5  	& 2.5 	& --   	& 2.5 	& 2.5   	& 2.5  \\
Muon Identification     			& 4     	& 4     	& 4     	& 4    	& --    	& 4    	& 4      	& 4     \\
Trigger									& 3.5	& 3.5	& 3.5	& 3.5	& -- 	& 3.5	& 3.5	& 3.5  \\
$Z\gamma$ 						& --     & --		& --		& --		& 8		&	--	&	--	&	--  \\
$V+jets$ lepton fake rate	& --	 	& 30		& 30		& -- 	& --		& --		& --		&	--  \\
Z-$p_{T}$ reweighting (s)    & --     & $\pm 1\sigma$ & -- & -- & -- & -- & -- & --  \\
Electron smearing (s)  			& $\pm 1\sigma$ & $\pm 1\sigma$ & $\pm 1\sigma$ & $\pm 1\sigma$ & -- & $\pm 1\sigma$ & $\pm 1\sigma$ & $\pm 1\sigma$ \\
Muon smearing (s) 				& $\pm 1\sigma$ & $\pm 1\sigma$ & $\pm 1\sigma$ & $\pm 1\sigma$ & -- & $\pm 1\sigma$ & $\pm 1\sigma$ & $\pm 1\sigma$ \\
\end{tabular}
\end{ruledtabular}
\end{center}
\end{table}

\begin{table}
\begin{center}
\caption{\label{tab:d0systttm} 
Systematic uncertainties on the signal and background contributions
for D0's $\tau\tau\mu$ +X 
channel.  Systematic uncertainties are listed by name; see the
original references for a detailed explanation of their meaning and on
how they are derived.  Shape uncertainties are labeled with the ``s''
designation. Cross section uncertainties on
the $gg\to H$ signal depend on the jet multiplicity, as described in
the main text. Uncertainties are relative, in percent, and are
symmetric unless otherwise indicated.
}\vskip 0.1cm
{\centerline{$\tau\tau\mu$ +X  channels relative uncertainties (\%)}}
\vskip 0.099cm
\begin{ruledtabular}
\begin{tabular}{ l  c  c  c  c  c  c  c  c  }  \\
Contribution & Dibosons & ~~$Z/\gamma^*$~~&~~~~$t\bar{t}~~~~$    & ~~Instrumental~~  & $gg\to H$ & $qq\to qqH$ & $VH$ \\ 
\hline
Luminosity/Normalization    &  6      &   6                     & 6            &  24   &   6  &   6  &   6  \\
Trigger &  3      &   3                     & 3           & --   &   3  &   3  &   3  \\
Cross Section (Scale/PDF)      &  7        &   6           & 10                       &  --  &   13-33/7.6-30 & 4.9 & 6.2  \\
PDF              &      2.5     &   2.5          & 2.5                 &  --   &   2.5   &   2.5     &   2.5      \\
Tau Id per $\tau$  (Type 1/2/3)    &  7/3.5/5  &   7/3.5/5                   & 7/3.5/5      &  --   &   7/3.5/5       &   7/3.5/5        &   7/3.5/5         \\
Tau Energy Scale &  1  &   1                  & 1    &  --   &   1     &   1        &  1      \\
Tau Track Match per $\tau$&  1.4  &   1.4                  & 1.4    &  --   &   1.4     &   1.4        &  1.4      \\
Muon Identification     &  2.9           &  2.9                       & 2.9            &  --   &   2.9      &   2.9     &   2.9       \\
\end{tabular}
\end{ruledtabular}
\end{center}
\end{table}

\begin{table*}
\begin{center}
\caption{\label{tab:d0lvjj}
Systematic uncertainties on the signal and background contributions for D0's
 $H\rightarrow W W^{*} \rightarrow \ell\nu jj$ electron and muon channels.  Systematic uncertainties are listed
 by name; see the original references for a detailed explanation of their meaning and on how they are
 derived.
Signal uncertainties are shown for $m_H=160$ GeV/$c^2$ for all channels except for $WH$,
shown for $m_H=115$ GeV/$c^2$.  Those affecting the shape of
the RF discriminant are indicated with ``Y.''
Uncertainties are listed as relative changes in normalization,
in percent, except for those also marked by ``S,'' where
the overall normalization is constant, and the value given
denotes the maximum percentage change from nominal in any region of the
distribution.}

\vskip 0.1cm
{\centerline{D0: $H\rightarrow W W^{*} \rightarrow \ell\nu jj$ Run~II channel relative uncertainties (\%)}}
\vskip 0.099cm
\begin{ruledtabular}
\begin{tabular}{llccccccl}

Contribution & Shape & $W$+jets & $Z$+jets & Top & Diboson & $gg\to H$ & $qq\to qqH$ & $WH$ \\ \hline
Jet energy scale & Y & $\binom{+6.7}{-5.4}^S$ & $<0.1$ & $\pm$0.7 & $\pm$3.3 & $\binom{+5.7}{-4.0}$ & $\pm$1.5 &$\binom{+2.7}{-2.3}$  \\
Jet identification & Y & $\pm 6.6^S$ & $<0.1$ & $\pm$0.5 & $\pm$3.8  & $\pm$1.0 & $\pm$1.1 & $\pm$1.0 \\
Jet resolution & Y & $\binom{+6.6}{-4.1}^S$ & $<0.1$ & $\pm$0.5 & $\binom{+1.0}{-0.5}$ & $\binom{+3.0}{-0.5}$ & $\pm 0.8$ & $\pm 1.0$ \\
Association of jets with PV & Y & $\pm 3.2^S$ & $\pm 1.3^S$ & $\pm$1.2 & $\pm$3.2 & $\pm$2.9 & $\pm$2.4 & $\binom{+0.9}{-0.2}$ \\
Luminosity & N & n/a & n/a & $\pm$6.1 & $\pm$6.1 & $\pm$6.1 & $\pm$6.1 &  $\pm$6.1 \\
Muon trigger  & Y & $\pm 0.4^S$ & $<0.1$ & $<0.1$ & $<0.1$ & $<0.1$ & $<0.1$ &  $<0.1$ \\
Electron identification & N & $\pm$4.0  & $\pm$4.0  & $\pm$4.0  & $\pm$4.0  & $\pm$4.0  & $\pm$4.0  & $\pm$4.0 \\
Muon identification  & N & $\pm$4.0  & $\pm$4.0  & $\pm$4.0  & $\pm$4.0  & $\pm$4.0  & $\pm$4.0  & $\pm$4.0  \\
ALPGEN tuning & Y & $\pm 1.1^S$ & $\pm 0.3^S$ & n/a & n/a & n/a & n/a & n/a \\
Cross Section & N & $\pm$6 & $\pm$6 &  $\pm$10 & $\pm$7 & $\pm$10 & $\pm$10 & $\pm$6 \\
Heavy-flavor fraction  & Y & $\pm$20 & $\pm$20 & n/a & n/a & n/a & n/a & n/a  \\
Signal Branching Fraction & N & n/a &n/a &n/a& n/a & 0-7.3 & 0-7.3 &  0-7.3 \\
PDF & Y & $\pm 2.0^S$ & $\pm 0.7^S$ & $<0.1^S$ & $<0.1^S$ & $<0.1^S$ & $<0.1^S$ & $<0.1^S$ \\
 &  &  &  &  &  &  &  &  \\
 &  & \multicolumn{ 3}{c}{Electron channel} & \multicolumn{ 3}{c}{Muon channel} &  \\
Multijet Background & Y  & \multicolumn{ 3}{c}{$\pm$6.5} & \multicolumn{ 3}{c}{$\pm$26} &  \\
\end{tabular}
\end{ruledtabular}
\end{center}
\end{table*}

\begin{table}
\begin{center}
\caption{\label{tab:cdfsystH4l} Systematic uncertainties on the signal and background contributions for CDF's
$H\rightarrow \ell^{\pm}\ell^{\mp}\ell^{\prime \pm}\ell^{\prime \mp}$ channel. This channel is sensitive to
gluon fusion production and $WH$, $ZH$ and VBF production.  Systematic uncertainties are listed by name (see
the original references for a detailed explanation of their meaning and on how they are derived). Uncertainties
are relative, in percent, and are symmetric unless otherwise indicated.  The uncertainties associated with the
different background and signal processed are correlated unless otherwise noted.  Boldface and italics indicate
groups of uncertainties which are correlated with each other but not the others within a line.  Shape uncertainties 
are labeled with an "s".}

\vskip 0.1cm
{\centerline{CDF: $H\rightarrow \ell^{\pm}\ell^{\mp}\ell^{\prime \pm}\ell^{\prime \mp}$ channel relative uncertainties (\%)}}
\vskip 0.099cm
\begin{ruledtabular}
\begin{tabular}{lcccccc} \\
Contribution               &   $ZZ$     &  $Z(/\gamma^*)$+jets  &  $gg \to H$ &     $WH$    &    $ZH$    & VBF       \\ \hline
{\bf Cross Section :}      &            &                       &             &             &            &           \\
Scale                      &            &                       &   7.0       &             &            &           \\
PDF Model                  &            &                       &   7.7       &             &            &           \\
Total                      & {\it 10.0} &                       &             &  {\bf 5.0}  & {\bf 5.0}  & 10.0      \\
$\mathcal{BR}(H\to VV)$    &            &                       &   3.0       &     3.0     &    3.0     & 3.0       \\
{\bf Acceptance :}         &            &                       &             &             &            &           \\
PDF Model                  &     2.7    &                       &             &             &            &           \\
Higher-order Diagrams      &     2.5    &                       &             &             &            &           \\
Jet Fake Rates             &            &     50.0              &             &             &            &           \\
\met\ Resolution           &   s        &                       &    s        &             &     s      &    s      \\
Lepton ID Efficiencies     & {\it 3.6}  &                       & {\it 3.6 }  &   {\it 3.6} &  {\it 3.6} & {\it 3.6} \\
Trigger Efficiencies       & {\it 0.4}  &                       & {\it 0.5 }  &   {\it 0.5} &  {\it 0.5} & {\it 0.5} \\
Luminosity                 & {\it 5.9}  &                       & {\it 5.9 }  &   {\it 5.9} &  {\it 5.9} & {\it 5.9} \\ 
\end{tabular}
\end{ruledtabular}

\end{center}
\end{table}



\clearpage\newpage

\begin{table}
\begin{center}

\caption{\label{tab:cdfsystttHLJ} Systematic uncertainties on the
signal and background contributions for CDF's $t\bar{t}H \to
\ell+$jets channels.  Systematic uncertainties are listed by name; see
the original references for a detailed explanation of their meaning
and on how they are derived.  Systematic uncertainties for $t\bar{t}H$
shown in this table are obtained for $m_H=115$ GeV/$c^2$.
Uncertainties are relative, in percent, and are symmetric unless
otherwise indicated.}
\begin{small}

\vskip 0.1cm
{\centerline{CDF: $t\bar{t}H$ $\protect \ell+\raisebox{.3ex}{$\not$}E_{T}$ 4 jets channel relative uncertainties (\%)}}
\vskip 0.099cm
\begin{tabular}{lcccccccccc}\hline\hline
& \multicolumn{2}{c}{1 tight, 1 loose}& \multicolumn{2}{c}{1 tight, $\ge2$ loose}& \multicolumn{2}{c}{2 tight, 0 loose}& \multicolumn{2}{c}{2 tight, $\ge1$ loose}& \multicolumn{2}{c}{$\ge3$ tight, $\ge0$ loose} \\
Contribution                                    & $t\bar{t}$         & $t\bar{t}H$        & $t\bar{t}$          & $t\bar{t}H$         & $t\bar{t}$         & $t\bar{t}H$        & $t\bar{t}$          & $t\bar{t}H$        & $t\bar{t}$           & $t\bar{t}H$          \\ \hline \noalign{\smallskip}
$t\bar{t}$ Cross Section                        &                    & 10                 &                     & 10                  &                    & 10                 &                     & 10                 &                      & 10                   \\
$t\bar{t}H$ Cross Section                       & 10                 &                    & 10                  &                     & 10                 &                    & 10                  &                    & 10                   & \\
Luminosity ($\sigma_{\mathrm{inel}}(p\bar{p})$) & 3.8                & 3.8                & 3.8                 & 3.8                 & 3.8                & 3.8                & 3.8                 & 3.8                & 3.8                  & 3.8                  \\
Luminosity Monitor                              & 4.4                & 4.4                & 4.4                 & 4.4                 & 4.4                & 4.4                & 4.4                 & 4.4                & 4.4                  & 4.4                  \\[1.1ex]
$B$-Tag Efficiency                              & $^{+1.79}_{-1.89}$ & $^{-0.23}_{-0.86}$ & $^{+4.77}_{-4.75}$  & $^{-1.74}_{-1.84}$  & $^{+9.09}_{-9.75}$ & $^{+7.50}_{-5.98}$ & $^{+14.42}_{-9.41}$ & $^{+5.14}_{-6.72}$ & $^{+14.79}_{-19.02}$ & $^{+15.46}_{-14.28}$ \\[2.2ex]
Mistag Rate                                     & $^{+1.89}_{-0.72}$ & $^{+1.09}_{-0.11}$ & $^{+12.41}_{-6.71}$ & $^{+5.14}_{-4.84}$  & $^{-0.27}_{+0.64}$ & $^{-0.14}_{+0.39}$ & $^{+9.61}_{-3.56}$  & $^{+1.92}_{+1.75}$ & $^{+2.99}_{-5.14}$   & $^{+1.13}_{-1.37}$   \\[2.2ex]
Jet Energy Scale                                & $^{+2.77}_{-4.38}$ & $^{-8.80}_{+8.06}$ & $^{+3.57}_{-0.33}$  & $^{-8.33}_{+11.92}$ & $^{+2.52}_{-3.80}$ & $^{-9.06}_{+7.42}$ & $^{+3.77}_{-0.48}$  & $^{-9.77}_{+8.77}$ & $^{+1.48}_{-2.61}$   & $^{-5.66}_{+6.74}$   \\[2.2ex]
ISR+FSR+PDF                                     & 0.36               & 3.04               & 0.38                & 0.75                & 1.29               & 2.73               & 3.86                & 5.28               & 0.33                 & 5.13                 \\[2.2ex]\hline\hline
\end{tabular}

\vspace*{1cm}

{\centerline{CDF: $t\bar{t}H$ $\protect \ell+\raisebox{.3ex}{$\not$}E_{T}$ 5 jets channel relative uncertainties (\%)}}
\vskip 0.099cm
\begin{tabular}{lcccccccccc}\hline\hline
                    & \multicolumn{2}{c}{1 tight, 1 loose}& \multicolumn{2}{c}{1 tight, $\ge2$ loose}& \multicolumn{2}{c}{2 tight, 0 loose}& \multicolumn{2}{c}{2 tight, $\ge1$ loose}& \multicolumn{2}{c}{$\ge3$ tight, $\ge0$ loose} \\
Contribution                                    & $t\bar{t}$           & $t\bar{t}H$        & $t\bar{t}$           & $t\bar{t}H$        & $t\bar{t}$           & $t\bar{t}H$        & $t\bar{t}$           & $t\bar{t}H$        & $t\bar{t}$           & $t\bar{t}H$          \\ \hline \noalign{\smallskip}
$t\bar{t}$ Cross Section                        &                      & 10                 &                      & 10                 &                      & 10                 &                      & 10                 &                      & 10                   \\
$t\bar{t}H$ Cross Section                       & 10                   &                    & 10                   &                    & 10                   &                    & 10                   &                    & 10                   &                      \\
Luminosity ($\sigma_{\mathrm{inel}}(p\bar{p})$) & 3.8                  & 3.8                & 3.8                  & 3.8                & 3.8                  & 3.8                & 3.8                  & 3.8                & 3.8                  & 3.8                  \\
Luminosity Monitor                              & 4.4                  & 4.4                & 4.4                  & 4.4                & 4.4                  & 4.4                & 4.4                  & 4.4                & 4.4                  & 4.4                  \\[1.1ex]
$B$-Tag Efficiency                              & $^{+1.25}_{-0.55}$   & $^{-1.96}_{+2.06}$ & $^{+1.99}_{-5.21}$   & $^{-0.99}_{+0.89}$ & $^{+8.69}_{-9.74}$   & $^{+5.80}_{-7.30}$ & $^{+11.36}_{-12.13}$ & $^{+4.48}_{-4.50}$ & $^{+14.94}_{-16.28}$ & $^{+12.96}_{-15.87}$ \\[2.2ex]
Mistag Rate                                     & $^{+2.81}_{-0.78}$   & $^{+1.96}_{-0.66}$ & $^{+12.47}_{-11.50}$ & $^{+1.19}_{-2.53}$ & $^{-1.94}_{+0.92}$   & $^{-0.57}_{-0.77}$ & $^{+10.70}_{-7.19}$  & $^{+0.87}_{-2.66}$ & $^{+4.02}_{-9.48}$   & $^{+1.15}_{-0.23}$   \\[2.2ex]
Jet Energy Scale                                & $^{+14.48}_{-11.71}$ & $^{-1.02}_{+2.51}$ & $^{+9.96}_{-12.79}$  & $^{-0.64}_{-1.34}$ & $^{+11.84}_{-13.49}$ & $^{-2.21}_{+0.66}$ & $^{+13.07}_{-9.15}$  & $^{-3.40}_{+1.48}$ & $^{+6.51}_{-7.57}$   & $^{-3.12}_{+2.45}$   \\[2.2ex]
ISR+FSR+PDF                                     & 3.42                 & 2.41               & 11.28                & 0.79               & 5.24                 & 2.30               & 3.89                 & 3.26               & 3.95                 & 2.88                 \\[2.2ex]\hline\hline
\end{tabular}

\vspace*{1cm}

{\centerline{CDF: $t\bar{t}H$ $\protect \ell+\raisebox{.3ex}{$\not$}E_{T}$ 6 or more jets channel relative uncertainties (\%)}}
\vskip 0.099cm
\begin{tabular}{lcccccccccc}\hline\hline
                    & \multicolumn{2}{c}{1 tight, 1 loose}& \multicolumn{2}{c}{1 tight, $\ge2$ loose}& \multicolumn{2}{c}{2 tight, 0 loose}& \multicolumn{2}{c}{2 tight, $\ge1$ loose}& \multicolumn{2}{c}{$\ge3$ tight, $\ge0$ loose} \\
Contribution                                    & $t\bar{t}$           & $t\bar{t}H$          & $t\bar{t}$           & $t\bar{t}H$         & $t\bar{t}$           & $t\bar{t}H$          & $t\bar{t}$           & $t\bar{t}H$         & $t\bar{t}$           & $t\bar{t}H$          \\ \hline \noalign{\smallskip}
$t\bar{t}$ Cross Section                        &                      & 10                   &                      & 10                  &                      & 10                   &                      & 10                  &                      & 10                   \\
$t\bar{t}H$ Cross Section                       & 10                   &                      & 10                   &                     & 10                   &                      & 10                   &                     & 10                   &                      \\
Luminosity ($\sigma_{\mathrm{inel}}(p\bar{p})$) & 3.8                  & 3.8                  & 3.8                  & 3.8                 & 3.8                  & 3.8                  & 3.8                  & 3.8                 & 3.8                  & 3.8                  \\
Luminosity Monitor                              & 4.4                  & 4.4                  & 4.4                  & 4.4                 & 4.4                  & 4.4                  & 4.4                  & 4.4                 & 4.4                  & 4.4                  \\[1.1ex]
$B$-Tag Efficiency                              & $^{+1.52}_{-1.47}$   & $^{-2.07}_{+1.85}$   & $^{+4.07}_{-1.53}$   & $^{-0.89}_{+2.99}$  & $^{+9.02}_{-8.39}$   & $^{+4.27}_{-8.07}$   & $^{+17.30}_{-8.32}$  & $^{+4.78}_{-3.91}$  & $^{+12.00}_{-14.59}$ & $^{+13.13}_{-12.00}$ \\[2.2ex]
Mistag Rate                                     & $^{+1.76}_{-2.29}$   & $^{+1.72}_{+0.21}$   & $^{+17.63}_{-16.95}$ & $^{+4.43}_{-3.03}$  & $^{-1.46}_{+2.68}$   & $^{-2.55}_{-1.33}$   & $^{+15.68}_{-12.32}$ & $^{+2.25}_{+0.98}$  & $^{+8.47}_{-11.76}$  & $^{-0.12}_{-2.05}$   \\[2.2ex]
Jet Energy Scale                                & $^{+25.07}_{-21.07}$ & $^{+12.17}_{-12.62}$ & $^{+17.29}_{-20.68}$ & $^{+11.78}_{-9.86}$ & $^{+25.58}_{-22.19}$ & $^{+10.81}_{-13.16}$ & $^{+26.49}_{-17.30}$ & $^{+10.02}_{-8.69}$ & $^{+23.29}_{-19.76}$ & $^{+8.58}_{-11.05}$  \\[2.2ex]
ISR+FSR+PDF                                     & 13.17                & 0.75                 & 17.33                & 2.32                & 12.38                & 1.42                 & 20.89                & 1.15                & 14.84                & 0.38                 \\[2.2ex]\hline\hline
\end{tabular}
\end{small}

\end{center}
\end{table}


\begin{table}[h]
\begin{center}
\caption{\label{tab:cdfsysttthmetjets} Systematic uncertainties on the
signal and background contributions for CDF's $t\bar{t}H$ 2-tag and
3-tag $\protect \raisebox{.3ex}{$\not$}E_{T}$+jets channels.  Systematic
uncertainties are listed by name; see the original references for a
detailed explanation of their meaning and on how they are derived.
Systematic uncertainties for $t\bar{t}H$ shown in this table are obtained
for $m_H=120$ GeV/$c^2$.  Uncertainties are relative, in percent, and are
symmetric unless otherwise indicated.}
\vskip 0.1cm
{\centerline{CDF: $t\bar{t}H$ $\protect \raisebox{.3ex}{$\not$}E_{T}$+jets 2-tag channel relative uncertainties (\%)}}
\vskip 0.099cm
\begin{ruledtabular}
\begin{tabular}{lccc}\\
Contribution              & non-$t\bar{t}$ & $t\bar{t}$ & $t\bar{t}H$   \\ \hline
Luminosity ($\sigma_{\mathrm{inel}}(p{\bar{p}})$)
                          & 0      & 3.8     & 3.8   \\
Luminosity Monitor        & 0      & 4.4     & 4.4   \\
Jet Energy Scale          & 0      & 2       & 11    \\
Trigger Efficiency        & 0      & 7       & 7     \\
$B$-Tag Efficiency        & 0      & 7       & 7     \\
ISR/FSR                   & 0      & 2       & 2     \\
PDF                       & 0      & 2       & 2     \\
$t{\bar{t}}$ Cross Section& 0      & 10      & 0     \\
$t{\bar{t}}b{\bar{b}}$ Cross Section  & 0    & 3       & 0    \\
Signal Cross Section      & 0      & 0       & 10    \\
Background Modeling       & 6      & 0       & 0     \\
Background $B$-tagging    & 5      & 0       & 0     \\
\end{tabular}
\end{ruledtabular}

\vskip 0.3cm
{\centerline{CDF: $t\bar{t}H$ $\protect \raisebox{.3ex}{$\not$}E_{T}$+jets 3-tag channel relative uncertainties (\%)}}
\vskip 0.099cm
\begin{ruledtabular}
\begin{tabular}{lccc}\\
Contribution              & non-$t\bar{t}$ & $t\bar{t}$ & $t\bar{t}H$   \\ \hline
Luminosity ($\sigma_{\mathrm{inel}}(p{\bar{p}})$)
                          & 0      & 3.8     & 3.8   \\
Luminosity Monitor        & 0      & 4.4     & 4.4   \\
Jet Energy Scale          & 0      & 3       & 13    \\
Trigger Efficiency        & 0      & 7       & 7     \\
$B$-Tag Efficiency        & 0      & 9       & 9     \\
ISR/FSR                   & 0      & 2       & 2     \\
PDF                       & 0      & 2       & 2     \\
$t{\bar{t}}$ Cross Section& 0      & 10      & 0     \\
$t{\bar{t}}b{\bar{b}}$ Cross Section  & 0    & 5       & 0    \\
Signal Cross Section      & 0      & 0       & 10    \\
Background Modeling       & 6      & 0       & 0     \\
Background $B$-tagging    & 10     & 0       & 0     \\
\end{tabular}
\end{ruledtabular}
\end{center}
\end{table}

\begin{table}[h]
\begin{center}
\caption{\label{tab:cdfsysttthalljets} Systematic uncertainties on the signal and
background contributions for CDF's $t\bar{t}H$ 2-tag and 3-tag all jets channels.
Systematic uncertainties are listed by name; see the original references for a
detailed explanation of their meaning and on how they are derived.  Systematic
uncertainties for $t\bar{t}H$ shown in this table are obtained for $m_H=120$
GeV/$c^2$.  Uncertainties are relative, in percent, and are symmetric unless
otherwise indicated.}
\vskip 0.1cm
{\centerline{CDF: $t\bar{t}H$ all jets 2-tag channel relative uncertainties (\%)}}
\vskip 0.099cm
\begin{ruledtabular}
\begin{tabular}{lccc}\\
Contribution              & non-$t\bar{t}$ & $t\bar{t}$ & $t\bar{t}H$   \\ \hline
Luminosity ($\sigma_{\mathrm{inel}}(p{\bar{p}})$)
                          & 0      & 3.8     & 3.8   \\
Luminosity Monitor        & 0      & 4.4     & 4.4   \\
Jet Energy Scale          & 0      & 11      & 20    \\
Trigger Efficiency        & 0      & 7       & 7     \\
$B$-Tag Efficiency        & 0      & 7       & 7     \\
ISR/FSR                   & 0      & 2       & 2     \\
PDF                       & 0      & 2       & 2     \\
$t{\bar{t}}$ Cross Section& 0      & 10      & 0     \\
$t{\bar{t}}b{\bar{b}}$ Cross Section  & 0    & 3       & 0    \\
Signal Cross Section      & 0      & 0       & 10    \\
Background Modeling       & 9      & 0       & 0     \\
Background $B$-tagging    & 5      & 0       & 0     \\
\end{tabular}
\end{ruledtabular}

\vskip 0.3cm
{\centerline{CDF: $t\bar{t}H$ all jets 3-tag channel relative uncertainties (\%)}}
\vskip 0.099cm
\begin{ruledtabular}
\begin{tabular}{lccc}\\
Contribution              & non-$t\bar{t}$ & $t\bar{t}$ & $t\bar{t}H$   \\ \hline
Luminosity ($\sigma_{\mathrm{inel}}(p{\bar{p}})$)
                          & 0      & 3.8     & 3.8   \\
Luminosity Monitor        & 0      & 4.4     & 4.4   \\
Jet Energy Scale          & 0      & 13      & 22    \\
Trigger Efficiency        & 0      & 7       & 7     \\
$B$-Tag Efficiency        & 0      & 9       & 9     \\
ISR/FSR                   & 0      & 2       & 2     \\
PDF                       & 0      & 2       & 2     \\
$t{\bar{t}}$ Cross Section& 0      & 10      & 0     \\
$t{\bar{t}}b{\bar{b}}$ Cross Section  & 0    & 6       & 0    \\
Signal Cross Section      & 0      & 0       & 10    \\
Background Modeling       & 9      & 0       & 0     \\
Background $B$-tagging    & 10     & 0       & 0     \\
\end{tabular}
\end{ruledtabular}
\end{center}
\end{table}

\begin{table}
\begin{center}
\caption{\label{tab:cdfsysttautau} Systematic uncertainties on the signal and background contributions for CDF's
$H\rightarrow \tau^+\tau^-$ channels.  Systematic uncertainties are listed by name; see the original references
for a detailed explanation of their meaning and on how they are derived. Systematic uncertainties for the Higgs
signal shown in these tables are obtained for $m_H=120$ GeV/$c^2$.  Uncertainties are relative, in percent, and
are symmetric unless otherwise indicated. Shape uncertainties are labeled with an "S".}
\vskip 0.1cm
{\centerline{CDF: $H \rightarrow \tau^+ \tau^- (e/\mu+\tau_{had})$ channel relative uncertainties (\%)}}
\vskip 0.099cm
\begin{ruledtabular}
\begin{tabular}{lccccccccccc}\\
Contribution & $Z/\gamma^* \rightarrow \tau\tau$ & $Z/\gamma^* \rightarrow ee$ & $Z/\gamma^* \rightarrow \mu\mu$  & $t\bar{t}$ & diboson  & fakes from SS
&W+jets & $WH$      & $ZH$  & VBF      & $gg\rightarrow H$ \\
\hline
PDF Uncertainty                                      &   -  &   -  &  -   &  -   &   -  &  -  &  -   &  1.2  &  0.9  &  2.2  &  4.9   \\
ISR/FSR 1 JET                                        &   -  &   -  &  -   &  -   &   -  &  -  &  -   &  6.7  &  8.7  &  8.8  &  3.6   \\
ISR/FSR $\ge$ 2 JETS                                 &   -  &   -  &  -   &  -   &   -  &  -  &  -   &  4.8  &  3.8  &  3.9  & 19.1   \\
JES (S) 1 JET                                        &  9.5 &  8.5 &  8.5 & 14.5 &  0.5 &  -  &  4.2 &  2.8  &  6.4  &  6.5  &  4.3   \\
JES (S) $\ge$ 2 JETS                                 & 18.9 & 22.3 & 22.3 &  1.3 & 10.7 &  -  & 15.4 &  5.1  &  3.9  &  3.7  & 14.5   \\
Normalization 1 JET                                  &  2.0 &  5.0 &  5.0 & 10.0 &  6.0 & 1.3 & 14.8 &  5.0  &  5.0  & 10.0  & 23.5   \\
Normalization $\ge$2 JETS                            &  2.0 &  5.0 &  5.0 & 10.0 &  6.0 & 2.5 & 14.8 &  5.0  &  5.0  & 10.0  & 33.0   \\
$\varepsilon_{trig}$ (e leg)                         &  0.3 &  0.3 &   -  &  0.3 &  0.3 &  -  &  -   &  0.3  &  0.3  &  0.3  &  0.3   \\
$\varepsilon_{trig}$ ($\mu$ leg)                     &  1.0 &   -  &  1.0 &  1.0 &  1.0 &  -  &  -   &  1.0  &  1.0  &  1.0  &  1.0   \\
$\varepsilon_{trig}$ ($\tau$ leg)                    &  3.0 &  3.0 &  3.0 &  3.0 &  3.0 &  -  &  -   &  3.0  &  3.0  &  3.0  &  3.0   \\
$\varepsilon_{ID}e$                                  &  2.4 &  2.4 &   -  &  2.4 &  2.4 &  -  &  -   &  2.4  &  2.4  &  2.4  &  2.4   \\
$\varepsilon_{ID}\mu$                                &  2.6 &   -  &  2.6 &  2.6 &  2.6 &  -  &  -   &  2.6  &  2.6  &  2.6  &  2.6   \\
$\varepsilon_{ID}\tau$                               &  3.0 &  3.0 &  3.0 &  3.0 &  3.0 &  -  &  -   &  3.0  &  3.0  &  3.0  &  3.0   \\
$\varepsilon_{vtx}$                                  &  0.5 &  0.5 &  0.5 &  0.5 &  0.5 &  -  &  -   &  0.5  &  0.5  &  0.5  &  0.5   \\
Luminosity                                           &  5.9 &  5.9 &  5.9 &  5.9 &  5.9 &  -  &  -   &  5.9  &  5.9  &  5.9  &  5.9   \\
\end{tabular}
\end{ruledtabular}

\end{center}
\end{table}


\begin{table}
\begin{center}
\caption{\label{tab:cdfsystVtautau} Systematic uncertainties on the signal and background contributions for CDF's
$WH \rightarrow \ell \nu \tau^+\tau^-$ and $ZH \rightarrow \ell^+ \ell^- \tau^+ \tau^-$ channels.  Systematic
uncertainties are listed by name; see the original references for a detailed explanation of their meaning and
on how they are derived. Systematic uncertainties for the Higgs signal shown in these tables are obtained for
$m_H=120$ GeV/$c^2$.  Uncertainties are relative, in percent, and are symmetric unless otherwise indicated.}
\vskip 0.1cm
{\centerline{CDF: $WH \rightarrow \ell \nu \tau^+\tau^-$ and $ZH \rightarrow \ell^+ \ell^- \tau^+ \tau^-$ $\ell\ell\tau_h+X$ channel relative uncertainties (\%)}}
\vskip 0.099cm
\begin{ruledtabular}
\begin{tabular}{lccccccccccccccc}\\
Contribution & $ZZ$ & $WZ$ & $WW$ & $DY(ee)$ & $DY(\mu\mu)$ & $DY(\tau\tau)$ & $Z\gamma$ & $t\bar{t}$ & $W\gamma$ & $W+jet$ & $WH$ & $ZH$ & $VBF$ & $gg \rightarrow H$\\
\hline
Luminosity                   & 5.9 & 5.9 & 5.9 & 5.9 & 5.9 & 5.9 & 5.9 & 5.9 & 5.9 & 5.9 & 5.9 & 5.9 & 5.9 & 5.9 \\
Cross Section                &11.7 &11.7 &11.7 & 5.0 & 5.0 & 5.0 &11.7 &14.1 &11.7 & 5.0 & 5.0 & 5.0 &10.0 &10.0 \\
Z-vertex Cut Efficiency      & 0.5 & 0.5 & 0.5 & 0.5 & 0.5 & 0.5 & 0.5 & 0.5 & 0.5 & 0.5 & 0.5 & 0.5 & 0.5 & 0.5 \\
Trigger Efficiency           & 1.1 & 1.1 & 1.0 & 1.0 & 1.0 & 1.1 & 1.1 & 1.0 & 0.8 & 1.0 & 1.2 & 1.2 & 1.2 & 1.1 \\
Lepton ID Efficiency         & 2.4 & 2.3 & 2.4 & 2.4 & 2.4 & 2.4 & 2.4 & 2.4 & 2.3 & 2.4 & 2.4 & 2.4 & 2.4 & 2.4 \\
Lepton Fake Rate             &10.7 & 8.0 &26.7 &26.0 &26.6 &15.1 &27.1 &22.4 &22.8 &28.7 & 2.9 & 2.3 &15.1 &13.6 \\
Jet Energy Scale             & 1.3 & 1.1 & 0.0 & 3.2 & 5.1 & 0.6 & 6.6 & 0.1 & 2.0 & 0.2 & 0.1 & 0.03& 0.6 & 0.4 \\
MC stat                      & 3.7 & 2.9 & 7.6 & 1.5 & 1.7 & 2.2 & 4.1 & 3.1 & 20.0& 3.1 & 1.5 & 1.4 & 3.8 & 9.4 \\
PDF Model                    &  -  &  -  &  -  &  -  &  -  &  -  &  -  &  -  &  -  &  -  & 1.2 & 0.9 & 2.2 & 4.9 \\
ISR/FSR Uncertainties        &  -  &  -  &  -  &  -  &  -  &  -  &  -  &  -  &  -  &  -  & 1.3 & 2.1 & 0.6 & 0.2 \\
\end{tabular}
\end{ruledtabular}

\vskip 0.3cm
{\centerline{CDF: $WH \rightarrow \ell \nu \tau^+\tau^-$ and $ZH \rightarrow \ell^+ \ell^- \tau^+ \tau^-$ $e\mu\tau_h+X$ channel relative uncertainties (\%)}}
\vskip 0.099cm
\begin{ruledtabular}
\begin{tabular}{lccccccccccccccc}\\
Contribution & $ZZ$ & $WZ$ & $WW$ & $DY(ee)$ & $DY(\mu\mu)$ & $DY(\tau\tau)$ & $Z\gamma$ & $t\bar{t}$ & $W\gamma$ & $W+jet$ & $WH$ & $ZH$ & $VBF$ & $gg \rightarrow H$\\
\hline
Luminosity                   & 5.9 & 5.9 & 5.9 & 5.9 & 5.9 & 5.9 & 5.9 & 5.9 & 5.9 & 5.9 & 5.9 & 5.9 & 5.9 & 5.9 \\
Cross Section                &11.7 &11.7 &11.7 & 5.0 & 5.0 & 5.0 &11.7 &14.1 &11.7 & 5.0 & 5.0 & 5.0 &10.0 &10.0 \\
Z-vertex Cut Efficiency      & 0.5 & 0.5 & 0.5 & 0.5 & 0.5 & 0.5 & 0.5 & 0.5 & 0.5 & 0.5 & 0.5 & 0.5 & 0.5 & 0.5 \\
Trigger Efficiency           & 1.4 & 1.4 & 1.1 & 1.1 & 1.3 & 1.1 & 1.4 & 1.1 & 1.0 & 0.7 & 1.3 & 1.3 & 1.2 & 1.2 \\
Lepton ID Efficiency         & 2.4 & 2.4 & 2.4 & 2.4 & 2.4 & 2.4 & 2.4 & 2.4 & 2.4 & 2.4 & 2.4 & 2.4 & 2.4 & 2.4 \\
Lepton Fake Rate             & 9.0 & 6.5 &26.6 &20.8 &31.4 &25.2 &39.4 &27.8 &19.3 &41.9 & 1.6 & 2.5 &28.5 &29.2 \\
Jet Energy Scale             & 0.0 & 0.3 & 2.2 & 0.0 & 0.8 & 1.5 & 0.5 & 0.8 & 0.0 & 0.0 & 0.2 & 0.1 & 1.7 & 0.0 \\
MC stat                      & 12.9& 7.2 & 20.9& 57.7& 12.6& 7.7 & 10.2& 12.4& 35.4& 25.8& 2.1 & 3.9 &13.0 &44.7 \\
PDF Model                    &  -  &  -  &  -  &  -  &  -  &  -  &  -  &  -  &  -  &  -  & 1.2 & 0.9 & 2.2 & 4.9 \\
ISR/FSR Uncertainties        &  -  &  -  &  -  &  -  &  -  &  -  &  -  &  -  &  -  &  -  & 0.6 & 0.2 & 0.1 & 0.0 \\
\end{tabular}
\end{ruledtabular}

\vskip 0.3cm
{\centerline{CDF: $WH \rightarrow \ell \nu \tau^+\tau^-$ and $ZH \rightarrow \ell^+ \ell^- \tau^+ \tau^-$ $\ell\tau_h\tau_h+X$ channel relative uncertainties (\%)}}
\vskip 0.099cm
\begin{ruledtabular}
\begin{tabular}{lccccccccccccccc}\\
Contribution & $ZZ$ & $WZ$ & $WW$ & $DY(ee)$ & $DY(\mu\mu)$ & $DY(\tau\tau)$ & $Z\gamma$ & $t\bar{t}$ & $W\gamma$ & $W+jet$ & $WH$ & $ZH$ & $VBF$ & $gg \rightarrow H$\\
\hline
Luminosity                   & 5.9 & 5.9 & 5.9 & 5.9 & 5.9 & 5.9 & 5.9 & 5.9 & 5.9 & 5.9 & 5.9 & 5.9 & 5.9 & 5.9 \\
Cross Section                &11.7 &11.7 &11.7 & 5.0 & 5.0 & 5.0 &11.7 &14.1 &11.7 & 5.0 & 5.0 & 5.0 &10.0 &10.0 \\
Z-vertex Cut Efficiency      & 0.5 & 0.5 & 0.5 & 0.5 & 0.5 & 0.5 & 0.5 & 0.5 & 0.5 & 0.5 & 0.5 & 0.5 & 0.5 & 0.5 \\
Trigger Efficiency           & 1.0 & 1.1 & 0.9 & 1.0 & 1.1 & 1.1 & 1.1 & 1.0 & 0.7 & 0.9 & 1.1 & 1.1 & 1.1 & 1.1 \\
Lepton ID Efficiency         & 3.3 & 3.3 & 3.3 & 3.3 & 3.3 & 3.3 & 3.3 & 3.3 & 3.3 & 3.3 & 3.3 & 3.3 & 3.3 & 3.3 \\
Lepton Fake Rate             &10.4 & 6.8 &38.1 &43.3 &39.9 &24.8 &32.8 &34.2 &28.8 &34.8 & 3.1 & 5.9 &28.1 &26.3 \\
Jet Energy Scale             & 5.5 & 0.0 & 0.0 & 3.3 & 1.6 & 1.2 & 1.6 & 0.0 & 0.0 & 1.1 & 0.1 & 0.6 & 1.8 & 1.7 \\
MC stat                      & 12.5& 8.1 & 16.9& 18.3& 12.5& 4.9 & 12.6& 14.7& 70.7& 8.7 & 2.0 & 3.3 & 9.4 &18.3 \\
PDF Model                    &  -  &  -  &  -  &  -  &  -  &  -  &  -  &  -  &  -  &  -  & 1.2 & 0.9 & 2.2 & 4.9 \\
ISR/FSR Uncertainties        &  -  &  -  &  -  &  -  &  -  &  -  &  -  &  -  &  -  &  -  & 1.2 & 0.5 & 0.4 & 0.04\\
\end{tabular}
\end{ruledtabular}
\end{center}
\end{table}

\begin{table}
\begin{center}
  \caption{ \label{tab:cdfallhadsyst} Systematic uncertainties on the
    signal and background contributions for CDF's $WH+ZH \rightarrow
    jjbb$ and $VBF \rightarrow jjbb$ channels.  Systematic
    uncertainties are listed by name; see the original references for
    a detailed explanation of their meaning and on how they are
    derived.  Uncertainties with provided shape systematics are
    labeled with ``s''.  Systematic uncertainties for $H$ shown in
    this table are obtained for $m_H=115$ GeV/$c^2$.  Uncertainties
    are relative, in percent, and are symmetric unless otherwise
    indicated.  The cross section uncertainties are uncorrelated with
    each other (except for single top and $t{\bar{t}}$, which are
    treated as correlated).  The QCD uncertainty is also uncorrelated
    with other channels' QCD rate uncertainties.
}
\vskip 0.1cm
{\centerline{CDF: $WH+ZH\rightarrow jjbb$ and $VBF \rightarrow jjbb$ channel relative uncertainties (\%)}}
\vskip 0.099cm
\begin{ruledtabular}
\begin{tabular}{lccccccc} \\
Contribution              & QCD & $t\bar{t}$ & single-top & diboson & $W/Z$+Jets & VH  & VBF \\ \hline
Jet Energy Correction     &     & 9 s        & 9 s        & 9 s     & 9 s        & 9 s & 9 s \\
PDF Modeling              &     &            &            &         &            & 2   & 2   \\
SecVtx+SecVtx             &     & 7.1        & 7.1        & 7.1     & 7.1        & 7.1 & 7.1 \\
SecVtx+JetProb            &     & 6.4        & 6.4        & 6.4     & 6.4        & 6.4 & 6.4 \\
Luminosity                &     & 6          & 6          & 6       & 6          & 6   & 6   \\
ISR/FSR modeling          &     &            &            &         &            & 3 s & 3 s \\
Jet Width                 &     & s          & s          & s       & s          & s   & s   \\
Trigger                   &     & 3.6        & 3.6        & 3.6     & 3.6        & 3.6 & 3.6 \\
QCD Interpolation         & s   &            &            &         &            &     &     \\
QCD MJJ Tuning            & s   &            &            &         &            &     &     \\
QCD NN Tuning             & s   &            &            &         &            &     &     \\
cross section             &     & 7          & 7          & 6       & 50         & 5   & 10  \\
\end{tabular}
\end{ruledtabular}
\end{center}
\end{table}




\begin{table}[h]
\begin{center}
\caption{\label{tab:cdfsystgg} Systematic uncertainties on the signal and background 
contributions for CDF's $H\rightarrow \gamma \gamma$ channels.  Systematic uncertainties 
are listed by name; see the original references for a detailed explanation of their 
meaning and on how they are derived.  Uncertainties are relative, in percent, and are 
symmetric unless otherwise indicated.}
\vskip 0.1cm
{\centerline{CDF: $H \rightarrow \gamma \gamma$ channel relative uncertainties (\%)}}
\vskip 0.099cm
\begin{ruledtabular}
\begin{tabular}{lcccc} \\
Channel & CC & CP & C$^\prime$C & C$^\prime$P \\ \hline
\bf{Signal Uncertainties :} & & & & \\
Luminosity & 6 & 6 & 6 & 6 \\
$\sigma_{ggH}/\sigma_{VH}/\sigma_{VBF}$ &
    14/7/5 & 14/7/5 & 14/7/5 & 14/7/5 \\
PDF & 5 & 2 & 5 & 2 \\
ISR/FSR & 3 & 4 & 2 & 5 \\
Energy Scale & 0.2 & 0.8 & 0.1 & 0.8 \\
Trigger Efficiency & 1.0 & 1.3 & 1.5 & 6.0 \\
$z$ Vertex & 0.07 & 0.07 & 0.07 & 0.07 \\
Conversion ID & -- & -- & 7 & 7 \\
Detector Material & 0.4 & 3.0 & 0.2 & 3.0 \\
Photon/Electron ID & 1.0 & 2.8 & 1.0 & 2.6 \\
Run Dependence & 3.0 & 2.5 & 1.5 & 2.0 \\
Data/MC Fits & 0.4 & 0.8 & 1.5 & 2.0 \\ \hline
\bf{Background Uncertainties :} & & & & \\
Fit Function & 2.8 & 0.9 & 6.1 & 3.3 \\
\end{tabular}
\end{ruledtabular}
\end{center}
\end{table}



\begin{table}[h]
\begin{center}
\caption{\label{tab:d0systgg} Systematic uncertainties on the signal and background contributions for D0's
$H\rightarrow \gamma \gamma$ channel. Systematic uncertainties for the Higgs signal shown in this table are
obtained for $m_H=125$ GeV/$c^2$.  Systematic uncertainties are listed by name; see the original references
for a detailed explanation of their meaning and on how they are derived.  Uncertainties are relative, in
percent, and are symmetric unless otherwise indicated.}
\vskip 0.1cm
{\centerline{D0: $H \rightarrow \gamma \gamma$ channel relative uncertainties (\%)}}
\vskip 0.099cm
\begin{ruledtabular}
\begin{tabular}{lcc}\\
Contribution &  ~~~Background~~~  & ~~~Signal~~~    \\
\hline
Luminosity~~~~                            &  6     &  6    \\
Acceptance                                &  --    &  2    \\
electron ID efficiency                    &  2     &  --   \\
electron track-match inefficiency         & 10     &  --   \\
Photon ID efficiency                      &  3     &   3   \\
Photon energy scale                       &  2     &   1   \\
Cross Section                             &  4     &  10   \\
Background subtraction                    &  15 &  -       \\
ONN Shape                              & 1-5  & -       \\
\end{tabular}
\end{ruledtabular}
\end{center}
\end{table}

\end{document}

Leftover citations

\bibitem{tevtop10}  The CDF and D0 Collaborations and the Tevatron Electroweak Working Group,
arXiv:1007.3178 [hep-ex] (2010), and arXiv:0903.2503~[hep-ex] (2009).



\bibitem{nnlo2}
K.~A.~Assamagan {\it et al.}  [Higgs Working Group Collaboration],
   ``The Higgs working group: Summary report 2003,''
  arXiv:hep-ph/0406152 (2004).

\bibitem{Brein}
O.~Brein, A.~Djouadi, and R.~Harlander, Phys. Lett. B {\bf 579}, 149 (2004).

\bibitem{Berger}
E. Berger and J. Campbell,  Phys. Rev. D {\bf 70} 073011 (2004).